\documentclass[twocolumn,twocolappendix]{aastex701}

\newcommand{\wltot}{\tilde{L}_{\rm tot}}
\newcommand{\lgal}{L_{\rm{gal}}}
\newcommand{\lcen}{L_{\rm{cen}}}
\newcommand{\lgrp}{L_{\rm{grp}}}
\newcommand{\Mh}{M_{\rm{h}}}
\newcommand{\fsat}{f_{\rm{sat}}}
\newcommand{\fsatb}{f_{\rm sat}^{\rm sf}}
\newcommand{\fsatr}{f_{\rm sat}^{\rm q}}
\newcommand{\dn}{\rm D_{\rm{n}}4000}
\newcommand{\psat}{P_{\rm sat}}

\newcommand{\wcen}{w_{\rm cen}}
\newcommand{\wcenc}{w_{\rm cen,c}}
\newcommand{\wcenr}{w_{\rm cen,r}}
\newcommand{\wcenb}{w_{\rm cen,b}}

\newcommand{\lsolhh}{{\rm L}_\odot h^{-2}}
\newcommand{\msol}{M_\odot}
\newcommand{\hmsol}{{\rm M}_\odot h^{-1}}
\newcommand{\mgal}{M_\ast}
\newcommand{\lsat}{L_{\rm sat}}
\newcommand{\lsatb}{L_{\rm sat}^{\rm sf}}
\newcommand{\lsatr}{L_{\rm sat}^{\rm q}}

\newcommand{\bsat}{B_{\rm sat}}
\newcommand{\bsatc}{B_{\rm sat,c}}
\newcommand{\bsatred}{B_{\rm sat,r}}
\newcommand{\bsatblue}{B_{\rm sat,b}}

\newcommand{\zphot}{z_{\rm phot}}
\newcommand{\zspec}{z_{\rm spec}}
\newcommand{\vmax}{V_{\rm max}}

\usepackage{graphicx}	
\usepackage{amsmath}	
\usepackage{comment} 
\usepackage{hyperref}
\usepackage{natbib}

\shorttitle{BGS Galaxy Groups}

\submitjournal{ApJ}

\begin{document}

\title[BGS Galaxy Groups]{Self Calibrated Galaxy Groups from the DESI DR1 Bright Galaxy Survey}




\begin{NoHyper}
    \correspondingauthor{I.~Williams}
\end{NoHyper}

\author[0000-0003-1552-6155, gname='Ian', sname='Williams']{I.~Williams}
\affiliation{Department of Physics and Center for Cosmology and Particle Physics, New York University, New York, NY 10003, USA}
\email[show]{ianwilliams@nyu.edu}

\author[0000-0003-3578-6149,gname='Jeremy', sname='Tinker']{J.~Tinker}
\affiliation{Department of Physics and Center for Cosmology and Particle Physics, New York University, New York, NY 10003, USA}
\email{jeremy.tinker@nyu.edu}

\author[0000-0002-3775-0484, gname='Kai', sname='Wang']{Kai~Wang}
\affiliation{Institute for Computational Cosmology, Department of Physics, Durham University, South Road, Durham DH1 3LE, UK}
\email{kai.wang@durham.ac.uk}

\author[gname='Jessica Nicole', sname='Aguilar']{J.~Aguilar}
\affiliation{Lawrence Berkeley National Laboratory, 1 Cyclotron Road, Berkeley, CA 94720, USA}
\email{jaguilar@lbl.gov}

\author[0000-0001-6098-7247, gname='Steven', sname='Ahlen']{S.~Ahlen}
\affiliation{Department of Physics, Boston University, 590 Commonwealth Avenue, Boston, MA 02215 USA}
\email{ahlen@bu.edu}

\author[0000-0003-0467-5438, gname='Florian', sname='Beutler']{F.~Beutler}
\affiliation{Institute for Astronomy, University of Edinburgh, Royal Observatory, Blackford Hill, Edinburgh EH9 3HJ, UK}
\email{florian.beutler@ed.ac.uk}

\author[0000-0001-9712-0006, gname='Davide', sname='Bianchi']{D.~Bianchi}
\affiliation{Dipartimento di Fisica ``Aldo Pontremoli'', Universit\`a degli Studi di Milano, Via Celoria 16, I-20133 Milano, Italy}
\affiliation{INAF-Osservatorio Astronomico di Brera, Via Brera 28, 20122 Milano, Italy}
\email{davide.bianchi1@unimi.it}

\author[gname='David', sname='Brooks']{D.~Brooks}
\affiliation{Department of Physics \& Astronomy, University College London, Gower Street, London, WC1E 6BT, UK}
\email{david.brooks@ucl.ac.uk}

\author[0000-0003-3044-5150, gname='Aurelio', sname='Carnero Rosell']{A.~Carnero Rosell}
\affiliation{Departamento de Astrof\'{\i}sica, Universidad de La Laguna (ULL), E-38206, La Laguna, Tenerife, Spain}
\affiliation{Instituto de Astrof\'{\i}sica de Canarias, C/ V\'{\i}a L\'{a}ctea, s/n, E-38205 La Laguna, Tenerife, Spain}
\email{acarnero@iac.es}

\author[0000-0001-7316-4573, gname='Francisco Javier', sname='Castander']{F.~J.~Castander}
\affiliation{Institut d'Estudis Espacials de Catalunya (IEEC), c/ Esteve Terradas 1, Edifici RDIT, Campus PMT-UPC, 08860 Castelldefels, Spain}
\affiliation{Institute of Space Sciences, ICE-CSIC, Campus UAB, Carrer de Can Magrans s/n, 08913 Bellaterra, Barcelona, Spain}
\email{fjc@ice.csic.es}

\author[gname='Todd', sname='Claybaugh']{T.~Claybaugh}
\affiliation{Lawrence Berkeley National Laboratory, 1 Cyclotron Road, Berkeley, CA 94720, USA}
\email{tmclaybaugh@lbl.gov}

\author[0000-0002-5954-7903, gname='Shaun', sname='Cole']{S.~Cole}
\affiliation{Institute for Computational Cosmology, Department of Physics, Durham University, South Road, Durham DH1 3LE, UK}
\email{shaun.cole@durham.ac.uk}

\author[0000-0002-1769-1640, gname='Axel ', sname='de la Macorra']{A.~de la Macorra}
\affiliation{Instituto de F\'{\i}sica, Universidad Nacional Aut\'{o}noma de M\'{e}xico,  Circuito de la Investigaci\'{o}n Cient\'{\i}fica, Ciudad Universitaria, Cd. de M\'{e}xico  C.~P.~04510,  M\'{e}xico}
\email{macorra@fisica.unam.mx}

\author[0000-0002-2890-3725, gname='Jaime E.', sname='Forero-Romero']{J.~E.~Forero-Romero}
\affiliation{Departamento de F\'isica, Universidad de los Andes, Cra. 1 No. 18A-10, Edificio Ip, CP 111711, Bogot\'a, Colombia}
\affiliation{Observatorio Astron\'omico, Universidad de los Andes, Cra. 1 No. 18A-10, Edificio H, CP 111711 Bogot\'a, Colombia}
\email{je.forero@uniandes.edu.co}

\author[0000-0001-9632-0815, gname='Enrique', sname='Gaztañaga']{E.~Gaztañaga}
\affiliation{Institut d'Estudis Espacials de Catalunya (IEEC), c/ Esteve Terradas 1, Edifici RDIT, Campus PMT-UPC, 08860 Castelldefels, Spain}
\affiliation{Institute of Cosmology and Gravitation, University of Portsmouth, Dennis Sciama Building, Portsmouth, PO1 3FX, UK}
\affiliation{Institute of Space Sciences, ICE-CSIC, Campus UAB, Carrer de Can Magrans s/n, 08913 Bellaterra, Barcelona, Spain}
\email{gaztanaga@gmail.com}

\author[0000-0003-3142-233X, gname='Satya ', sname='Gontcho A Gontcho']{Satya~{Gontcho A Gontcho}}
\affiliation{University of Virginia, Department of Astronomy, Charlottesville, VA 22904, USA}
\email{satya@virginia.edu}

\author[gname='Gaston', sname='Gutierrez']{G.~Gutierrez}
\affiliation{Fermi National Accelerator Laboratory, PO Box 500, Batavia, IL 60510, USA}
\email{gaston@fnal.gov}

\author[0000-0003-1197-0902, gname='ChangHoon', sname='Hahn']{C.~Hahn}
\affiliation{Department of Astronomy, University of Texas at Austin, 2515 Speedway, TX 78712, USA}
\email{changhoon.hahn@utexas.edu}

\author[0000-0002-6024-466X, gname='Mustapha', sname='Ishak']{M.~Ishak}
\affiliation{Department of Physics, The University of Texas at Dallas, 800 W. Campbell Rd., Richardson, TX 75080, USA}
\email{mishak@utdallas.edu}

\author[0000-0002-0000-2394, gname='Stephanie', sname='Juneau']{S.~Juneau}
\affiliation{NSF NOIRLab, 950 N. Cherry Ave., Tucson, AZ 85719, USA}
\email{stephanie.juneau@noirlab.edu}

\author[gname='Robert', sname='Kehoe']{R.~Kehoe}
\affiliation{Department of Physics, Southern Methodist University, 3215 Daniel Avenue, Dallas, TX 75275, USA}
\email{kehoe@physics.smu.edu}

\author[0000-0001-6356-7424, gname='Anthony', sname='Kremin']{A.~Kremin}
\affiliation{Lawrence Berkeley National Laboratory, 1 Cyclotron Road, Berkeley, CA 94720, USA}
\email{akremin@lbl.gov}

\author[0000-0002-6731-9329, gname='Claire', sname='Lamman']{C.~Lamman}
\affiliation{The Ohio State University, Columbus, 43210 OH, USA}
\email{lamman.1@osu.edu}

\author[0000-0003-1838-8528, gname='Martin', sname='Landriau']{M.~Landriau}
\affiliation{Lawrence Berkeley National Laboratory, 1 Cyclotron Road, Berkeley, CA 94720, USA}
\email{mlandriau@lbl.gov}

\author[0000-0001-7178-8868, gname='Laurent', sname='Le Guillou']{L.~Le~Guillou}
\affiliation{Sorbonne Universit\'{e}, CNRS/IN2P3, Laboratoire de Physique Nucl\'{e}aire et de Hautes Energies (LPNHE), FR-75005 Paris, France}
\email{llg@lpnhe.in2p3.fr}

\author[0000-0002-3677-3617, gname='Alexie', sname='Leauthaud']{A.~Leauthaud}
\affiliation{Department of Astronomy and Astrophysics, UCO/Lick Observatory, University of California, 1156 High Street, Santa Cruz, CA 95064, USA}
\affiliation{Department of Astronomy and Astrophysics, University of California, Santa Cruz, 1156 High Street, Santa Cruz, CA 95065, USA}
\email{alexie@ucsc.edu}

\author[0000-0003-4962-8934, gname='Marc', sname='Manera']{M.~Manera}
\affiliation{Departament de F\'{i}sica, Serra H\'{u}nter, Universitat Aut\`{o}noma de Barcelona, 08193 Bellaterra (Barcelona), Spain}
\affiliation{Institut de F\'{i}sica d’Altes Energies (IFAE), The Barcelona Institute of Science and Technology, Edifici Cn, Campus UAB, 08193, Bellaterra (Barcelona), Spain}
\email{mmanera@ifae.es}

\author[0000-0002-1125-7384, gname='Aaron', sname='Meisner']{A.~Meisner}
\affiliation{NSF NOIRLab, 950 N. Cherry Ave., Tucson, AZ 85719, USA}
\email{aaron.meisner@noirlab.edu}

\author[gname='Ramon', sname='Miquel']{R.~Miquel}
\affiliation{Instituci\'{o} Catalana de Recerca i Estudis Avan\c{c}ats, Passeig de Llu\'{\i}s Companys, 23, 08010 Barcelona, Spain}
\affiliation{Institut de F\'{i}sica d’Altes Energies (IFAE), The Barcelona Institute of Science and Technology, Edifici Cn, Campus UAB, 08193, Bellaterra (Barcelona), Spain}
\email{rmiquel@ifae.es}

\author[0000-0002-2733-4559, gname='John', sname='Moustakas']{J.~Moustakas}
\affiliation{Department of Physics and Astronomy, Siena University, 515 Loudon Road, Loudonville, NY 12211, USA}
\email{jmoustakas@siena.edu}

\author[0000-0001-9070-3102, gname='Seshadri', sname='Nadathur']{S.~Nadathur}
\affiliation{Institute of Cosmology and Gravitation, University of Portsmouth, Dennis Sciama Building, Portsmouth, PO1 3FX, UK}
\email{seshadri.nadathur@port.ac.uk}

\author[0000-0002-4637-2868, gname='Enrique', sname='Paillas']{E.~Paillas}
\affiliation{Instituto de Estudios Astrof\'isicos, Facultad de Ingenier\'ia y Ciencias, Universidad Diego Portales, Av. Ej\'ercito Libertador 441, Santiago, Chile}
\email{enrique.paillas@udp.cl}

\author[0000-0002-0644-5727, gname='Will', sname='Percival']{W.~J.~Percival}
\affiliation{Department of Physics and Astronomy, University of Waterloo, 200 University Ave W, Waterloo, ON N2L 3G1, Canada}
\affiliation{Perimeter Institute for Theoretical Physics, 31 Caroline St. North, Waterloo, ON N2L 2Y5, Canada}
\affiliation{Waterloo Centre for Astrophysics, University of Waterloo, 200 University Ave W, Waterloo, ON N2L 3G1, Canada}
\email{will.percival@uwaterloo.ca}

\author[0000-0001-7145-8674, gname='Francisco', sname='Prada']{F.~Prada}
\affiliation{Instituto de Astrof\'{i}sica de Andaluc\'{i}a (CSIC), Glorieta de la Astronom\'{i}a, s/n, E-18008 Granada, Spain}
\email{fprada@iaa.es}

\author[0000-0001-6979-0125, gname='Ignasi', sname='Pérez-Ràfols']{I.~P\'erez-R\`afols}
\affiliation{Departament de F\'isica, EEBE, Universitat Polit\`ecnica de Catalunya, c/Eduard Maristany 10, 08930 Barcelona, Spain}
\email{ignasi.perez.rafols@upc.edu}

\author[0000-0002-3500-6635, gname='Corentin', sname='Ravoux']{C.~Ravoux}
\affiliation{Universit\'{e} Clermont-Auvergne, CNRS, LPCA, 63000 Clermont-Ferrand, France}
\email{corentin.ravoux@clermont.in2p3.fr}

\author[0000-0002-7522-9083, gname='Ashley J.', sname='Ross']{A.~J.~Ross}
\affiliation{Center for Cosmology and AstroParticle Physics, The Ohio State University, 191 West Woodruff Avenue, Columbus, OH 43210, USA}
\affiliation{Department of Astronomy, The Ohio State University, 4055 McPherson Laboratory, 140 W 18th Avenue, Columbus, OH 43210, USA}
\affiliation{The Ohio State University, Columbus, 43210 OH, USA}
\email{ross.1333@osu.edu}

\author[gname='Graziano', sname='Rossi']{G.~Rossi}
\affiliation{Department of Physics and Astronomy, Sejong University, 209 Neungdong-ro, Gwangjin-gu, Seoul 05006, Republic of Korea}
\email{graziano@sejong.ac.kr}

\author[0000-0002-0394-0896, gname='Rossana', sname='Ruggeri']{R.~Ruggeri}
\affiliation{Queensland University of Technology,  School of Chemistry \& Physics, George St, Brisbane 4001, Australia}
\email{rossana.ruggeri@qut.edu.au}

\author[0000-0002-1609-5687, gname='Lado', sname='Samushia']{L.~Samushia}
\affiliation{Abastumani Astrophysical Observatory, Tbilisi, GE-0179, Georgia}
\affiliation{Department of Physics, Kansas State University, 116 Cardwell Hall, Manhattan, KS 66506, USA}
\email{lado@phys.ksu.edu}

\author[0000-0002-9646-8198, gname='Eusebio', sname='Sanchez']{E.~Sanchez}
\affiliation{CIEMAT, Avenida Complutense 40, E-28040 Madrid, Spain}
\email{eusebio.sanchez@ciemat.es}

\author[0000-0002-0408-5633, gname='Christoph', sname='Saulder']{C.~Saulder}
\affiliation{Max Planck Institute for Extraterrestrial Physics, Gie\ss enbachstra\ss e 1, 85748 Garching, Germany}
\email{csaulder@mpe.mpg.de}

\author[gname='David', sname='Schlegel']{D.~Schlegel}
\affiliation{Lawrence Berkeley National Laboratory, 1 Cyclotron Road, Berkeley, CA 94720, USA}
\email{djschlegel@lbl.gov}

\author[0000-0002-3461-0320, gname='Joseph Harry', sname='Silber']{J.~Silber}
\affiliation{Lawrence Berkeley National Laboratory, 1 Cyclotron Road, Berkeley, CA 94720, USA}
\email{jhsilber@lbl.gov}

\author[0000-0002-2949-2155, gname='Małgorzata', sname='Siudek']{M.~Siudek}
\affiliation{Institute of Space Sciences, ICE-CSIC, Campus UAB, Carrer de Can Magrans s/n, 08913 Bellaterra, Barcelona, Spain}
\affiliation{Instituto de Astrof\'{\i}sica de Canarias, C/ V\'{\i}a L\'{a}ctea, s/n, E-38205 La Laguna, Tenerife, Spain}
\email{msiudek@iac.es}

\author[0000-0003-1704-0781, gname='Gregory', sname='Tarlé']{G.~Tarl\'{e}}
\affiliation{University of Michigan, 500 S. State Street, Ann Arbor, MI 48109, USA}
\email{gtarle@umich.edu}

\author[0009-0001-2732-8431, gname='Rajeev', sname='Vaisakh']{R.~Vaisakh}
\affiliation{Department of Physics, Southern Methodist University, 3215 Daniel Avenue, Dallas, TX 75275, USA}
\email{vvaisakh@smu.edu}

\author[gname='Benjamin Alan', sname='Weaver']{B.~A.~Weaver}
\affiliation{NSF NOIRLab, 950 N. Cherry Ave., Tucson, AZ 85719, USA}
\email{benjamin.weaver@noirlab.edu}


\begin{abstract}
We present a galaxy group catalog built from Data Release 1 of the Dark Energy Spectroscopic Instrument (DESI) Bright Galaxy Survey, using the Self Calibrated Group Finder. The algorithm uses clustering data from the observed sample to calibrate the Group Finder’s free parameters, rather than simulations. The catalog contains 2,229,741 galaxies from $\sim2680~\mathrm{deg}^2$ of sky with 3-pass coverage and $r$-band fluxes $<19.5$. The catalog classifies all galaxies in the flux-limited sample as centrals or satellites and estimates host dark matter halo masses. We show the satellite fraction $\fsat$, luminosity halo mass relation, stellar halo mass relation and halo occupation distributions (HODs) from the catalog. We find these non-parametric HODs can be fit by standard HOD models, with satellite power-law index $\alpha$ values around 0.75 for star-forming galaxies and around 1.1 for quiescent objects. In agreement with recent results, we find that the quenched fraction of central galaxies depends mostly on halo mass and is nearly independent of central galaxy stellar mass for galaxies more massive than $10^{10.3}$ $\hmsol$ and halos more massive than $10^{12.2}$ $\hmsol$. In an appendix, we introduce a new Hybrid Nearest Neighbor (HNN) technique to mitigate systematics from fiber incompleteness. HNN uses photometric redshifts and multiple on-sky neighbors to assign redshifts to unobserved targets. Calibrated on DESI Survey Validation 3 data, our technique yields correct redshifts $\sim 50\%$ of the time and produces unbiased halo occupation statistics.
\end{abstract}



\section{Introduction}
\label{sec:intro}

In the modern cosmological paradigm, galaxies form in the centers of collapsed dark matter halos. Both galaxies and halos then grow via mergers and accretion into the diverse masses and morphologies we observe today. Because of these intertwined histories, the \textit{galaxy-halo connection} is a critical piece of any theory of galaxy formation, as well as a tool to improve our ability to probe cosmology from galaxy surveys. 

Many approaches have been used to explore the galaxy-halo connection (see \citealt{wechsler_tinker:18} for a review). Hydrodynamic simulations offer a theory-first model, by attempting to resolve or otherwise include as much physics as possible. On the other end of the spectrum is data-driven modeling, where the most common model is the Halo Occupancy Distribution (HOD), a probabilistic statement about which galaxies occupy which halos \citep{roman_etal:01, zheng_etal:05, zheng_etal:07, zheng_etal:09, zehavi_etal:11}. 

Here we take an intermediate approach: halo-based galaxy group finding. Galaxy group finders attempt to determine which galaxies in a sample are in bound gravitational systems together. When focused on massive clusters, methods such as MaxBCG \citep{koester_etal:07_methods,koester_etal:07} and redMaPPer \citep{Rykoff_etal:14,Rykoff_etal:16,to_etal:20} have been applied to photometric surveys, enabling a variety of applications. Percolation-based methods such as friend-of-friends \citep{huchra_geller:82} have also been applied to many galaxy samples, including the Sloan Digital Sky Survey \citep{berlind_etal:06_groups} and the Galaxy and Mass Assembly survey \citep{robotham_etal:11}. In contrast, \textit{halo-based} group finders employ our knowledge of the halo mass function and other halo properties to aid in group finding. As with most HODs, halo-based group finders operate in a central-satellite paradigm, asserting that one galaxy sits in the center of the dark matter potential well, and other group members are satellites residing in subhalos. As a result, halo-based group finders provide group membership \textit{and} an assignment of dark matter halos to galaxies, making them models of the galaxy-halo connection across several decades of halo mass. Compared to the HOD, group finders offer a non-parametric model, avoiding the need to specify a fixed functional form. Instead of using a 1-dimensional summary statistic to capture the relation (as in HODs), group catalogs use the actual observed distribution of galaxies in specifying the galaxy-halo connection.

The first halo-based group finder was introduced in \cite{yang_etal:05}. The method made use of the \textit{abundance matching} technique, ordering galaxies/groups by luminosity (or stellar mass) and assigning halo masses from an assumed halo mass function. Many group finder variations have been developed; see \cite{campbell_etal:15} for a review. Other variations include a 2nd-turnaround radius based group finder \citep{tully:15}, a halo-based method for processing the entire DR8 Legacy Survey using mostly photometric redshifts \citep{Yang.ExtendedGF.21}, and a neural-network based approach to group finding \citep{Ma.2025}. 

One of the advantages of the \cite{yang_etal:05} algorithm is that there is only one free parameter: $B_{\rm sat}$, a threshold to determine if a galaxy is a member of a nearby group. In \cite{yang_etal:05} this was calibrated by mock catalogs to balance completeness and purity of the groups. However, this low-parameter approach led to several issues: 1) systematic misclassifications of quiescent satellite galaxies, 2) an inability to differentiate the halo masses for subclasses of galaxies at fixed group luminosity, and 3) insufficient scatter in the halo mass at fixed group luminosity. By introducing additional free parameters specific to subpopulations of galaxies and calibrating on observed data, the Self Calibrated Group Finder (SCGF) is able to deal with these issues (\citealt{Tinker_Alg}, hereafter Paper I). In Paper II \citep{Tinker_SDSS}, the method was initially applied to the Sloan Digital Sky Survey (SDSS) NYU Value-Added Galactic Catalog \citep{blanton_etal:05_vagc}. Here we apply to the next-generation dataset: the Dark Energy Spectroscopic Instrument (DESI) Bright Galaxy Survey (BGS). 

As the first Stage-IV spectroscopic galaxy survey, DESI is providing the largest redshift-space map of the nearby universe to date. BGS, which we use here, is a flux-limited survey with an $r$-band magnitude limit of 19.5, almost two magnitudes deeper than SDSS, achieving a median redshift of $z\sim0.2$ across the final survey footprint of $\sim 17,000 \rm~deg^2$ \citep{Hahn_2023}. However, due to the limited patrol radius of the fiber positioners, the program is unable visit every target in dense regions of the sky, creating a non-trivial \textit{fiber incompleteness} systematic. Here we develop a new methodology to handle this important systematic, which is stronger than in SDSS. Our goal is to produce a galaxy group catalog capable of exploring the galaxy-halo connection into the fainter regimes made possible by DESI BGS. 

Throughout this work we adopt the DESI fiducial cosmology, from Planck 2018: flat $\Lambda$CDM, with $\Omega_m=0.315$ and $H_0=100h~\rm km~s^{-1}Mpc^{-1}$ \citep{Planck:2018vyg}. This assumed cosmology is used for distance-redshift conversions, computing the halo mass function, and inherent in the halo definitions. For conciseness and visual clarity we will often write absolute magnitudes without the $-5\log h$ adjustment for these units, but it is always present unless otherwise noted.

This report is organized as follows: In \S\ref{sec:data} we describe the data used to construct the catalog, including details on how the issue of fiber incompleteness is dealt with in our catalog. In \S\ref{sec:caldata} we present new measurements that we use to calibrate free parameters in the group finder. In \S\ref{sec:methods} we review the algorithm used for construction the catalog. In \S\ref{sec:results} we present our results as a few key statistics from our catalog. We also compare the results to a mock, and comment on future work. 

\section{Data}
\label{sec:data}

DESI is a robotic, fiber-fed, highly multiplexed spectroscopic surveyor operating at the 4-m Mayall telescope at Kitt Peak National Observatory \citep{Abareshi_2022}. DESI is conducting an eight-year survey of about $17,000\deg^2$ \citep{SurveyOps.Schlafly.2023}, leading to 63 million spectroscopically confirmed galaxies and quasars \citep{Spectro.Pipeline.Guy.2023}, much higher than the initial forecast of 39 million. This is made possible because DESI can obtain simultaneous spectra of almost 5000 targets over a $3^\circ$ field every pointing. The first data release, which this work is built on, is public, containing over 18 million unique objects \citep{DESI2024.I.DR1}.  Cosmological inference from both the first \citep{DESI2024.VI.KP7A, DESI2024.VII.KP7B} and second \citep{DESI.DR2.DR2, DESI.DR2.BAO.cosmo} data releases hinted at the possibility of an evolving dark energy equation of state, especially when combined with supernovae data. For galaxy science, DESI provides a enormous new dataset to enable the next generation of studies. For a general overview of DESI science, see \cite{Snowmass2013.Levi,DESI2016a.Science}.

DESI BGS targets are defined and selected from the photometric Legacy Survey DR9 \citep{LS.Overview.Dey.2019}, as described in \cite{Hahn_2023}. We use the following photometric and meta properties from this catalog, as determined by \textit{The Tractor} software \citep{Tractor.Lang.2016}: $r$ and $g$ band fluxes, right ascension and declination, \texttt{FRACFLUX}\_\texttt{I} ($I=G,R,Z$),  \texttt{TARGETID}, \texttt{PHOTSYS}, \texttt{MASKBITS}, \texttt{SURVEY}, \texttt{PROGRAM} and \texttt{MORPHTYPE}. We use the DESI BGS DR1 spectroscopic targets in \citep{DESI2024.I.DR1}. We use the BGS ANY ``Full" catalog\footnote{Available at: https://data.desi.lbl.gov/public/dr1/survey/ catalogs/dr1/LSS/iron/LSScats/v1.5/BGS\_ANY\_full.dat.fits}, which contains galaxy BGS targets within the DR1 footprint (see \citealt{LSSCatalogs.Ross.2025} for details). The  \texttt{MW\_TRANSMISSION\_R} and \texttt{MW\_TRANSMISSION\_G} columns in this catalog is used to correct the fluxes for Galactic extinction. In addition to the redshifts $z$ as extracted by \texttt{Redrock} \citep{Redrock.Bailey.2024, Spectro.Pipeline.Guy.2023}, we use some derived galaxy properties from the \texttt{FastSpecFit} Value-Added Catalog \citep{fastspecfit}: the $g$ and $r$ band $K$-corrected absolute magnitudes $M_g^{0.1}$ and $M_r^{0.1}$, \texttt{$\dn\_MODEL$}, and the stellar masses $M_\star$\footnote{If we find disagreement between the redshifts listed in \texttt{FastSpecFit} and the DESI LSS catalogs, we ignore the \texttt{FastSpecFit} derived quantities and fill these missing values using the same methods as our unobserved galaxy handling.}. No evolution corrections have been applied to magnitudes appearing anywhere in this work. 

For information on the DESI instrument, see \cite{DESI2016b.Instr,DESI2022.KP1.Instr, Corrector.Miller.2023, FocalPlane.Silber.2023, FiberSystem.Poppett.2024}. For details on the imaging, pipelines and software, see \cite{BASS.Zou.2017,LS.Overview.Dey.2019,TS.Pipeline.Myers.2023,Redrock.Bailey.2024,RedrockQSO.Brodzeller.2023,AstrometricCalib.Kent.2023,SGA.Moustakas.2023,LSSCatalogs.Ross.2025}. For information on the DESI survey operations see \cite{SurveyOps.Schlafly.2023}.

In our handing of fiber incompleteness, \S\ref{sec:fiberincompleteness}, we also use photometric redshifts derived from the Legacy Survey DR9 and DR10 (\cite{Zhou_2023}). In order to validate the method, we use data the 1\% Survey, part of DESI's survey validation program. For more details, see \cite{MWSPrelim.AllendePrieto.2020,BGSPrelim.RuizMacias.2020,LGRPrelim.Zhou.2020,ELGPrelim.Raichoor.2020,QSOPrelim.Yeche.2020,VIGalaxies.Lan.2023,VIQSO.Alexander.2023,MWS.TS.Cooper.2023,BGS.TS.Hahn.2023,LRG.TS.Zhou.2023,ELG.TS.Raichoor.2023,QSO.TS.Chaussidon.2023,DESI2023a.KP1.SV,DESI2023b.KP1.EDR}.

\subsection{Sample Definition}
\label{sec:sampledef}
\begin{figure*}
    \centering
    \includegraphics[width=1.0\linewidth]{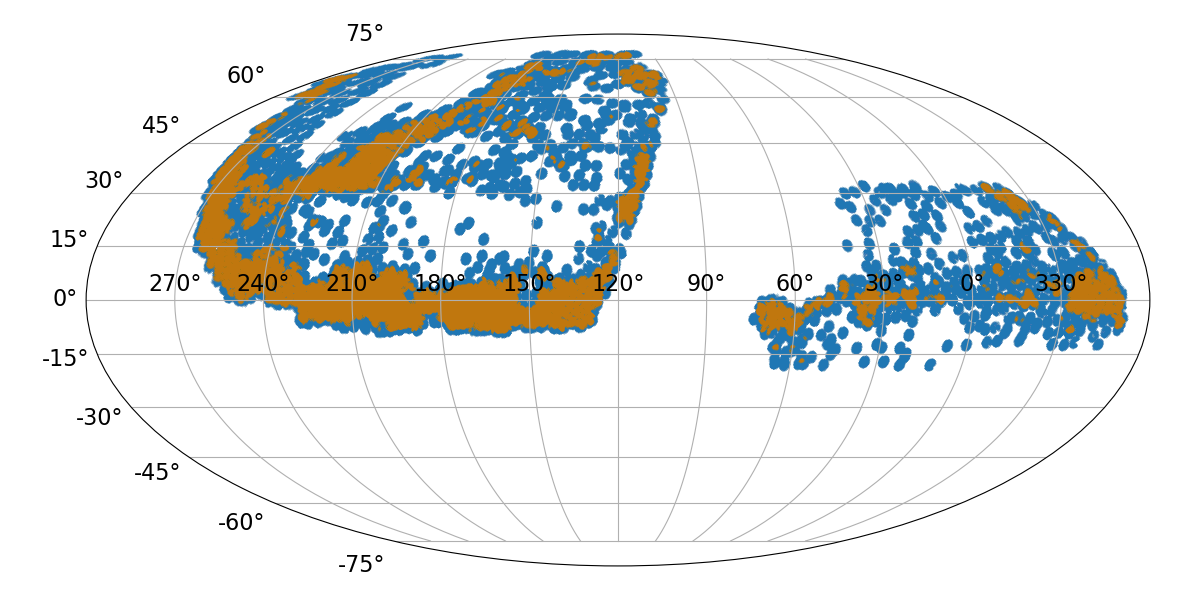}
    \caption{\textbf{Footprint.} Shown in blue is the DESI DR1 footprint. In this catalog, we only utilize regions where 3+ passes have been made. Those regions are displayed in orange, totaling $\sim 2680~\rm deg^2$.}
    \label{fig:footprint}
\end{figure*}

Although the footprint of DR1 is $>9000~\rm deg^2$, we restrict our catalog to regions where 3+ passes have been made (see Fig.~\ref{fig:footprint}). Thus to include a target in our catalog, it must be with $97.7'$ of 3+ DR1 tile centers\footnote{The quantity \texttt{NTILE} from the DESI LSS Catalogs specifies how many times a galaxy could have been targeted by a fiber. Our method of defining the footprint is \textit{not} equivalent to cutting to \texttt{NTILE} $\ge 3$}. This cut reduces this footprint to $2680~\rm deg^2$. We then cut to the targets in the BRIGHT sample (which applies a magnitude limit of $r<19.5$ in the DECaLS region and $r<19.54$ in the BASS/MzLS region), removing the lower completeness FAINT sample. These bulk selections bring the galaxy count to 2,286,412.

The DESI photometric and spectroscopic pipelines already remove many extended objects that are not galaxies, but we also apply several quality cuts. In particular, because our group catalog includes $\sim20\%$ targets without spectra, spurious galaxy targets that would have been cleaned up once a spectra is taken would instead be included in our catalog. We mitigate this in two ways.

First, we examine targets that fall within the ellipses defined by the Siena Galaxy Atlas (SGA, \citealt{SGA.Moustakas.2023}). We match the ellipse centers to this selection of targets using 3" and keep those targets. Other targets in the SGA ellipses are removed if they have an apparent $g-r<0.65$. This procedure keeps the fiber placed on the central part of the SGA galaxy, but removes blue/white targets elsewhere within the ellipse, which are usually HII regions in a high angular-size galaxies. From manual inspection, this step occasionally removes some faint blue galaxies, but mostly cleans up some badly shredded spiral galaxies. This removes 10,718 targets. After a few additional cuts from visual inspection, the count is 2,275,683.

The second cut aims to deal with overlapping targets where accurately determining of the nature of the objects is difficult. We remove targets with \texttt{fracflux} $>0.65$ in 2 or more of the \textit{grz} bands. From visual inspection, this cut removes some shredded targets too small for SGA, but mostly removes overlapping targets (star-galaxy or galaxy-galaxy) much of which appear unrelated. The removal of uncorrelated object pairs should induce no systematic effect on the resultant population; it is random down-sampling. We assume the removal of legitimate merging galaxies and small shredded targets should have a subdominant effect on the population of galaxies compared to the effects of fiber incompleteness (see \S\ref{sec:fiberincompleteness}). This step removes 28,954 targets.

After these cuts, we have 2,246,729. remaining galaxies. We then remove all unobserved targets that were photometrically classified with \texttt{MORPHTYPE} of `PSF', which are typically quasars or stars that were still selected as BGS targets. We then demand that spectroscopically observed targets be classified as galaxies by \texttt{Redrock}. After these steps, 2,232,260 remain. Note that because the redshift fitting code occasionally produces multiple good fits as expressed by the $\Delta \chi^2$ (\texttt{DELTACHI2}) property, we treat spectroscopic galaxies with $\Delta \chi^2<40$ as if no spectra were taken. With this taken into account, we find that 461,409 (20.7\%) of the remaining objects are missing spectra. These ``lost galaxies'' are assigned redshifts from a procedure described in \S\ref{sec:fiberincompleteness} and Appendix \ref{apdx:lostgals}. Afterwards, we apply a redshift cut to $0.001 < z < 0.5$, and also remove galaxies with luminosities greater than $4\times10^{11} \lsolhh$ ($M_r^{0.1}\lesssim-24.5$), as most of these are due to bad redshift assignments from our lost galaxy handling or the spectral fitting. After these last cuts, the final catalog has 2,229,741 galaxies.

\subsection{Identifying Quiescent Galaxies}
\label{sec:quiescent}
Quiescent galaxies exhibit many differences from star-forming ones in their observational properties, environments, and their galaxy-halo connection \citep{baldry_etal:04, baldry_etal:06, balogh_etal:04, blanton_moustakas:09, skibba_sheth:09, zehavi_etal:11, hearin_watson:13, Tinker_SDSS}. As such, our group finding methods have different parameters for handling these two populations, as will be described in \S\ref{sec:groupfinding}. We classify galaxies as quiescent or star-forming based on a luminosity dependent cut in the spectroscopic quantity $\dn$, similar to Paper II (we specifically use $\dn\_MODEL$ from \texttt{FastSpecFit}). This approach has be shown to identify dust-reddened, star-forming galaxies better than purely photometric cuts, which are prone to asymmetric biases for edge-on dusty galaxies. Spectroscopic quantities like $\dn$ or Star-Formation Rates are generally noisier than in SDSS; we compare overlapping samples in more detail in Appendix~\ref{apdx:sdss_compare}. We find that derived Star-Formation Rates have too large uncertainties to easily incorporate into the classification.

\begin{figure*}s
    \centering
    \includegraphics[width=1.0\linewidth]{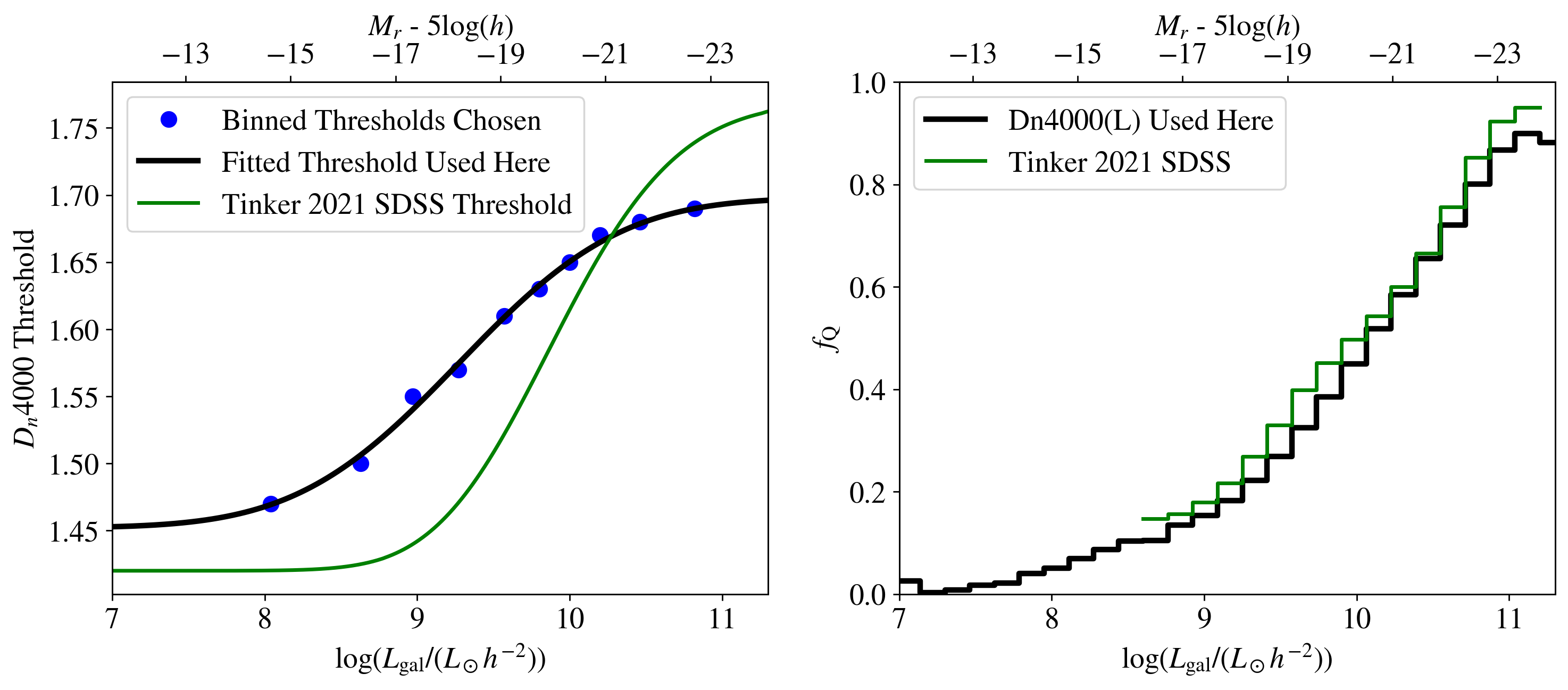}
    \caption{\textbf{Quiescent Classification and Fraction.} \textit{Left}: Galaxies with $\dn$ values greater than the black line are considered quiescent, unless their $M_g^{0.1}-M_r^{0.1}$ color is $<0.65$. The blue dots show the threshold values chosen from inspecting the bimodality of $\dn$ in $\lgal$ bins, which were then fit to the black line. The green line shows the threshold used in \protect Paper II. \textit{Right}: The resultant quiescent fraction of spectroscopic galaxies in the catalogs. The results from Paper II are shown in green. The green curves in both plots were modified using the SDSS to BGS magnitude conversion function shown in Appendix \ref{apdx:sdss_compare} to aid in the comparison.}
    \label{fig:quiescent_classification}
\end{figure*}

We also impose a hard cut: if the color $M_g^{0.1}-M_r^{0.1} < 0.65$, well into the blue side of the bimodality irrespective of the luminosity $\lgal$, we classify the galaxy as star-forming. Otherwise we use the $\lgal$ dependent cut on $\dn$ as shown in black on the left figure of Fig.~\ref{fig:quiescent_classification}. The cutoff values were determined by examining histograms of $\dn$ values in 10 $\lgal$ bins spanning $8 <\log \lgal < 11$. Although we used 2-component Gaussian Mixture Models, the fits were too poor to utilize and so the values were chosen from visual inspection. Our fit appears flatter than the SDSS version of this curve used in Paper II, driven by the lower signal-to-noise of DESI spectra. Overall these choices identify somewhat less quiescent galaxies than in Paper II at most luminosities; the precise $\lgal$ dependence can be seen in on the right-hand side of Fig.~\ref{fig:quiescent_classification}.  The functional form of the threshold $D_{\rm crit}$ is:

\begin{equation}
  \label{eq:quiescent_threshold}
  D_{\rm crit} = 1.452 + 0.123\left[1+{\rm erf}\left(\frac{\log\lgal-9.275}{1.186}\right)\right].
\end{equation}

The right-hand side of Fig.~\ref{fig:quiescent_classification} shows the quiescent fraction $f_Q(L)$ of the spectroscopic samples from both catalogs. We find using a $L$-dependent $g-r$ cut produces a similar $f_Q$ curve. At the low luminosity end, the quiescent fraction is very low and the bimodality in $\dn$ is quite weak, making it difficult to determine a good cutoff. In this regime there is likely increased contamination, which is important for the smaller quiescent sample. For this reason, we do not separate the quiescent from star-formers in our projected clustering measurements in the fainter bins, as described later in \S\ref{sec:calibration-clustering}.  

\subsection{Fiber Incompleteness}
\label{sec:fiberincompleteness}

The fiducial 5-year DESI BGS design achieves a fiber assignment efficiency of $\sim80\%$ by the end of survey, but has been improved to $\sim85\%$ thanks to extensions and improvements \citep{Hahn_2023}\footnote{Note that the fiber incompleteness related statistics in \cite{Hahn_2023} are not directly comparable to this work for the reasons described in \S\ref{sec:sampledef}.}. The unobserved targets, which we refer to as ``lost galaxies'', are not a random sample of BGS targets, but are preferentially located in overdense regions, carrying with them significant selection effects. As such, dropping them from the analysis would induce major systematic effects on group finding. In Appendix~\ref{apdx:lostgals} we present a study of fiber incompleteness induced systematics, and describe the method we employ for assigning redshifts to the lost galaxies.

Before using our method, we take the unobserved BGS targets and match at 1'' with the NYU Value-Added Galactic Catalog (NYU-VAGC) to borrow spectroscopic redshifts where available \citep{blanton_etal:05_vagc}. With this cross-catalog matching, we provide redshifts to 21,530 (5.1\%) of the lost galaxies. We also copy $M_{\star}$ and whether the galaxy was classified as quiescent or star-forming from the NYU-VAGC catalog. The remaining lost galaxies receive redshifts from the Hybrid Nearest Neighbor technique we have developed, described in Appendix~\ref{apdx:lostgals}. The technique utilizes on-sky neighbor information and photometric redshift estimates \citep{2021MNRAS.501.3309Z} to assign redshifts to all lost galaxies.

After redshifts are assigned, other spectroscopic quantities must be assigned. To calculate absolute magnitudes, we need $K$-corrections, which are usually calculated per-object in \texttt{FastSpecFit} using spectral information. To get $r$ and $g$ band $K$-corrections for lost galaxies, we use the $k$-nearest neighbor technique. We build a kd-tree with a data vector of ($M_r, 10\times(g-r)$, $50\times z)$, where $M_r$ is the absolute $r$-band magnitude (with no $K$-correction). We lookup the nearest $k=50$ points in this space and choose the one closest to the mean ($k_g,k_r$) values from those 50 points. With these values $M_r^{0.1}$ and $M_g^{0.1}$ can be calculated. To provide $\dn$ and $\mgal$, we adopt the same approach, building a kd-tree with a data vector of ($M_r^{0.1}$, 5*$(M_g^{0.1}-M_r^{0.1})$). The look-ups also use $k=50$, choosing the point nearest the mean of ($5\times\dn$, $\mgal$). All of these choices were calibrated to yield the most accurate look-ups when tested on withheld data.

\section{Calibration Data}
\label{sec:caldata}

The SCGF has ten free parameters as described in \S\ref{sec:methods}. As with the application to SDSS in Paper II, we use a combination of projected clustering and total satellite luminosity ($\lsat$) measurements to constrain these parameters and produce a better model of how quiescent vs. star-forming galaxies occupy halos.

\subsection{Projected Clustering}
\label{sec:calibration-clustering}

Here we present new projected clustering measurements from DESI BGS DR1. We measure clustering in $r$-band magnitude bins split by quiescent vs. star-forming, using the spectroscopic BGS BRIGHT ($r<19.5$) DR1 sample. We construct volume limited samples using the information shown in Table~\ref{tab:clustering}. We follow common DESI techniques for clustering measurements, using angular up-weighting, pair-inverse probabilities (PIP) weightings, redshift failure weights, and linear imagining systematics weights fit separately each subsample \citep{LSSCatalogs.Ross.2025}. Error bars were estimated using the jackknife technique with 128 angular-space jackknives. The $\pi_{\rm max}$ is 40 Mpc/$h$. 

The results are shown in Fig.~\ref{fig:wp_fit}, along with the group catalog results described later. Note that in the faintest two bins we do not separate quiescent from star-forming galaxies. This is due to low quiescent galaxy counts and larger uncertainties in identifying quiescent galaxies. In these faintest two bins, which only extend to $z<0.02595$ and 0.04067, the Local Void may be lowering the amplitude in the 2-halo regime, which may be the cause of the dip at $\sim$10 Mpc~$h^{-1}$ \citep{Abbas_sheth:07}. Fortunately, this appears to have no effect on the inferred group finder parameters.

To check the sensitivity of these small-scale measurements to the weighting choices made, we calculated the clustering with several alternative weighting schemes. We measured the projected clustering restricted to the 3-pass footprint using the same methods, but without angular up-weighting, which is designed to deal with pairs of galaxies with a 0\% probability to observe. We also made the measurements without imaging systematics weights. We find agreement across these choices. 

\begin{table}
    \centering
    \begin{tabular}{c|c|c|c}
        $M_r^{0.1}-5\log(h)$ & $z_{\rm max}$ & $N_{\rm gal}^{\rm sf}$ & $N_{\rm gal}^{\rm q}$ \\
        \hline
        \hline
        -23 to -22 & 0.49523 & 60,660 & 194,543 \\
        -22 to -21 & 0.33694 & 385,685 & 628,733  \\
        -21 to -20 & 0.22620 & 683,535 & 615,861  \\
        -20 to -19 & 0.14977 & 548,905 & 274,748 \\
        -19 to -18 & 0.09792 & 274,503 & 76,485  \\
        -18 to -17 & 0.06336 & 121,189 & 20,767  \\
        -17 to -16 & 0.04067 & 48,717 & 5,879  \\
        -16 to -15 & 0.02595 & 19,768 & 1,484 
    \end{tabular}
    \caption{Description of the volume limited samples used in this work, which are derived from the BGS BRIGHT $r<19.5$ flux limit. Due to low quiescent galaxy counts in the faintest two bins, we work with combined samples for those two bins.}
    \label{tab:clustering}
\end{table}

\subsection{The Total Satellite Luminosity Statistic, \texorpdfstring{$\lsat$}{Lsat}}
\label{sec:calibration-lsat}

The group finder operates in a halo model paradigm where groups are composed of a central galaxy located at the center of the dark matter halo, and some number of satellite galaxies located in substructure within the halo. As discussed in Paper I and II, including measurements of the total satellite luminosity, $\lsat$, is a critical improvement to the SCGF. The amount of substructure within a halo, and thus its total $\lsat$, scales almost linearly with $\Mh$ \citep{tinker_etal:21_lsat}. Moreover, star-forming and quiescent galaxies have different $\lsat$ at fixed $\mgal$ \citep{alpaslan_tinker:20}. For these reasons, including $\lsat$ data helps calibrate our free parameters in a complimentary fashion to the clustering measurements.

We use the $\lsat$ measurements of \cite{tinker_etal:19_lsat}, as was done in Paper II. In brief, regions of the DESI Legacy Imagings Survey are stacked around centrals galaxies from the a preliminary version of this group catalog. A background is subtracted from the stacks (see \cite{tinker_etal:19_lsat} for details). This procedure allows $\lsat$ measurements to be made in bins of $\lcen$ for quiescent and star-forming galaxies separately, without redshift information on the photometric-only satellites (which are much fainter than the BGS spectroscopic sample). As with Paper II, we use the ratio of $\lsatr/\lsatb$ instead of the raw measurements. The black bars in Fig.~\ref{fig:lsatratio_fit} show these data, along with group finder results as explained below. Note that because we re-use the measurements of \cite{tinker_etal:19_lsat}, we are not yet fully using the statistics of DESI BGS, as those measurements were originally made around spectroscopic centrals chosen from SDSS, which has a different footprint and is not as deep as BGS. In a future version we plan to replace these measurements and fully use the constraining power of BGS.

\section{Group Finding Methods}
\label{sec:methods}

Here we explain how the Self Calibrated Group Finder works. First we give the algorithm at fixed group finder parameters, which is based on the method of \cite{yang_etal:05} and \cite{tinker_etal:11}, and then we explain the process for exploring the group finder parameter space and choosing the parameters used in the final catalog. See Paper I for additional details.

\subsection{Group Finding at Fixed Parameters}
\label{sec:groupfinding}

We use the group finding algorithm described in Paper I. Here we provide a summary of those procedures, noting minor changes made as part of this work. We define halos as virialized, spherical objects with mass 

\begin{equation}
    M_h = M_{\Delta} = \frac{4}{3}\pi R^3_{\Delta} \Delta \bar\rho_m ,
\end{equation}

\noindent where $\bar\rho_m$ is the mean matter density of the universe in comoving units, and $\Delta=200$. We assume halos follow the universal Navarro-Frenk-White (NFW) profile \citep{nfw:96}, 

\begin{equation}
    \rho_h(r) = \frac{\rho_0}{(r/r_s)(1+r/r_s)^2)} ,
\end{equation}

\noindent where $\rho_0$ is a normalization constant fixed by the mass of the halo, and the scale radius $r_s$ is a free parameter. The concentration of a halo is define as $c_\Delta=R_\Delta / r_s$. Because we cannot measure individual halo concentration, we assume all halos follow the mean concentration-mass relation of \cite{maccio_etal:08}. At the time group finding begins, the sample has already been made into a spectroscopically complete, flux-limited sample via the techniques outlined in \S\ref{sec:fiberincompleteness}.

In brief, the method is:

\begin{enumerate}
\item Initialize by labeling all galaxies as centrals (groups of size 1), so $\lgrp=\lgal$.
\item Inverse-abundance match centrals using their weighted luminosity $\wcen w_f \lgrp$ to a halo\footnote{We use luminosity rather than stellar mass in the abundance matching because of our flexible weighting scheme. Abundance matching on stellar mass is usually preferred because quiescent galaxies have lower luminosities at fixed stellar mass, leading to underestimated halo masses. However, our weighting scheme can capture these differences. This makes using luminosity, which has lower uncertainty than stellar mass, preferable. The mutual information study of \cite{alpaslan_tinker:20} found that luminosity contains more information than stellar mass about halos, especially once split into quiescent and star-forming samples, strengthening this choice.}. 
\item In order of decreasing halo mass, for each central galaxy, determine if nearby lower-luminosity galaxies are satellites. Satellites are removed from this list.
\item For each group, sum the luminosity across members, $\lgrp$.
\item If the value of $\fsat$ is converged, then exit. Otherwise, return to step 2.
\end{enumerate}

\noindent {1.}~\textbf{Initialization.} All galaxies are initially treated as centrals, and so can be thought of as groups of size 1, with $\lgrp=\lgal$. This is just a starting point to allow a first assignment of halos so the group finding can begin. Despite potentially large uncertainties in the assigned halo masses during this first iteration, the algorithm quickly produces better mass estimates and the final halo masses are very weakly affected by how the initialization proceeds.

\vspace{0.2cm}
\noindent {2.}~\textbf{Weighted inverse-abundance matching.} Abundance matching is a method for assigning galaxies to dark matter halos (\citealt{kravtsov_etal:04, vale_ostriker:06, conroy_etal:06}, see \citealt{wechsler_tinker:18} for a review). Here we proceed in the opposite direction, assigning a halo to each central galaxy (group). The inverse abundance matching is defined as

\begin{equation}
  \label{eq:amtot}
    \int_{\wltot}^\infty n(\wltot^\prime)\,d\wltot^\prime =
\int_{\Mh}^\infty n(\Mh^\prime)\,d\Mh^{\prime},
\end{equation}

\noindent which expresses that a group of total weighted luminosity $\wltot$ occupies a halo of mass $\Mh$. For the halo mass function $n(M^\prime_h)$, we use \cite{tinker_etal:08_mf} at $z=0$. The total weighted group luminosity is defined as $\wltot = \wcen w_f \lgrp$, where $\lgrp$ is the sum of luminosities of group members. The first weighting factor $\wcen$ is dependent on the central galaxy's properties, motivated by past measurements in how quiescent and star-forming galaxies occupy halos (it is defined mathematically at the end of this section). The second factor $w_f$ is part of correcting the flux-limited sample: because BGS is flux-limited, the abundance matching is carried out in narrow redshift bins (0.05 width), with $1/V_{\rm MAX}$ weighting, as as described in Paper II. Because satellite luminosity is lost from the flux limit between the redshift bin edges, we follow \cite{yang_etal:05} and apply a multiplicative correction $w_f$. We calculate

\begin{equation}
    \frac{\int_0^{\inf} \Phi(L) dL}{\int_{L_{\rm lim}}^{\inf} \Phi(L) dL},
\end{equation}

\noindent  where $\Phi(L)$ is the DESI BGS $r$-band Luminosity Function \citep{Moore.2025} and $L_{\rm lim}$ is a limiting luminosity calculated with the flux limit of 19.5 at various redshifts. We fit these results to get the correction factor 

\begin{equation}
    w_f(z) = 10^{0.561(z/0.311)^{2.077}}.
\end{equation}

\noindent Due to the narrowness of the redshift bins, $w_f$ typically affects halos masses only by a few percent. Note that groups with no satellites will only have $\lgrp = \lcen$, which will produce very little scatter in the resultant luminosity to halo mass relation. Paper II included an additional weighting based on other galaxy properties to allow for additional scatter at fixed $\lcen$, but we do not include this weight here.

\vspace{0.2cm}

\noindent {3.}~\textbf{Assign satellites.} Starting with the most massive halo, we check if neighboring galaxies should be considered satellites. We assume satellites in a group are distributed roughly similar to the dark matter, and calculate how confident we are separately in projected and redshift space, informed by our knowledge of dark matter halos. 

\vspace{0.2cm}
\noindent For a candidate galaxy of luminosity $\lgal$ with projected separation $R_p$ and line-of-sight separation $\Delta z$ from a central galaxy, we classify it as a satellite if $\psat>0.5$, where
    
\begin{equation}
  \label{eq:psat}
  \psat = \left[1-\left(1+\frac{P_{\rm proj}(R_p)P_{z}(\Delta z)}{\bsatc(\lgal)}\right)^{-1}\right].
\end{equation}

\noindent Here the two contributions $P_{\rm proj}(R_p)$ and $P_z(\Delta z)$ are defined as \citep{Bartelmann:96, yang_etal:05}:

\begin{equation}
    P_{\rm proj}(R_p) = 2 r_s \bar{\delta} f(R_p / r_s) ;
\end{equation}

\begin{equation}
    f(x) = \left\{
    \begin{array}{lll}
    \frac{1}{x^{2}-1}\left(1-\frac{{\ln
    {\frac{1+\sqrt{1-x^2}}{x}}}}{\sqrt{1-x^{2}}}\right)   &  \mbox{if   $x<1$}  \\
    \frac{1}{3}   &   \mbox{if   $x=1$}   \\   
    \frac{1}{x^{2}-1}\left(1-\frac{{\rm
          atan}\sqrt{x^2-1}}{\sqrt{x^{2}-1}}\right) & \mbox{if $x>1$}
    \end{array} \right.\,;
\end{equation}

\begin{equation}
    \bar{\delta} = \frac{200}{3} \frac{c^3_{200}}{\ln (1 + c_{200}) - c_{200}/(1+c_{200})} ~;
\end{equation}

\begin{equation}
    \label{eq:pz}
    P_z(\Delta z) = \frac{c}{\sqrt{2\pi} \sigma_v} \exp \Bigg [ \frac{-(c \Delta z )^2}{2 \sigma_v}   \Bigg] .
\end{equation} 

\noindent The projected distance $R_p$ contributes via $P_{\rm proj}(R_p)$, the projected NFW surface density, and the redshift space distance $c\Delta z$ ($\Delta z = z_{\rm cen} - z_{\rm gal}$) contributes via $P_z(\Delta z)$, the Gaussian form that follows from a isothermal, isotropic velocity distribution of satellites (see Paper I for further explanation). Finally, $\bsatc(\lcen)$ is a parameter that controls how easily galaxies are classified as satellites (low $\bsat$ increases satellite count) and is defined in the next section.

\vspace{0.2cm}
\noindent {4.}~\textbf{Re-sum group luminosities.} With satellites assigned we can now recalculate the groups' luminosities $\lgrp$ by summing over all member galaxies.

\vspace{0.2cm}
\noindent {5.}~\textbf{Exit Criteria.} After just 2 iterations, the halo assignment for field galaxies and small groups are already quite stable. For richer systems, continuing to iterate is more important as the satellites contribute a significant fraction of the group's total luminosity, and thus strongly affect halo assignment. While running the group finder in the MCMC loop for calibration (explained in the next section), we allow it exit after the $i^{\rm th}$ iteration if $|\fsat^{i} - \fsat^{i-1}| < 0.001$, typically 4 or 5 iterations. For the final catalog, we run 10 iterations, which converges to $|\fsat^{10} - \fsat^{9}| < 0.00003$.

\vspace{0.2cm}

Note that the assumed cosmology affects both the halo mass function along with the definition of halo mass and all related quantities. Tests in Paper I, \cite{tinker_etal:11} and \cite{yang_etal:07_catalog} generally found low sensitivity of different cosmological parameters on group finder results, though more extensive tests would be advisable if the group catalog presented here was applied towards cosmology.

In the above procedure, there were two quantities undefined: 1) $\wcen$, the weight that multiplies $\lgrp$ for the purposes of abundance matching, and 2) $\bsat$, which defines the threshold for considering a neighboring galaxy to be a satellite. The motivations and definitions for these parameters are the same as in Paper I, which we summarize here.

The basic satellite assignment equations of \cite{yang_etal:05} do not allow intrinsic properties of a neighboring galaxy to affect it's likelihood to be a satellite. Our parameterized satellite threshold parameter $\bsatc(\lgal)$ allows color and luminosity information to affect this. We define it in the following way:

\begin{equation}
  \label{eq:bsat}
  \bsatc(\lgal) = \beta_{0,c} + \beta_{L,c}(\log\lgal-9.5),
\end{equation}

\noindent where the subscript $c={\rm r,b}$ refers to r for quiescent and b for star-forming. Four free parameters, $\beta_{0,r}$, $\beta_{L,r}$, $\beta_{0,b}$, and $\beta_{L,b}$ are thus introduced. The projected clustering data at the scale of halos strongly constrains these values as shown in Paper I and II.

To allow the group finder to give different halos to quiescent and star-forming galaxies at fixed luminosity, we use a weight factor $\wcenc(\lgal)$ with the following definition:

\begin{equation}
  \label{eq:wcen}
  \log \wcenc(\lgal) =\frac{- \omega_{0,c}}{2}\left[1+{\rm erf}\left(\frac{
      \log\lgal-\omega_{L,c}}{\sigma_{\omega, c}} \right)\right], 
\end{equation}

\noindent where again $c=r,b$ for quiescent and star-forming galaxies\footnote{In Paper II the manuscript contained a typo: a minus sign was missing from the right-hand side of the $\log \wcen$ definition.}. This flexible form introduces six free parameters, which is largely constrained by the $\lsat$ data discussed in \S\ref{sec:calibration-lsat}.

\subsection{Exploring the Group Finder Parameter Space}
\label{sec:exploring_params}

To find suitable values for the ten parameters, we follow methods similar to Paper II. Using the Markov-Chain Monte Carlo implementation \textsc{emcee} \citep{emcee}, we explore the parameter space and evaluate a goodness of fit as follows:

\begin{itemize}
    \item Run the group finder until $\fsat$ is converged to $<0.001$, yielding a central/satellite classification for each galaxy and a halo mass for each group. 
    \item Tabulate the mean halo occupation of volume-limited subsamples matching our projected clustering measurements, \S\ref{sec:calibration-clustering}.
    \item For each galaxy subsample, populate the $z=0$ halo catalog of the Bolshoi-Planck simulation \citep{klypin_etal:16} using the tabulated halo occupation, assuming Poisson satellite variance. 
    \item For each mock, calculate $w_p(r_p)$ using \textsc{corrfunc} \citep{corrfunc} and then $\chi^2_{w_p}$ by comparing to the projected clustering measurements.
    \item For each group, convert their $\Mh$ to $\lsat$ via a tabulated lookup table (\textit{not} the observed $\lgrp -\lcen$)
    \item Using these $\lsat$, compute an average $\lsatr / \lsatb$ in bins of $\lcen$ to matching the $\lsat$ measurements, \S\ref{sec:calibration-lsat} and calculate $\chi^2_{\lsatr / \lsatb}$.
    \item Sum the total goodness of fit $\chi^2_{\rm tot}$.
\end{itemize}

We have 2 luminosity bins where the galaxies were not separated by color due to low red counts, and 6 where they were, creating the 14 different $w_p(r_p)$ contributions in the overall goodness of fit, along with the $\lsat$ contribution:

\begin{equation} 
    \chi^2_{\rm tot}=\sum_{i=1}^{14} \chi^2_{w_p,i}+\chi^2_{\lsatr / \lsatb}.
\end{equation} \label{eq:chisqr}

With this procedure we explore the parameter space for 200,000 iterations. We select parameters for our final catalog as the set that produced the lowest $\chi^2_{\rm tot}$ from models that give relations in middle 68\% of the quantities shown in Fig.~\ref{fig:gfparams}, $\log \wcen^{\rm q} / \wcen^{\rm sf}$, $\bsat^{\rm q}$ and $\bsat^{\rm sf}$. This catalog has a reduced $\chi^2_{\rm tot}$ of $\sim0.72$. With these parameters, we re-run the group finder with 10 iterations to produce the final catalog, which we present below.

\section{Results}
\label{sec:results}

Here we present the results of the final group catalog.  The group finding parameters used were determined using the procedure defined in the previous section. In addition to the results shown here, we fit Halo Occupation Distribution models to the group catalog and show the results in Appendix~\ref{apdx:hods}.

\subsection{Fit to Calibration Data}
\label{sec:fit-to-cal}

\begin{figure*}
    \centering
    \includegraphics[width=1.0\linewidth]{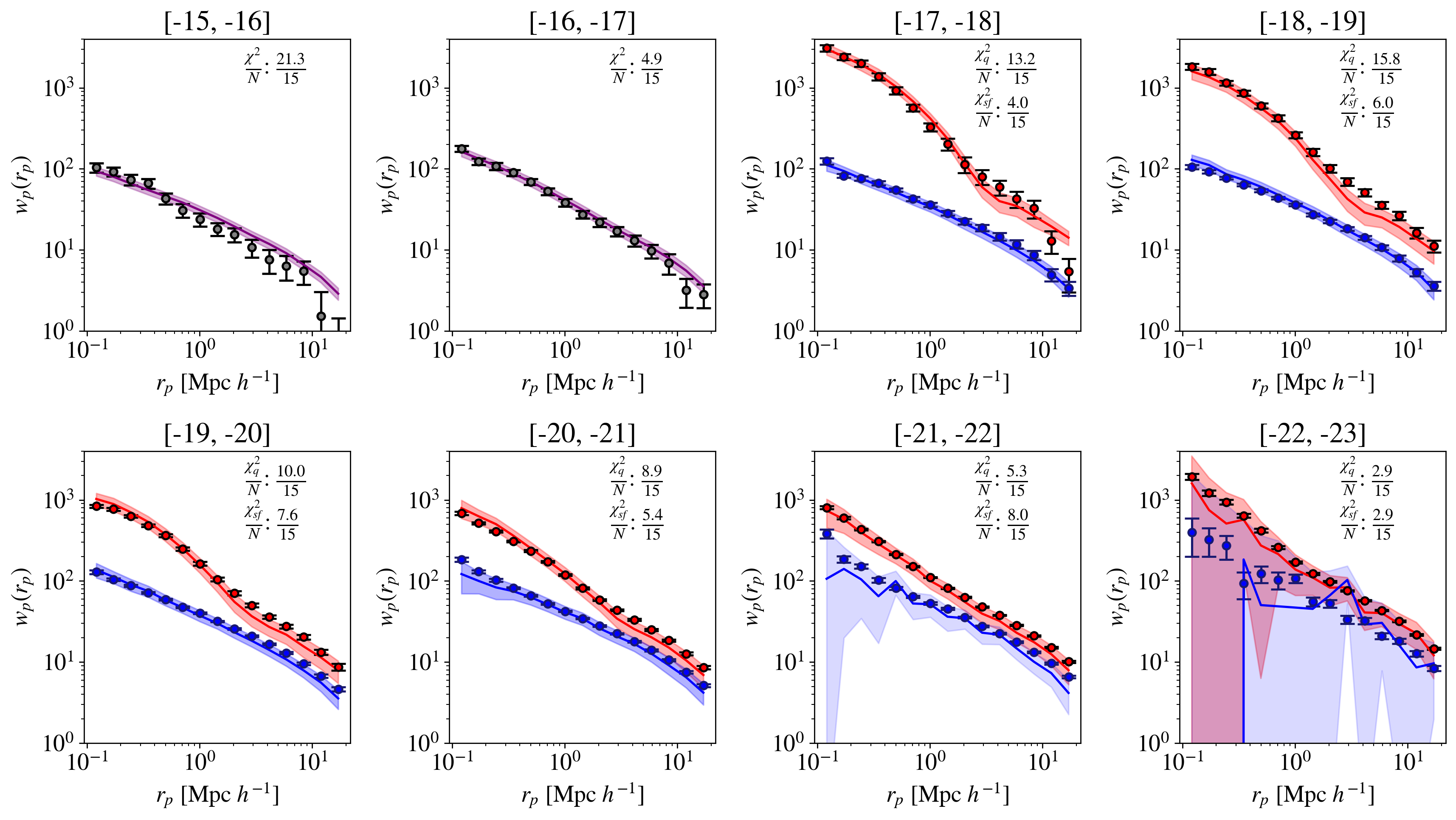}
    \caption{\textbf{Projected clustering fit}. Here we show how well the group finder's galaxy halo connection is able to build mocks matching our measured $w_p(r_p)$. The red or blue dots with black error bars are the measured projected clustering of the quiescent or star-forming samples of BGS DR1, in magnitude-wide bins. The results of mock building from the final catalog are shown in the red and blue lines, with shaded regions as estimated error bars. The first two bins do not split the sample into quiescent and star-forming. The limited volume of the box used when building mocks creates large error bars for the brightest bin, especially for star-formers which are rare at this magnitude. Overall the fit is good, with a mild under-prediction of the quiescent two-halo term in the model. Note that when calculating $\chi^2$, we use a diagonal covariance matrix.}
    \label{fig:wp_fit}
\end{figure*}

Fig.~\ref{fig:wp_fit} shows the projected clustering $w_p(r_p)$ from the final catalog's mocks compared to the DESI BGS DR1 measurements described in $\S$~\ref{sec:calibration-clustering}. The shaded $1\sigma$ error bars on the mocks' clustering are calculated by adding the error from the measurement (scaled by the measurement's volume over the mock's volume) to an assumed 0.1 dex systematic uncertainty (representing the mock building procedure's uncertainty). The model matches well. The greatest difference is a mild under-prediction of the red clustering in the 2-halo regime in all but the brightest bin.

\begin{figure}
    \centering
    \includegraphics[width=1.0\linewidth]{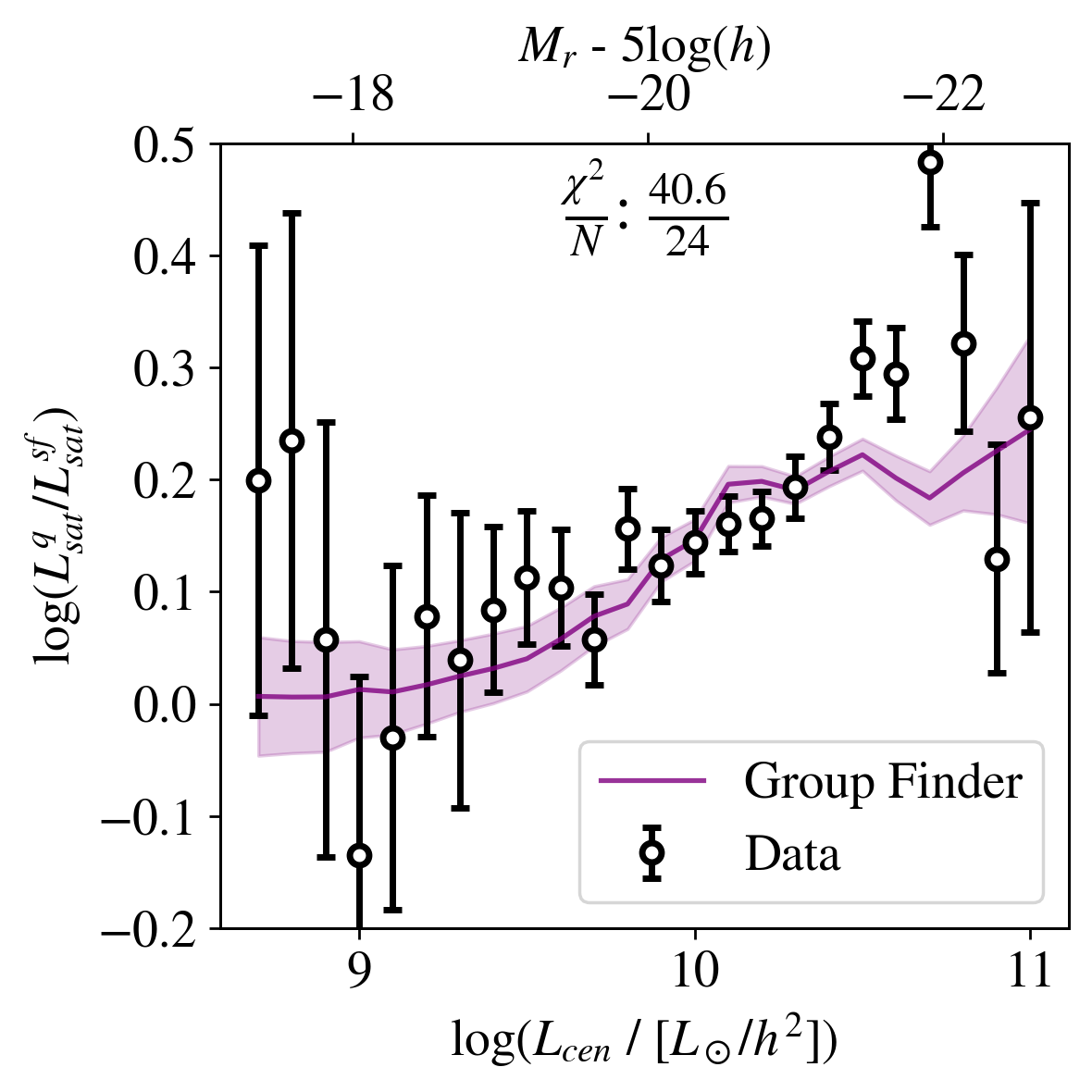}
    \caption{\textbf{$\lsat$ ratio fit}. Here we show how successful the group finder was in assigning different mass halos to quiescent vs. star-forming centrals in luminosity bins. The assigned halo masses of central galaxies are converted to $\lsat$ values by looking up the mean $\lsat$ found in $\Mh$ bins from a mock, not using the limited spectroscopic satellites found. The model's fit to the measurements is somewhat low in the high $\lgal$ regime, likely driven by the more precise clustering data. Note that when calculating $\chi^2$, we use a diagonal covariance matrix.}
    \label{fig:lsatratio_fit}
\end{figure}

Fig.~\ref{fig:lsatratio_fit} shows the $\lsatr/\lsatb$ ratio comparison to the measurements. The ratio from the group finder is not quite high enough at the high luminosity end. This indicates that the halos assigned to quiescent galaxies are not quite massive enough in comparison to star-forming ones at those luminosities. It is possible that the new projected clustering measurements with their superior statistics pull the parameters away from the preferences of the $\lsat$ data.

\subsection{Parameter Constraints}
\label{sec:param-constraints}

\begin{figure}
    \centering
    \includegraphics[width=1.0\linewidth]{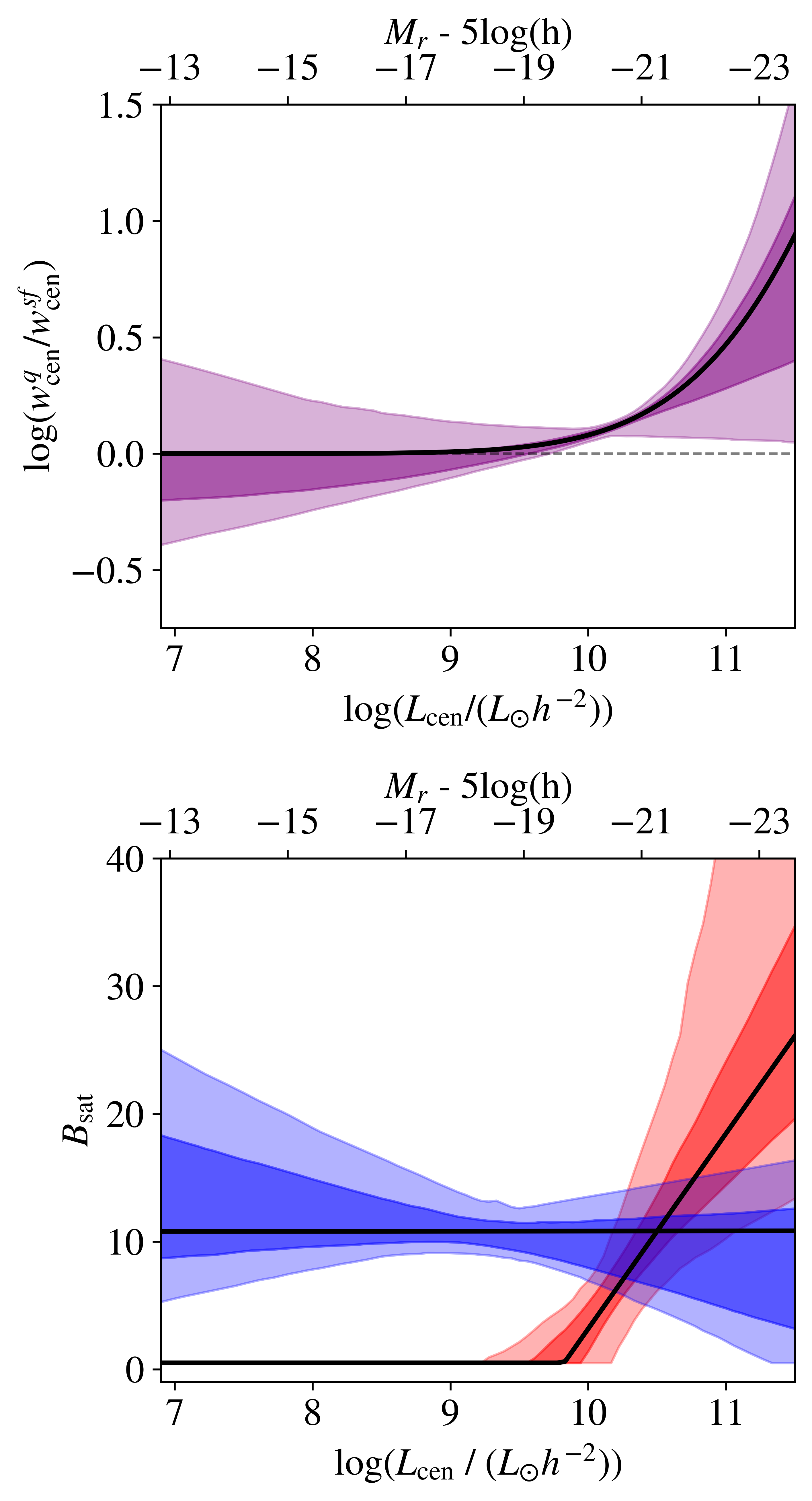}
    \caption{\textbf{Group finder parameter constraints.} \textit{Top:} The top plot shows, at fixed luminosity, how quiescent vs star-forming galaxies must be weighted differently in the abundance matching to receive appropriate halos. On the bright end, quiescent galaxies need more massive halos, while at lower $\lcen$ the differences are smaller and more weakly constrained. \textit{Bottom:} The bottom plot shows how the threshold for satellites, $\bsat$, changes for different luminosities of the two samples (high $\bsat$ means less satellites). For both plots the shaded regions show the middle 68\% and 95\% regions of the $y$-axis quantity, as evaluated using the MCMC chains. The black lines are the parameter set used in the final catalog, which maximize the likelihood within the inner shaded region. The shaded regions are computed using percentiles at each $\log \lgal$ bin, so that no single model may actually exist at the edges of the shaded regions.}
    \label{fig:gfparams}
\end{figure}

We use the MCMC chains to explore constraints on our 10 free parameters. Results are shown in Table~\ref{tab:mcmc_params}. Due to a degeneracy within the parameters of $\wcen$, those constraints are not as strong as the parameters of $\bsat$. To aid in understanding how these parameters effect group finding, we show Fig.~\ref{fig:gfparams}. The upper panel shows $\log(\wcenr/\wcenb)$ as a function of $\lgal$, which demonstrates how quiescent vs star-forming central galaxies at fixed luminosity are weighted to receive different halos. The lower plot shows the effective $\bsatc$ values. The shaded regions $1\sigma$ error bars were calculated from the MCMC chains, and the black lines are the values for the final catalog. 

The $\wcen$ constraints show little preference for whether star-forming or quiescent centrals at the same $\lcen$ should be assigned more massive halos when dimmer than -19. Brighter than this, quiescent centrals get increasingly more massive halos. These results, driven by the $\lsat$ data, can be seen in the LHMR shown later. The $\lsat$ data used here is not very constraining at -19 or fainter, and so there is considerable uncertainty at the low L end. 

Low $\bsat$ values correspond to more easily classifying a galaxy as a satellite. To match the high clustering amplitude of quiescent galaxies (especially fainter ones), the group finder needs to assign as many quiescent galaxies as possible to be satellites, up to $M_r^{0.1} \sim -20$. In this minimal $\bsatred = 0.5$ regime, most quiescent galaxies with a brighter neighbor become satellites. Meanwhile the $\bsatblue$ does not vary so strongly with $\lgal$, staying around $\sim10$, which is the standard value used in the older group finder versions. Compared to Paper II, these results are similar for the quiescent sample, but favor lower $\bsatblue$ values for faint galaxies. The change is driven by the inclusion of clustering data at lower luminosities, which are mostly star-forming galaxies. The higher $\bsatblue$ for faint galaxies found in Paper II underpredicts new faint galaxy clustering data, and so the new parameters have accommodated this by flattening $\bsatblue$ to $\sim 10$.

\begin{table*}[ht]
    \centering
    \caption{\textbf{Group Finder Parameter Constraints.} We show the middle 68\% confidence interval from the MCMC posteriors for each parameter, along with the parameter values used in the the final catalog. Due to a degeneracy in $\wcen$, some confidence intervals are wide.}
    \label{tab:mcmc_params}
    \begin{tabular}{llll}
        \hline
        \hline
        Parameter & Description & Final Catalog  & [68\% CI] \\
        \hline
        $\omega_{L,b}$         & Location of the erf in weighting star-forming centrals & 13.57    &  [12.83, 27.13]  \\
        $\sigma_{\omega,b}$    & Smoothness of the erf in weighting star-forming centrals & 2.11    &  [1.75, 10.91]  \\
        $\omega_{L,r}$         & Location of the erf in weighting quiescent centrals & 15.33    &  [13.85, 41.39]  \\
        $\sigma_{\omega,r}$    & Smoothness of the erf in weighting quiescent centrals & 3.30    &  [2.43, 24.41]  \\
        $\omega_{0,b}$         & Amplitude of the erf in weighting star-forming centrals & 14.86    &  [9.33, 62.40]  \\
        $\omega_{0,r}$         & Amplitude of the erf in weighting quiescent centrals & 4.37    &  [3.22, 17.34]  \\
        $\beta_{0,r}$          & Constant contribution to $\bsat$ for quiescent galaxies & -4.74    &  [-9.00, -0.26]  \\
        $\beta_{L,r}$          & Linear in $\log \lgal$ contribution to $\bsat$ for quiescent galaxies & 14.45    &  [10.11, 21.66]  \\
        $\beta_{0,b}$          & Constant contribution to $\bsat$ for star-forming galaxies & 9.40    &  [9.25, 11.47]  \\
        $\beta_{L,b}$          & Linear in $\log \lgal$ contribution to $\bsat$ for star-forming galaxies  & -2.18    &  [-3.29, 0.87]  \\
        \hline
    \end{tabular}
\end{table*}

\subsection{Satellite Fraction}
\label{sec:fsat}

The satellite fraction $\fsat$ in a luminosity or stellar mass bin is defined as the count of satellites within the bin over the total number of galaxies within the bin, with $1/\vmax$ weighting. Due to the narrowness of the bins, the $1/\vmax$ weighting is a minor correction for $\fsat(\lgal)$, but more important for $\fsat(M_\star)$, because the luminosity variance within the bin is greater. 

\begin{figure*}
    \centering
    \includegraphics[width=1.0\linewidth]{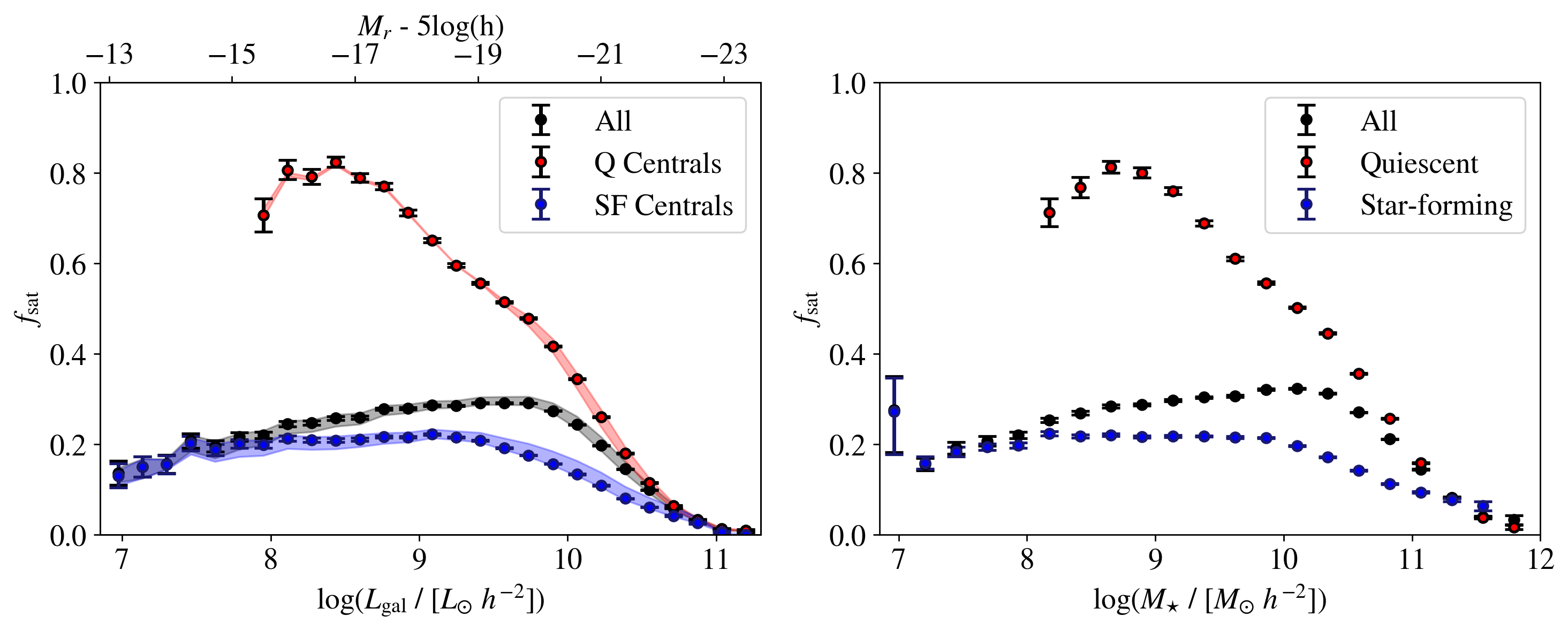}
    \caption{\textbf{Satellite Fraction.} The left plot shows $\fsatb$ and $\fsatr$ in the final catalog as a function of $\lgal$, while the right shows them as a function of $\mgal$. The measurements generally agree with Paper II, continuing to lower magnitudes. The error bars are statistical error from bootstrapping and the shaded regions are group finder parameter uncertainty estimates from the MCMC chains. The faint (and low mass) turndown in $\fsatr$ is likely a result of confusion between quiescent and star-forming galaxies in that regime.}
    \label{fig:fsat}
\end{figure*}

Fig.~\ref{fig:fsat} shows $\fsat(\lgal)$ and $\fsat(\mgal)$ split for quiescent and star-forming galaxies. We omit bins with $<100$ galaxies. The results are in good agreement with Paper II. At the faintest luminosities, we find that $\fsatr$ decreases, which may be indicative of wrongly classifying galaxies as quiescent. The opposite direction effect of quiescent galaxies contaminating the star-forming sample would be minor because there are so few of them in this regime. As discussed in \S\ref{fig:quiescent_classification}, we expect more misclassifications at low $\lgal$.

The error bars are statistical uncertainty estimates from bootstrapping the galaxies in the final catalog 3000 times. The shaded regions are represent group finder parameter uncertainty computed as the middle 68\% of values from the catalogs in the MCMC chains.  

\subsection{LHMR and SHMR }
\label{sec:lhmr}

\begin{figure}
    \centering
    \includegraphics[width=1.0\linewidth]{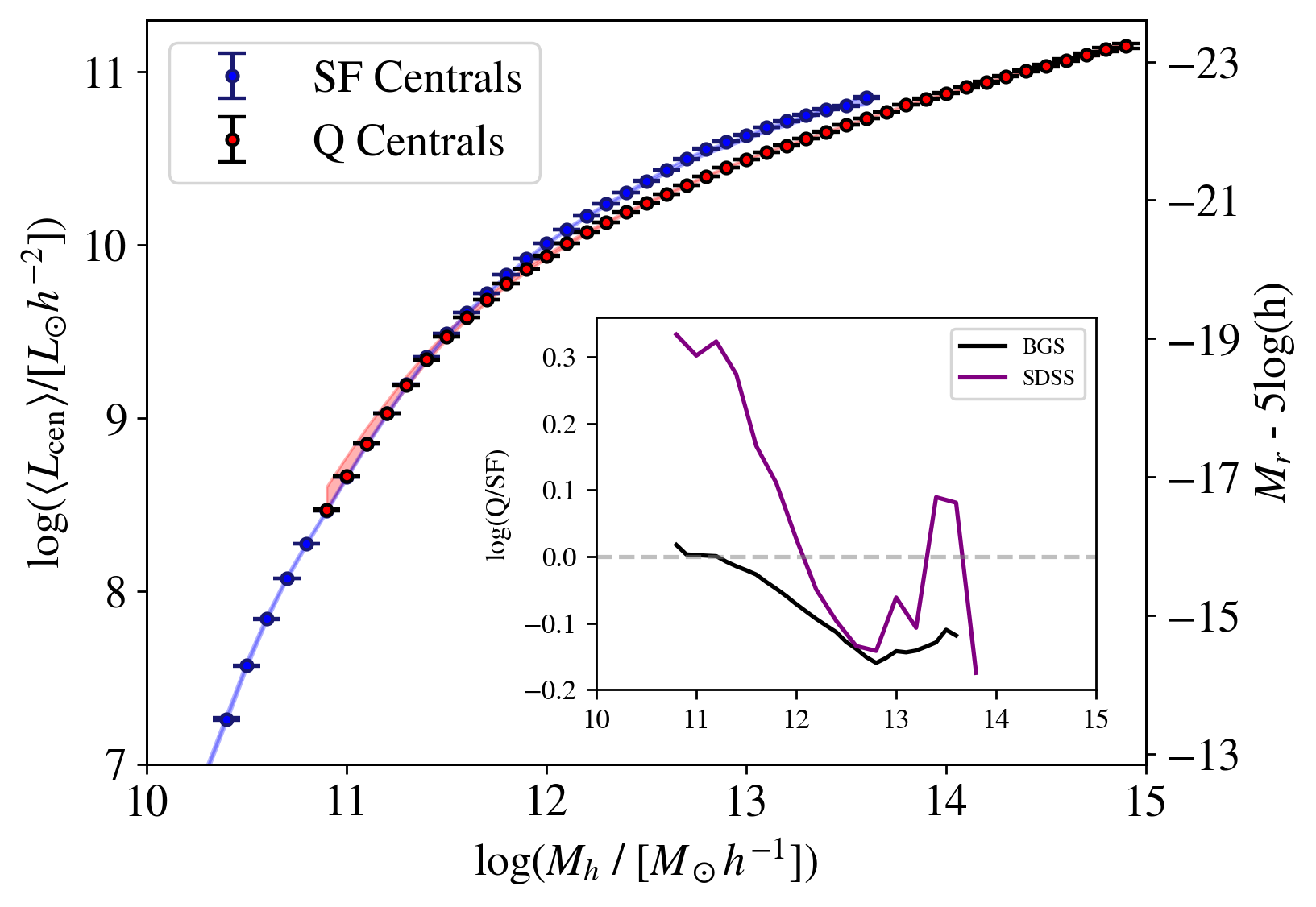}
    \caption{\textbf{Luminosity Halo Mass Relation.} The mean luminosity of quiescent and star-forming galaxies in halo mass bins in the final catalog. The error bars are from bootstrapping and the shaded regions are group finder parameter error estimates from the MCMC chains. The inset shows the log ratio of the quiescent value over the star-forming one in halo mass bins, along with the results of Paper II. The differences are driven by the inclusion of the $\chi$ parameter in Paper II.}
    \label{fig:lhmr}
\end{figure}

Fig.~\ref{fig:lhmr} shows the luminosity to halo mass relation (LHMR) extracted by the group finder. The error bars show the inner 68\% percentile range from bootstrapping, which is very small, and the shaded regions shown the same percentile range but from the MCMC chains. We omit bins with $<100$ galaxies for all plots in this section. 

The results are qualitatively similar to the SDSS ``no-$\chi$" results in Paper II. Above $\Mh \sim 10^{12} \msol$ the BGS catalog has slightly brighter galaxies in equivalent halos than in the SDSS catalog partially from the photometric differences between BGS and SDSS (discussed more in $\S$\ref{apdx:sdss_compare}). If the BGS magnitudes are shifted to be more ``SDSS-like" using an empirical relation fit from the overlapping samples, the LHMR differences lessen by about half.

\begin{figure}
    \centering
    \includegraphics[width=1.0\linewidth]{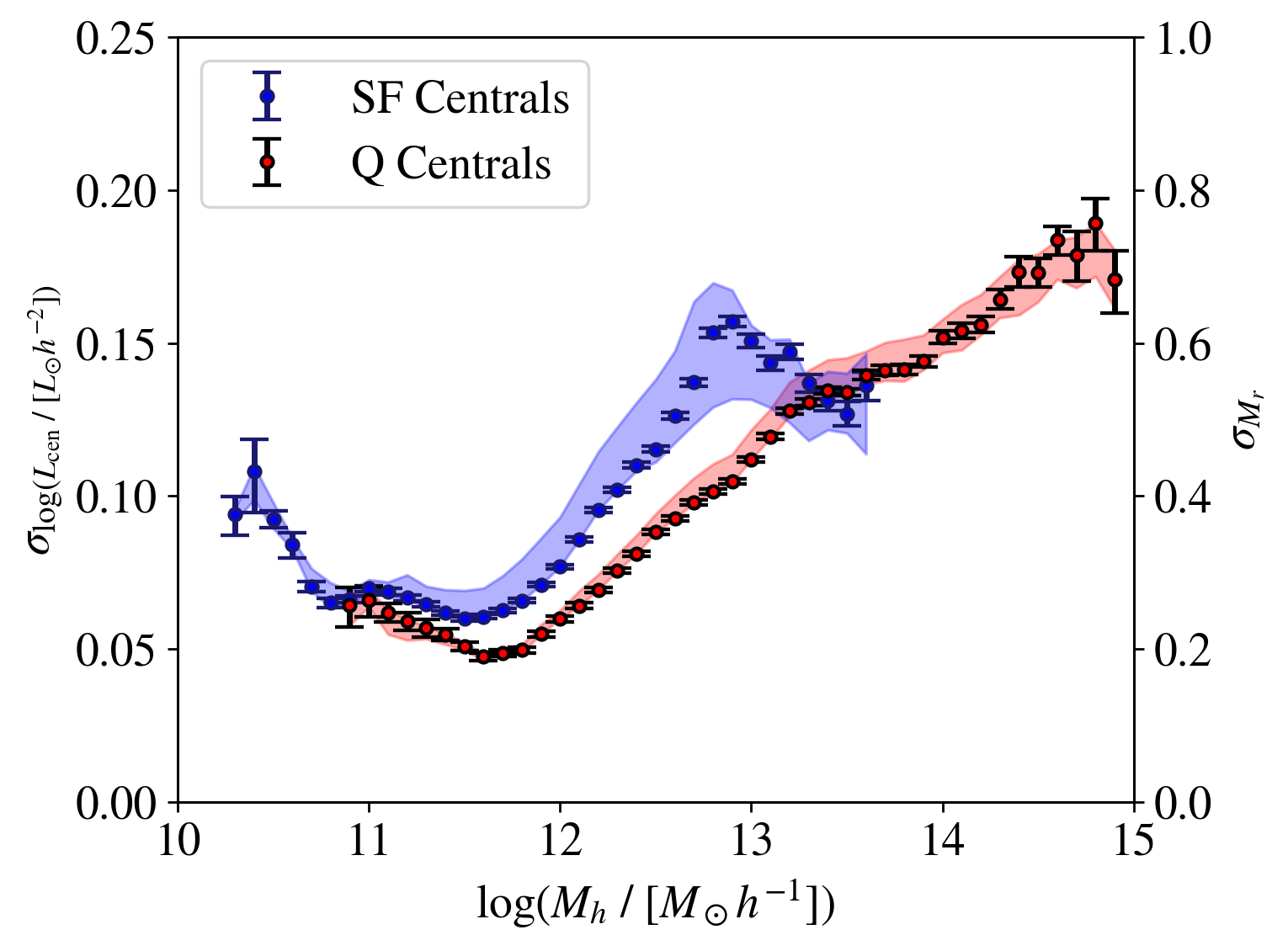}
    \caption{\textbf{Scatter in the LHMR.} The scatter of quiescent and star-forming galaxies in the final catalog. The scatter, especially at low-masses, is underestimated for reasons discussed in Paper II - when no satellites are observed, raw abundance matching on $\lcen$ will produce very little scatter. By only including centrals with observed satellites, the low-mass dip is largely removed and the figure looks more like Fig.~16 of Paper II. The error bars are from bootstrapping and the shaded regions are from the MCMC chains.}
    \label{fig:lhmr-scatter}
\end{figure}

Fig.~\ref{fig:lhmr-scatter} shows the scatter in the LHMR found by the finder, with similarly constructed error bars and shaded regions. The LHMR scatter is significantly underestimated. This is because we abundance matched on $\lgrp$, which in an absence of satellites is just $\lcen$. At low masses then, this forces an artificially tight relationship between $\lcen$ and $\Mh$. In Paper II additional scatter was allowed in the model by the inclusion of a galaxy-concentration $\chi$-parameter (see that work for details). This is why the LHMR is more similar to the ``no-$\chi$" version in Paper II than the primary results. Future work includes scatter-related improvements.

\begin{figure*}
    \centering
    \includegraphics[width=1.0\linewidth]{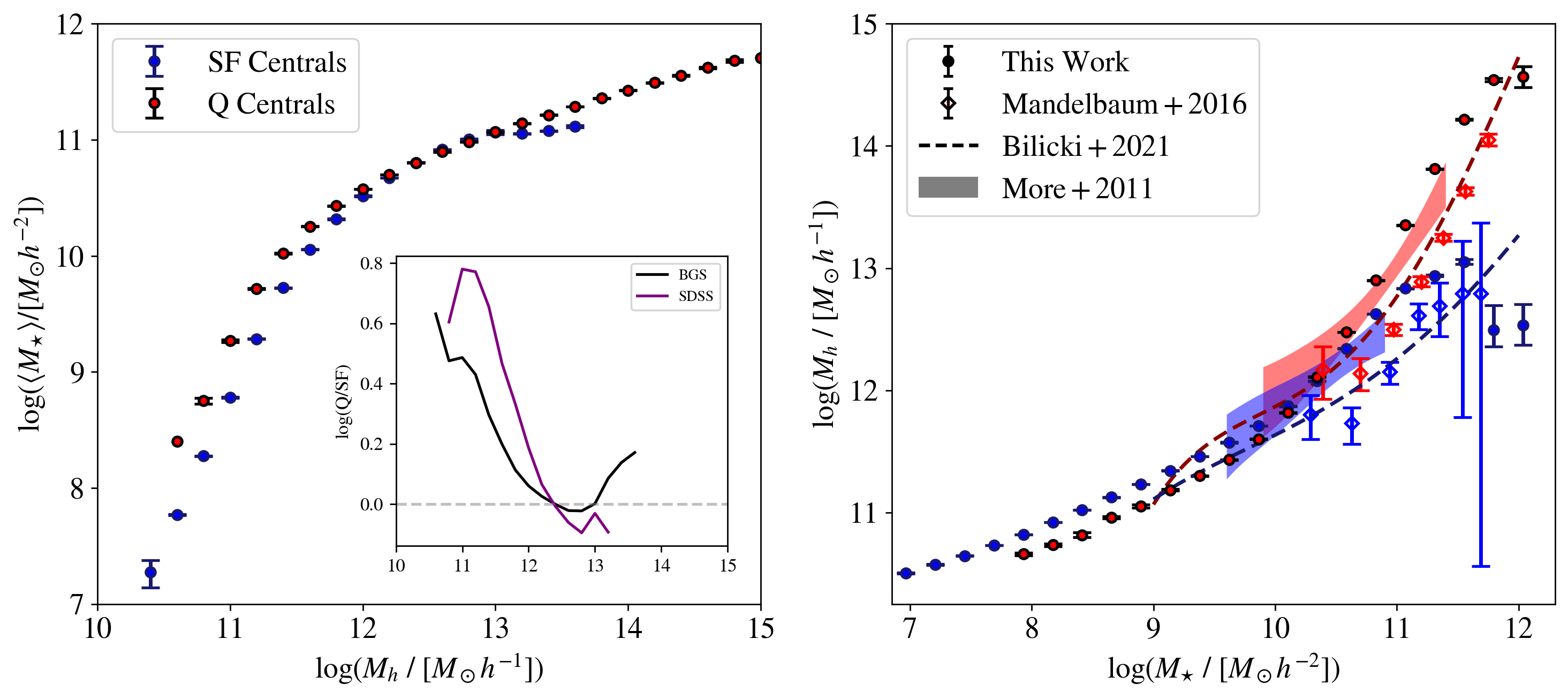}
    \caption{\textbf{Stellar Halo Mass Relation.} \textit{Left:}~The figure shows the mean stellar mass of galaxies in halo mass bins for star-forming and quiescent samples. The error bars are from bootstrapping. \textit{Right:}~The axes are inverted to show the mean halo mass in bins of stellar mass instead, along with results from other observational works. The diamonds with error bars are from galaxy-galaxy lensing \citep{mandelbaum_etal:16}. the dashed lines are also gravitational lensing \citep{Bilicki_etal_2021}, and the shaded regions are from satellite kinematics \citep{more_etal:11}.} 
    \label{fig:shmr}
\end{figure*}

\begin{figure}
    \centering
    \includegraphics[width=1.0\linewidth]{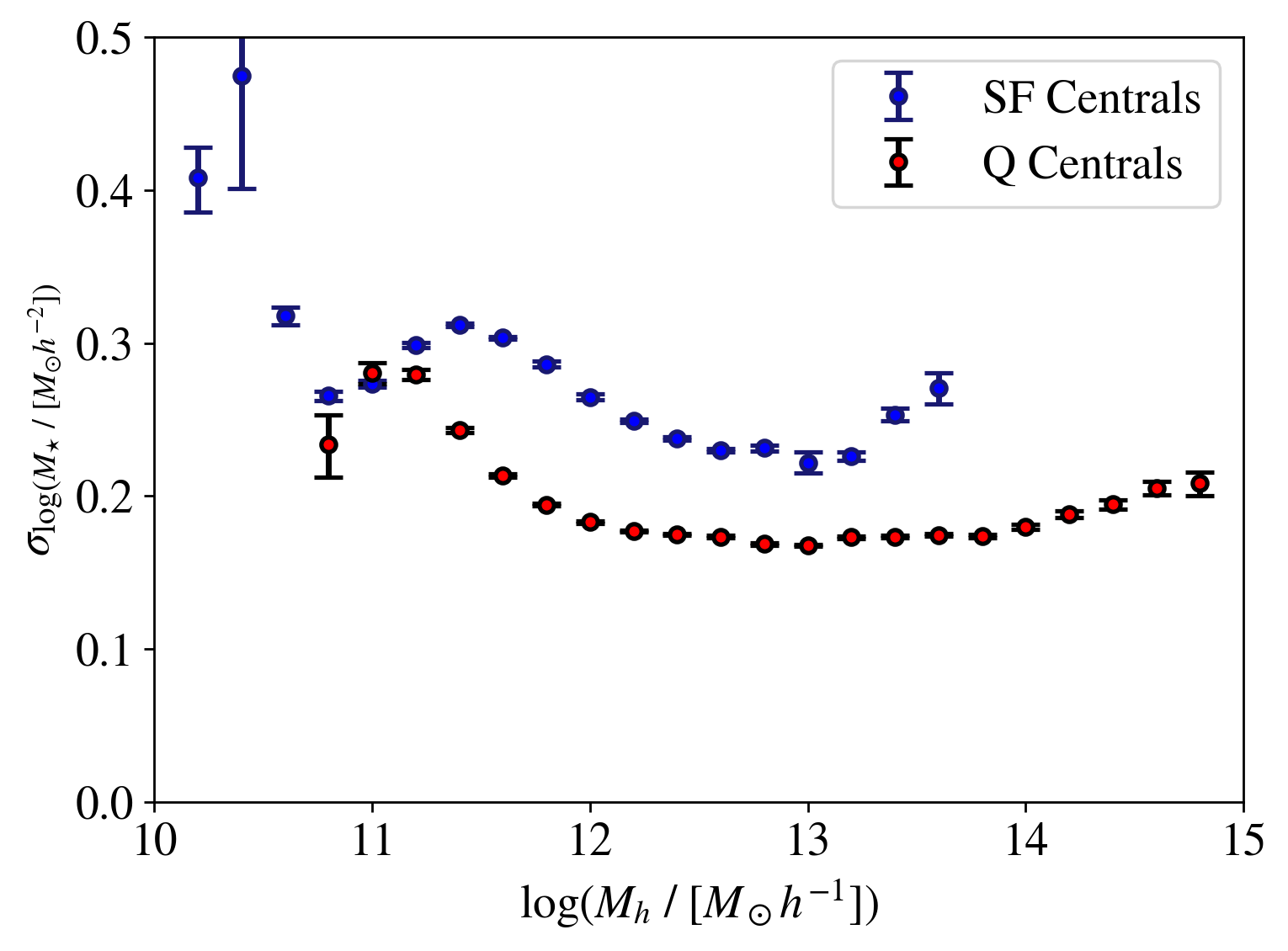}
    \caption{\textbf{Scatter in the SHMR.} The figure shows the scatter of quiescent and star-forming galaxies in the final catalog. These values are larger than the LHMR scatter, partially due to uncertainty in the stellar masses themselves. The error bars are from bootstrapping. As with the LHMR scatter, the star-forming sample shows higher scatter at most halo masses.}
    \label{fig:shmr-scatter}
\end{figure}

\begin{figure}
    \centering
    \includegraphics[width=1.0\linewidth]{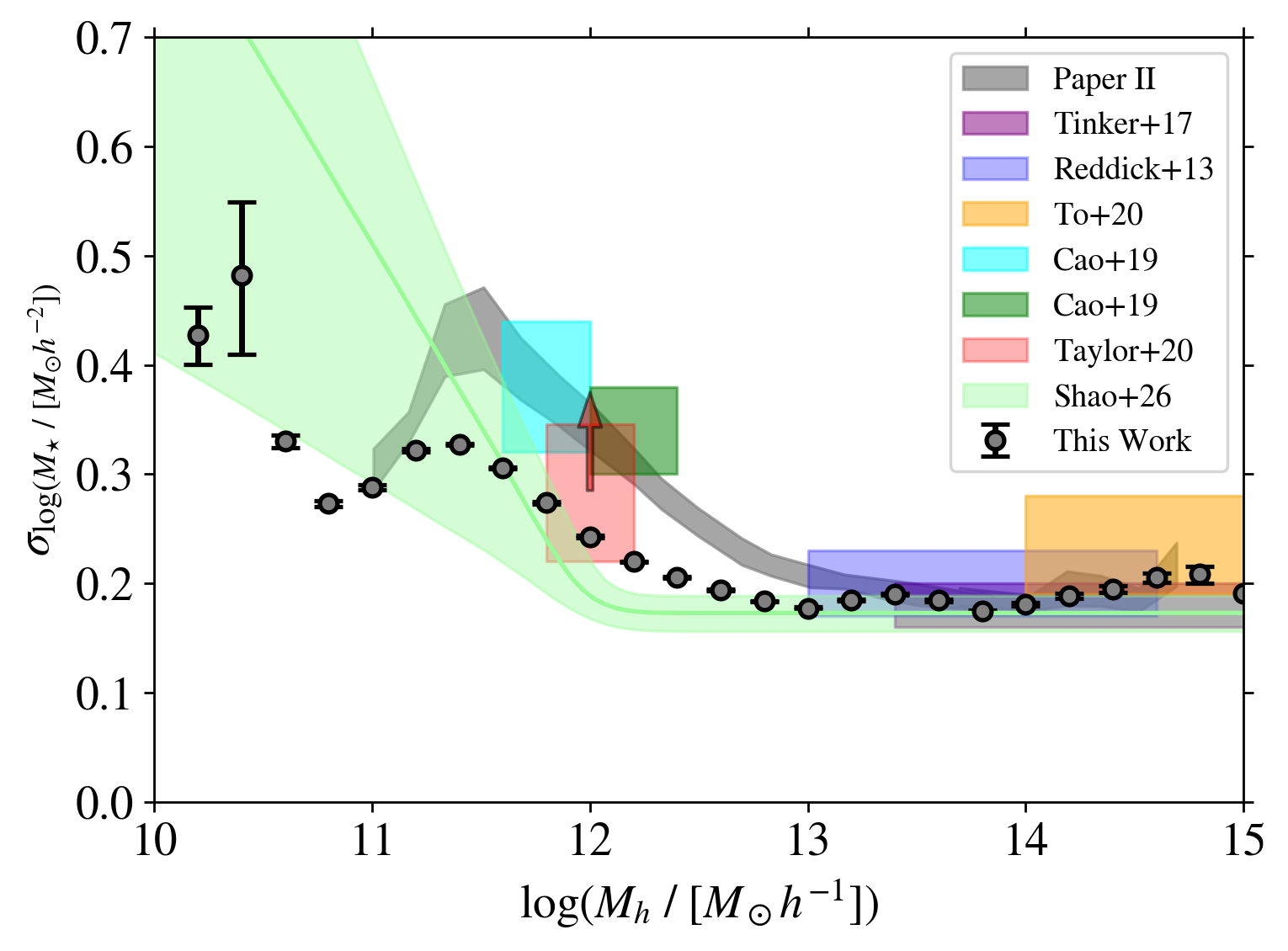}
    \caption{\textbf{Scatter in the SHMR Comparison}. Shown in black are our measurements of the overall SHMR scatter with bootstrapped error bars. The colored regions show values from various observational studies. The low mass measurements are from \cite{cao_etal:19} and \cite{Taylor_etal:2020}. The high halo mass measurements come from \cite{Tinker_etal_2017}, \cite{reddick_etal:13} and \cite{to_etal:20}. The recent results from \cite{Shao_etal_2026} are shown in bright green. Our results matches Paper II and others in high mass halos, but we have insufficient scatter in less massive halos because of the lack of $\chi$-parameter included here.}
    \label{fig:shmr-scatter-litcompare}
\end{figure}

We also show these relations with stellar masses instead of luminosities. Fig.~\ref{fig:shmr} shows the Stellar Halo Mass Relation (SHMR) and Fig.~\ref{fig:shmr-scatter} shows the scatter. We compare our SHMR to other measurements: galaxy-galaxy lensing \citep{mandelbaum_etal:16}, gravitational lensing \citep{Bilicki_etal_2021}, and satellite kinematics \citep{more_etal:11}. Note that the satellite kinematic results were updated in \cite{Lange_etal_2019}, but those results are only expressed in terms of luminosity, not stellar mass, so we do not include them here. For a recent summary of SHMR constraints and comparisons to simulations and semi-analytic models, see \cite{Wang_Peng_2025}. The scatter in stellar masses is larger than luminosities for two somewhat artificial reasons: 1) derived stellar masses are intrinsically more noisy than luminosities;  2) the abundance matching is performed on luminosities, not stellar mass. To compare this scatter to other values in the literature, we combine the quiescent and star-forming samples into an overall sample in Fig.~\ref{fig:shmr-scatter-litcompare}. Abundance matching studies of SDSS data give $\sigma_{\log\mgal}\sim 0.20$, close to the values here \citep{reddick_etal:13, Tinker_etal_2017}.

\subsection{Quenching in the \texorpdfstring{$\mgal-\Mh$ Plane}{Mgal-Mh Plane}}
\label{sec:quench-mgal-mh}

\begin{figure}
    \centering
    \includegraphics[width=1.0\linewidth]{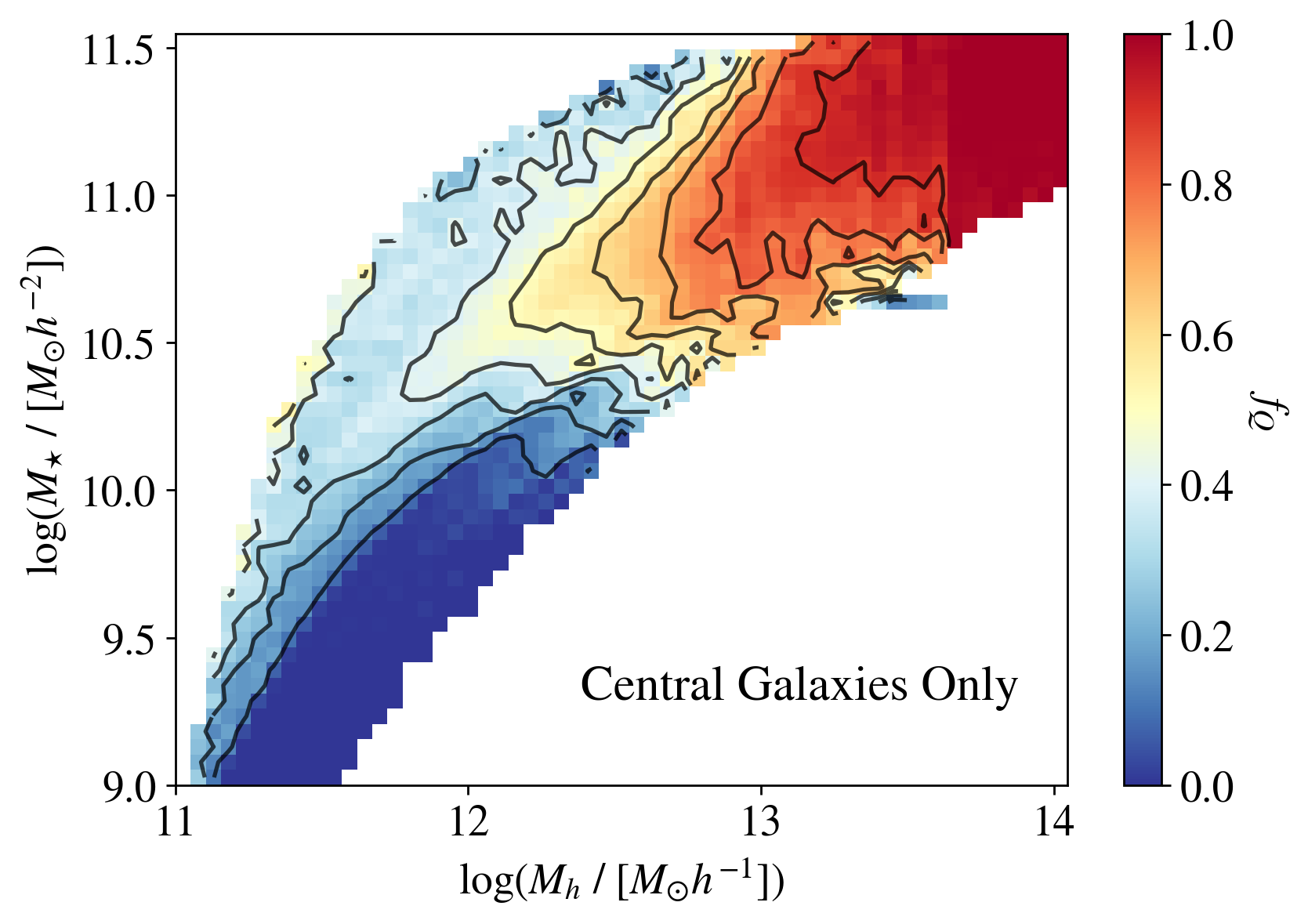}
    \caption{\textbf{Quiescent Fraction in the $\mgal-\Mh$ plane.} The figure is colored coded by the fraction of galaxies that are quiescent in each $\mgal,\Mh$ bin. At $\log\mgal \gtrsim 10.3$ and $\log \Mh \gtrsim 12.2$, $\Mh$ strongly correlates with $f_Q$, a marker of halo quenching. The low $\Mh$, high $\mgal$ regime contains many edge-on galaxies, likely biasing $\lgal$, $\mgal$, and $\Mh$, and so should be treated with skepticism.}
    \label{fig:quenched-heatmap}
\end{figure}

In Fig.~\ref{fig:quenched-heatmap} we show the quiescent fraction in the  $\mgal-\Mh$ plane for central galaxies. Above a characteristic scale of $\log\mgal \sim 10.3$ and $\log \Mh \gtrsim 12.2$, it is clear that $\Mh$ plays the dominant role in quenching the central galaxy, evident by the largely vertical contours. In this regime, at fixed $\mgal$, $\Mh$ strongly correlates with $f_Q$. Fixing $\Mh$ shows a weaker dependence of $f_Q$ on $\mgal$. This can be interpreted as a marker of halo quenching \citep{zu_mandelbaum:16}. 

As was recently pointed out in \cite{Zhao_Peng:26}, some previous studies using abundance matching failed to produce a clear halo quenching signal due to inaccurate assigned halo masses for quenched vs. star-forming galaxies. Because our model has the freedoms needed to find different SHMRs for quiescent and star-forming populations and the calibration data is sufficiently constraining, we see the halo quenching signal more clearly. In \cite{Zhao_Peng:26}, by also deriving halo masses in a way that matches weak-lensing derived SHMRs for quiescent and star-forming galaxies separately, a strong halo quenching signal was found at $\log \Mh \gtrsim 12.1$.

The low $\Mh$, high $\mgal$ regime should be treated with caution. The galaxies at the top edge of the plot with $\log \Mh < 12$ are largely edge-on disk galaxies, and appear to have extremal mass-light ratios. Dust extinction may bias$\lgal$ \citep{Maller_etal:09}, and thus $\Mh$. It is likely that $\mgal$ is also biased, but probably not in exactly the same way, producing extremal mass-light ratios and positioning those galaxies in the wrong location on this plot. For this reason any apparent trends in that regime should be treated with skepticism.

\section{Summary and Discussion}
\label{sec:summary_discussion}

\subsection{Summary}
\label{sec:summary}
 
In this work we have applied the Self Calibrated Group Finder to DESI BGS DR 1. The resultant catalog provides a central-satellite breakdown and an estimated halo mass for each group, including galaxies with no observed satellites. As with the previous application to SDSS in Paper II, free parameters in the group finder were calibrated by comparing predictions of the group finder to measurements of clustering and deep $\lsat$ measurements around central galaxies. A new data-driven technique was developed to overcome biases introduced by the $\sim20\%$ fiber incompleteness of DESI BGS.

The galaxy-halo connection of our catalog is able to build mocks that are statistically acceptable fits to the calibration data, with an overall reduced $\chi^2$ of $0.72$.  Compared to previous work, the model produces $\fsat$ values two mags fainter for quiescent galaxies, where we find a value of $\sim0.8$, and four mags fainter for star-forming galaxies, where we continue to see $\fsat\sim0.2$. Our LHMR retains the key feature that quiescent galaxies occupy more massive halos when $M_r^{0.1} \lesssim -20$. Our SHMR is similar to Paper II, with star-forming galaxies in more massive halos at fixed $\mgal$ up to $\log\mgal\sim10.4$, and similar occupancy above this. The scatter in our LHMR and SHMR is underestimated for low-mass halos, where the results of Paper II are currently preferable, but results are similar for halos $> 10^{13} \msol$. The model points towards halo quenching in galaxies $\log \mgal>10.5$, by means of a strong correlation between $\Mh$ and $f_Q$ at fixed $\mgal$.

Generally, when comparing this work to the results in Paper II, there are three key differences to consider: 1) systematic photometric differences, which are more extreme for brighter galaxies; 2) spectral quality differences leading to differences in our quiescent vs star-forming populations; and 3) no $\chi$-parameter utilized to allow individual secondary galaxy property to affect assigned halos, reducing LHMR scatter. The first difference on its own would roughly lead to a rescaling of results, complicating comparisons, but is not problematic. The second difference should lead to less sharp distinctions between the two populations in some cases. The third difference is studied directly in Paper II, which also presents no-$\chi$ results, which are much closer to our results here, as expected.

\subsection{Usage Advice}
\label{sec:advice}

To aid in the usability of this group catalog, here we provide a list of important details, warnings, and other usage advice. 

\begin{itemize}
    \item \textbf{Halo masses.} As with all halo-based group finders, the mass estimates should be treated with caution. While a $\sim0.2$~dex error is typical, it can be much larger for certain subpopulations and can easily bias results when utilized in ways that have not been explicitly checked in mock studies such as those of Paper I. Galaxies with no satellites and lost galaxy centrals will have larger uncertainties in $\Mh$.
    \item \textbf{Lost galaxies.} Our handling of lost galaxies places galaxies at the wrong redshift around 50\% of the time. Although we studied several induced biases and have partially ameliorated them, many other biases could still be induced. For some statistics, filtering out this population will be preferable, but this must be evaluated on a case-by-case basis as the removal of this non-random subset also induces biases.
    \item \textbf{High redshift.} As with any flux-limited sample, the decreasing target density at high redshift means that observed population are often bright centrals with no observed satellites, triggering the issues in $\Mh$ previously described. Although we include BGS targets to $z=0.5$, the usefulness of the high-$z$ regime is limited.
    \item \textbf{Spurious objects.} Although we have attempted to remove as many non-galactic objects as possible from the catalog, inevitably some objects will avoid our cuts. We advise extra care especially when examining extremal edges of the data: the brightest/faintest galaxies, $z > 0.3$ and z $< 0.01$, etc.
    \item \textbf{Rich clusters.} The halo model is insufficient to describe unrelaxed systems. Because rich clusters are far more likely to be unrelaxed, our halo-model based description of these systems may be misleading and should be used with caution.
    \item \textbf{Groups on the footprint edge. } When a galaxy is near the edge of the footprint, other group members may lie outside our sample and therefore be excluded from the catalog. This also leads to a systematic underestimation of the halo mass assigned to these galaxies. Due to our 3-pass footprint definition and the DR 1 footprint, the geometry of our footprint contains small holes that can trigger this effect.
    \item \textbf{Quiescent vs Star-forming.} Classifying quiescent galaxies is difficult at low luminosities, especially fainter than -17. Classifications should be used with caution in this regime.
    
\end{itemize}

\subsection{Future Work}
\label{sec:future}
For upcoming DESI data releases we plan to advance the group finding methodology used here in several ways. The primary goal is to go beyond halo masses and assign additional halo properties, such as halo age and angular momentum, as predicted by $N$-body simulations. Doing so would create a much more detailed model of the galaxy-halo connection that includes assembly bias predictions \citep{sheth_tormen_assemblybias:04, Gao_etal:05}. Enacting this requires 1) augmenting the abundance matching step in the group finder, and 2) additional calibration data. The sheer number of redshifts in the DESI BGS make it possible to construct calibration data after binning the galaxies in multiple dimensions based on different galaxy properties. 

The most straightforward way to incorporate additional galaxy and halo properties is to use the \textit{Conditional Abundance Matching} framework \citep{hearin_etal:14, hearin_etal:15, watson_etal:15}. 

This approach will allow for many galaxy properties to influence halo properties. This will naturally add scatter to relationships like the LHMR and SHMR, fixing one of the model deficiencies in this work. These planned improvements are one of the reasons we used the 10-parameter model and not the 14-parameter model used in Paper II (which adds LHMR scatter via galaxy concentration measurements).

\section*{Acknowledgments}
\label{sec:ack}

\noindent We would like to thank Mike Blanton, David Hogg, Ashley Ross, Ravi Sheth, and John Moustakas for helpful conversations. A special thanks to Anthony Carr and Fei Qin and for feedback on preliminary versions of this catalog. Thank you to Zhiwei Shao for providing SHMR scatter data.
\vspace{0.2cm}

\noindent This material is based upon work supported by the National Science Foundation under Grant Number 2307790.
\vspace{0.2cm}

\noindent This material is based upon work supported by the U.S. Department of Energy (DOE), Office of Science, Office of High-Energy Physics, under Contract No. DE–AC02–05CH11231, and by the National Energy Research Scientific Computing Center, a DOE Office of Science User Facility under the same contract. Additional support for DESI was provided by the U.S. National Science Foundation (NSF), Division of Astronomical Sciences under Contract No. AST-0950945 to the NSF’s National Optical-Infrared Astronomy Research Laboratory; the Science and Technology Facilities Council of the United Kingdom; the Gordon and Betty Moore Foundation; the Heising-Simons Foundation; the French Alternative Energies and Atomic Energy Commission (CEA); the Secretariat of Science, Humanities, Technology and Innovation (SECIHTI) of Mexico; the Ministry of Science, Innovation and Universities of Spain (MICIU/AEI/10.13039/501100011033), and by the DESI Member Institutions: \url{https://www.desi.lbl.gov/collaborating-institutions}. Any opinions, findings, and conclusions or recommendations expressed in this material are those of the author(s) and do not necessarily reflect the views of the U. S. National Science Foundation, the U. S. Department of Energy, or any of the listed funding agencies.

The authors are honored to be permitted to conduct scientific research on I'oligam Du'ag (Kitt Peak), a mountain with particular significance to the Tohono O’odham Nation.
\vspace{0.2cm}

\noindent The MultiDark Database used in this paper and the web application providing online access to it were constructed as part of the activities of the German Astrophysical Virtual Observatory as result of a collaboration between the Leibniz-Institute for Astrophysics Potsdam (AIP) and the Spanish MultiDark Consolider Project CSD2009-00064. The Bolshoi and MultiDark simulations were run on the NASA’s Pleiades supercomputer at the NASA Ames Research Center. The MultiDark-Planck (MDPL) and the BigMD simulation suite have been performed in the Supermuc supercomputer at LRZ using time granted by PRACE.

\vspace{0.2cm}

\section*{Data Availability}
\label{sec:data_avail}
 
\noindent The group catalog is publicly available as a value added catalog in DESI Data Release 1. See  \href{https://data.desi.lbl.gov/doc/releases/dr1/#value-added-catalogs}{https://data.desi.lbl.gov/doc/releases/dr1/}.

\vspace{0.2cm}

The data associated with the figures in this paper is publicly available on Zenodo at \href{https://zenodo.org/records/23288597}{https://zenodo.org/records/23288597}.

\vspace{0.2cm}

\noindent The code used to create this work, including the group finder itself, is available publicly at \href{https://github.com/ianw89/SelfCalGroupFinder}{https://github.com/ianw89/SelfCalGroupFinder}, tagged at \texttt{v1.0}.

\vspace{0.2cm}

\noindent Future updates and errata relevant to this work will be posted on \href{https://ianw89.github.io/}{https://ianw89.github.io/} and \href{https://www.galaxygroupfinder.net/}{https://www.galaxygroupfinder.net/}.

\bibliographystyle{aasjournalv7}
\bibliography{example}{}

\appendix

\section{Handling Unobserved Galaxies}
\label{apdx:lostgals}

\subsection{Fiber Incompleteness}

In fiber-fed spectroscopic galaxy surveys, some targets are not able to be observed due to instrumental limitations. In the case of DESI, there are two physical limitations \citep{DESI2016b.Instr, KP3s6-Bianchi}. Firstly, each of the 5000 robotic positioners can only access a small region of the focal plane. In a single exposure, about 15\% of the focal plane is accessible by two fibers; the rest is only accessible by a single fiber. Secondly, the finite physical size of the positioners prevents two very close targets from being observed simultaneously. These limitations imply that in overdense regions of the sky, many targets cannot be observed in a single exposure. Although the DESI survey visits sky regions several times, residual unobserved targets remain in some regions. The incompleteness of the observed sample induced from these limitations is called \textit{fiber incompleteness}.

Fiber incompleteness is the dominant systematic in constructing a BGS group catalog. This is because unobserved targets are a biased sample, residing in overdense areas of the sky. In DR1 and DR2, $\sim 20\%$ of targets in regions of the sky with 3+ passes are lost in this way. Even by the end of the fiducial BGS plan, when 5 passes will have been made over the entire footprint, we expect $\gtrsim 10\%$ incompleteness to remain. 

In SDSS, incompleteness from physical fiber size was present as well, although the effects were less severe than in DESI BGS. Numerous corrections for these effects have been suggested in the literature. However, many of these corrections are for two-point statistics and cannot be utilized for the purposes of building a group catalog, where the individual objects are of interest. In SDSS, the most popular one-point correction is \textit{nearest neighbor}. In this scheme, galaxies lost to fiber incompleteness are assigned redshifts from their nearest on-sky neighbor \citep{Blanton_03, blanton_etal:05_vagc}. This technique worked well for SDSS in many applications. However, at the depth of DESI BGS, there are far more interlopers that confuse this scheme. In this appendix we show how this solution biases group finder statistics, and present a new technique we call \textit{Hybrid Nearest Neighbor} (HNN), which assigns more accurate redshifts and recovers significantly less biased group finder statistics.

\subsection{Methods}

In order to understand how constructing a group catalog from an 80\% complete sample differs from a fully complete sample, we construct a fiducial ``Truth'' catalog and various ``Test'' catalogs. The truth catalog is built from the DESI One-Percent Survey (SV3), part of the DESI Early Data Release \citep{DESI2023b.KP1.EDR}. In this survey, 20 regions of the sky were observed by 10 or 11 bright tiles (and a similar number of dark tiles) in a rosette pattern, creating a highly complete inner region which we cut to. The sample is 106,965 targets from 124 $\mathrm{deg}^2$ with 97.9\% spectroscopic completeness. The group finding method from the main text is used\footnote{Because this fiber incompleteness study needed to be conducted prior to the MCMC of group finder parameters, here we used a fiducial set of group finding parameters, which were not the parameters of the final catalog. After the final catalog was built, we repeated the analysis presented in this appendix using the final parameters to ensure that the conclusions of the fiber incompleteness study did not change.} (see \S\ref{sec:groupfinding}). 

The test catalogs are built by taking a sky-cut of an internal release of a DESI BGS (which will become DR2) that matches the footprint of the SV3 Truth catalog\footnote{Two of the twenty SV3 regions do not have 3 passes in this internal data release, and so we omit those two regions from both the Truth and Test catalogs. These two regions are centered at (RA,DEC) = $(194.7,28.3)$ and $(194.7,24.7)$.}. This test sample is $\sim 104,553$ targets at 78.4\% completeness, where 94\% of the unobserved galaxies have known redshifts in SV3. From this sample we build test catalogs using our HNN technique (along with other methods for comparative purposes). Otherwise the group finding methods are kept exactly the same as the truth catalog.

There are two areas of information that can be utilized in assigning redshifts:
\begin{enumerate}
    \item \textbf{Photometric Redshifts.} Photometric information on the lost galaxies is largely folded into their photometric redshift estimates, $z_{\mathrm{phot}}$. We make use of photometric redshifts from the 9.1 (North) and 10.1 (South) photo-z catalogs available from the DESI Legacy Imaging Survey \citep{Zhou_2023}. Their uncertainty, from examining BGS DR1 spectroscopic galaxies, is roughly $\sigma=0.012+0.031z_{\rm true}$.
    \item \textbf{Neighbor information.} Because galaxy positions are correlated with their on-sky neighbors, additional distance information is available by considering the redshifts of on-sky neighbors. This information is not used in the $\zphot$ estimates.
\end{enumerate}
 
Photometric redshifts at low $z$ are not accurate enough on their own to be reliably used in group finding. If $z_{\mathrm{phot}}$ is within 0.005 ($\sim1500$ km/s) of the spectroscopic redshift, then they would be usable our group finder, as this is the order of large peculiar velocities in rich clusters which the group finder routinely handles. Unfortunately the available $z_{\mathrm{phot}}$ are only this close $\sim 33\%$ of the time, though they are rarely catastrophically wrong ($95\%$ fall within $z_{\mathrm{phot}}\pm0.055$). In position space, the median uncertainty corresponds to about $\sim$ 60 Mpc uncertainty on top of peculiar velocities. As a consequence, simply using the photometric redshifts in group finding artificially lowers the satellite fraction as it is likely to be placed too far from other structures. However, since $z_\mathrm{{phot}}$ is \textit{roughly} correct, it effectively brackets the viable redshift range to a much smaller region than the entire $0.001<z<0.5$ range. 

In \cite{Yang.ExtendedGF.21}, which performed group finding on the entire Legacy Survey where most redshifts are be photometric, the roughness of $\zphot$ and suppression of $\fsat$ was dealt with by loosening the redshift-space satellite finding in the group finder itself. They update Eq.~\ref{eq:pz} to a wider Gaussian based on the uncertainty in the $\zphot$ instead. This, combined with lowering the $\bsat$ parameter to encourage more satellites to be found, allowed the group finder to operate on large numbers of photometric-only galaxies but recover a reasonable $\fsat$. Here we opt for a different tactic, leaving the core group finding algorithm unchanged and assigning redshifts as a pre-processing step. 

\begin{figure*}
    \centering
    \includegraphics[width=1\linewidth]{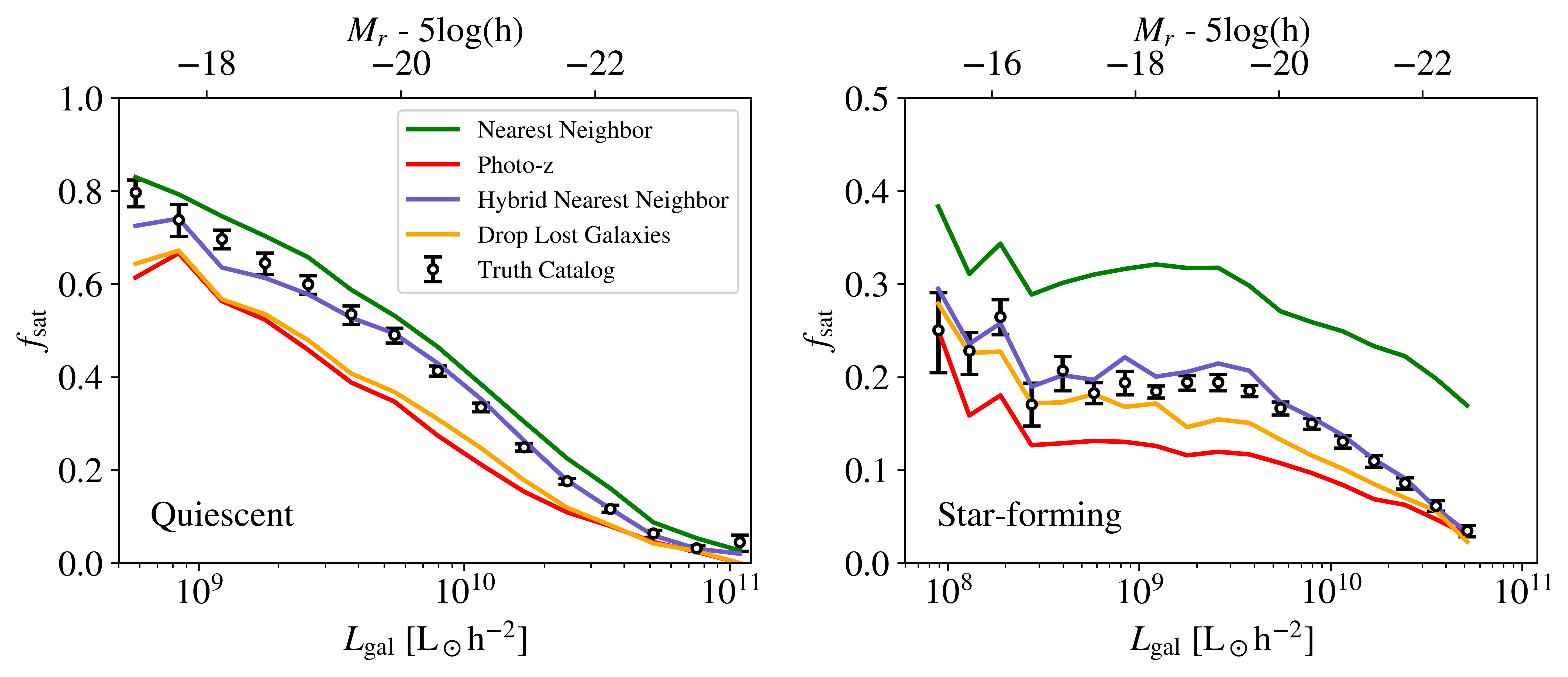}
    \caption{\textbf{Satellite fraction.} The two plots show the satellite fraction $\fsat$ split by quiescent and star-forming samples. The truth catalog from SV3 is shown in black, with bootstrapped error bars. Our new HNN method, used in the main text, is shown in purple. It largely recovers the correct satellite fraction. Results from catalogs constructed using various naive techniques are shown in orange, red, and green.}
    \label{fig:fsat-sv3}
\end{figure*}

To assign redshifts to the lost galaxies\footnote{Note that here and throughout this work, lost galaxies include galaxies that were observed spectroscopically but produced sufficiently poor redshift fittings as to be considered unobserved. We use the $\Delta \chi^2 < 40$ criteria commonly used throughout the DESI Collaboration for this purpose \citep{LSSCatalogs.Ross.2025, DESI2024.I.DR1}.}, we make use of both $\zphot$ and on-sky neighbor information in tandem. The HNN technique is designed to to use a mix of assigned redshifts from on-sky neighbors and photometric redshifts. The algorithm only takes a neighbor's redshift if it is close to the lost galaxy's $\zphot$ and if observed galaxy-galaxy pairs with similar properties (e.g., angular distance between them) are typically at the same redshift. The full algorithm is as follows.\\
\\
\noindent{For each lost galaxy:}
\begin{enumerate}
    \item Sort their nearest $N$ on-sky observed neighbors\footnote{Unlike the main sample which is limited to $r<19.5$, we apply no apparent magnitude cut for the neighbors considered. We are simply increasing the pool of neighbor redshifts available to include the FAINT sample, which is similar to the BRIGHT sample but deeper by $\sim0.8$ mag.} by projected distance to the lost galaxy.
    \item For the $n^{\mathrm{th}}$ on-sky neighbor, compute a score $S_n$ based the how often similar galaxy pairs are at the same redshift, and how close the $\zphot$ of the lost galaxy is to the neighbor's $z$. 
    \item Choose the neighbor with the highest score, $S_{\mathrm{max}}$.
    \item If $S_{\mathrm{max}} > 1.0$, use this neighbor's redshift. Otherwise, use the lost galaxy's $z_{\mathrm{phot}}$. 
\end{enumerate}

This procedure assigns a redshift to every lost galaxy, either from an on-sky neighbor or $\zphot$. The scoring function used for neighbor $n$ is
\begin{equation}
S_n =A~\exp\bigg\{\frac{-|z_{\mathrm{phot}}-z_n|^2}{10^{-\sigma}}\bigg\} + B~\mathrm{pairscore}(\theta_n, z_n, c_n, r_t, c_t) .
\end{equation}

Here, $z_n$ and $\theta_n$ are the neighbor's redshift and angular distance from the target, and $z_{\mathrm{phot}}$ and $r_t$ are the target's photo-z and $r$-band apparent magnitude. $c_t$ and $c_n$ are binary color values (quiescent vs star-forming) for the target and neighbor. The classification $c_n$ is determined using the procedure described in~\S\ref{sec:quiescent}, but $c_t$ is determined using a $g-r$ cut instead because we have no spectral information. To determine $c_t$ we first $k$-correct $M_r$ and $M_g$ using the same lookup table described in~\S\ref{sec:fiberincompleteness} but with $\zphot$ instead of $\zspec$, which we do not have. Then, we compute a color $M_g^{0.1}-M_r^{0.1}$ and compare to a $\lgal$-dependent cut, which was determined analogously to the $\dn$-based cut described in the main text. The equation for color threshold for lost galaxy quiescent classification is 

\begin{equation}
(g-r)_{\rm crit} = 0.718 + 0.066 \Bigg(1 + \mathrm{erf}\bigg(\frac{\log\lgal-9.461}{0.728}\bigg) \Bigg).
\end{equation}

\noindent The absolute magnitude and quiescent classification will be re-computed later once a new redshift is assigned, as discussed in~\S\ref{sec:fiberincompleteness}. 

$A$ and $B$ are order unity constants controlling the relative importance of the two terms and overall neighbor vs. $\zphot$ use. The width of the Gaussian $10^{-\sigma}$ is similar to the error in the $\zphot$ for $\sigma \sim 2-3$, which can be slightly different for quiescent vs. star-forming galaxies. For this reason, four separate versions of the constants $A$, $B$, and $\sigma$ exist for each color combination of the target and neighbor (red-red, red-blue, blue-red, blue-blue). 


The pairscore function returns a value between 0 and 1 that is the fraction of observed galaxy pairs with the ``same" redshift. To implement this, a 5-dimensional ($\theta_n, z_n, c_n, r_t, c_t$) binned matrix was built from observed galaxy pairs in SV3. A ``same redshift" score was computed for each pair using using a rounded tophat of width $\sim0.005$ ($\sim1500 \rm~km/s$, a peculiar velocity for a rich cluster), and then these values are inserted into the matrix. We also apply Gaussian smoothing with $\sigma=0.5$, half of a cell width. To get the actual values when pairscore is called, linear interpolation is used. 

The parameter values used are shown in Table~\ref{tab:hnn_params}, with the number of neighbors considered $N=8$. The parameters are such that both terms must contribute to be able to use a neighbor's redshift -- their distance in redshift space shouldn't exceed the uncertainty on $\zphot$, and their angular separation must be close enough that similar galaxy pairs are fairly often at the same redshift. When applied to our main catalog sample, HNN with these parameters uses neighbor redshifts for $59\%$ of lost galaxies. 

\begin{table}[]
    \centering
    \begin{tabular}{c|c||c|c|c}
        \textbf{Target} & \textbf{Neighbor} & $A$ & $B$ & $\sigma$ \\
        \hline
        \hline
        Star-forming & Star-forming & 0.95 & 1.1 & 3.0  \\
        Quiescent & Star-Forming & 0.98 & 1.2 & 2.6     \\
        Star-Forming & Quiescent & 0.95 & 1.1 & 3.0     \\
        Quiescent & Quiescent & 0.98 & 1.35 & 2.5       \\
    \end{tabular}
    \caption{Parameters used for the Hybrid Nearest Neighbor redshift assignments. These parameters balances the effects of the redshift assignment on a variety of group finder summary statistics while still assigning more correct redshifts than basic methods. }
    \label{tab:hnn_params}
\end{table}

In the results below we also compare HNN to several naive methods. The orange lines are results when the lost galaxies are dropped from the catalog. The green lines are Nearest Neighbor, which always sets the redshift of a lost galaxy to its nearest on-sky spectroscopic neighbor's. The red lines are when the $\zphot$ is used for every lost galaxy.

\subsection{Results}

We use the following metrics to gauge the effectiveness of the method: the satellite fraction $f_{\mathrm{sat}}(L)$, the luminosity function of the lost galaxies $\Phi_L(L)$, the purity and completeness of the central/satellite assignments with respect to the truth catalog, the accuracy of assigned redshifts, and the projected clustering $w_{p}(r_p)$. Ideal lost galaxy handling would assign accurate redshifts to all the lost galaxies and recover unbiased group catalog statistics. We find that attempting to maximize accurate redshift assignments can worsen biases in the group catalog statistics. We prioritize minimizing biases over maximizing correct redshifts.

The satellite fraction, split into the quiescent and star-forming subsamples, is shown in Fig.~\ref{fig:fsat-sv3}. Dropping lost galaxies biases $f_{\mathrm{sat}}$ in the expected way, as the unobserved targets are from overdense regions, and therefore more likely to satellites. Similarly, using $\zphot$ suppresses $\fsat$ due to the insufficient accuracy of $\zphot$. The basic nearest neighbor technique, which worked well in SDSS, overestimates $\fsat$ by associating lost galaxies with interloping on-sky neighbors. The HNN technique balances these effects by mixing neighbor and $\zphot$ usage.

\begin{figure}
    \centering
    \includegraphics[width=1\linewidth]{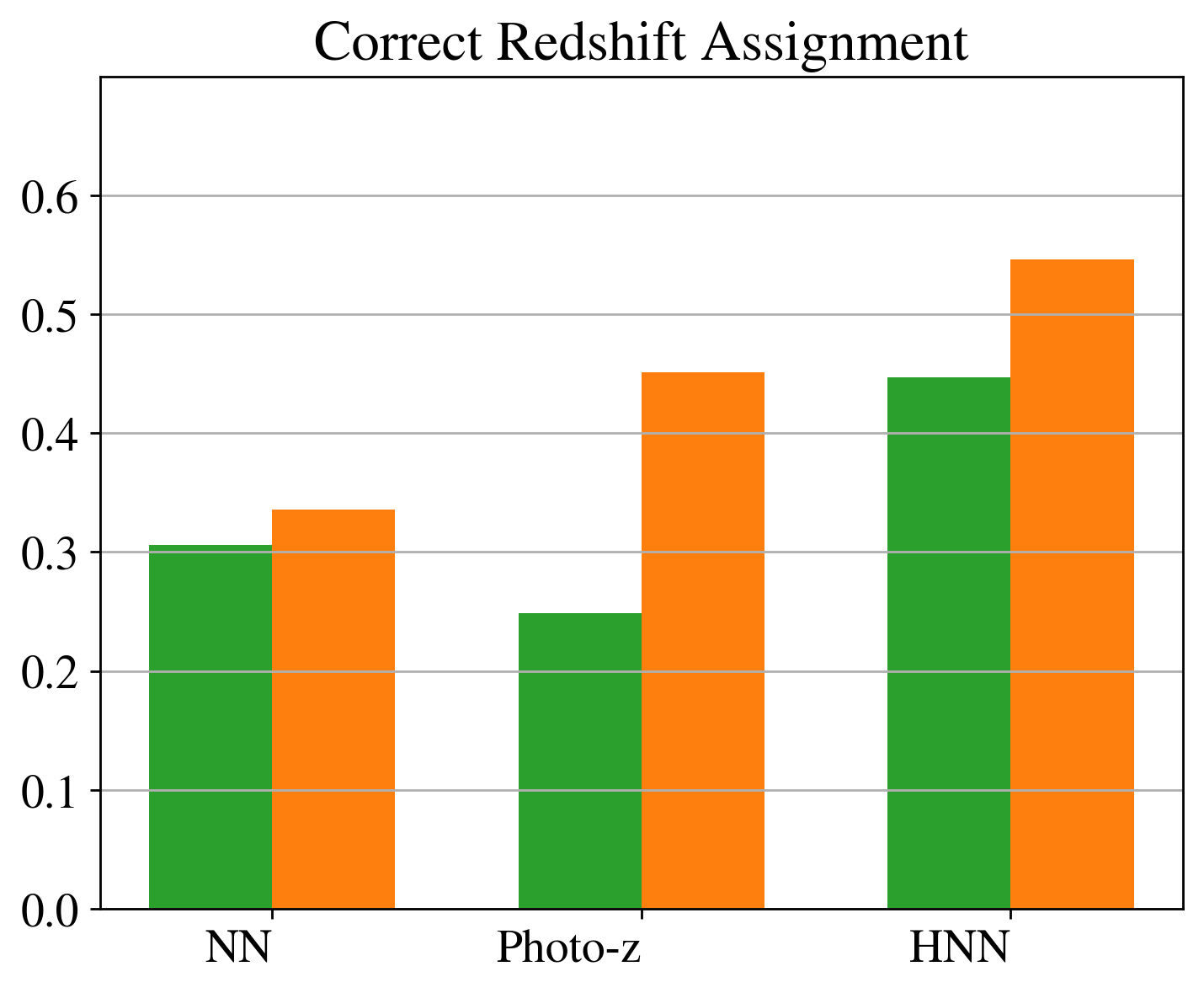}
    \caption{\textbf{Correctness of Redshift Assignments.} The green bars show the fraction of lost galaxies whose assigned redshifts approximately match the known spectroscopic redshift from SV3; $\Delta z < 0.005$. The orange bars show the fraction with $\Delta z < 0.01$. The HNN method shifts some lost galaxies from their $\zphot$ towards a neighbor, which is responsible for the increase in correct redshifts.}
    \label{fig:correct-redshifts}
\end{figure}

In Fig.~\ref{fig:correct-redshifts} we show the fraction of correct assigned redshifts in the green bars for HNN along with basic nearest neighbor and using only $\zphot$. We consider an assigned redshift correct if $\Delta z < 0.005$, because this is the order of a large peculiar velocity, which the group finder is able to handle. The orange bars show the fraction with $\Delta z < 0.01$, which can be thought of as the fraction of correct assignments and near misses. The HNN method effectively converts near misses from $\zphot$ into more exact redshifts by using the neighbor information.

\begin{figure*}
    \centering
    \includegraphics[width=1\linewidth]{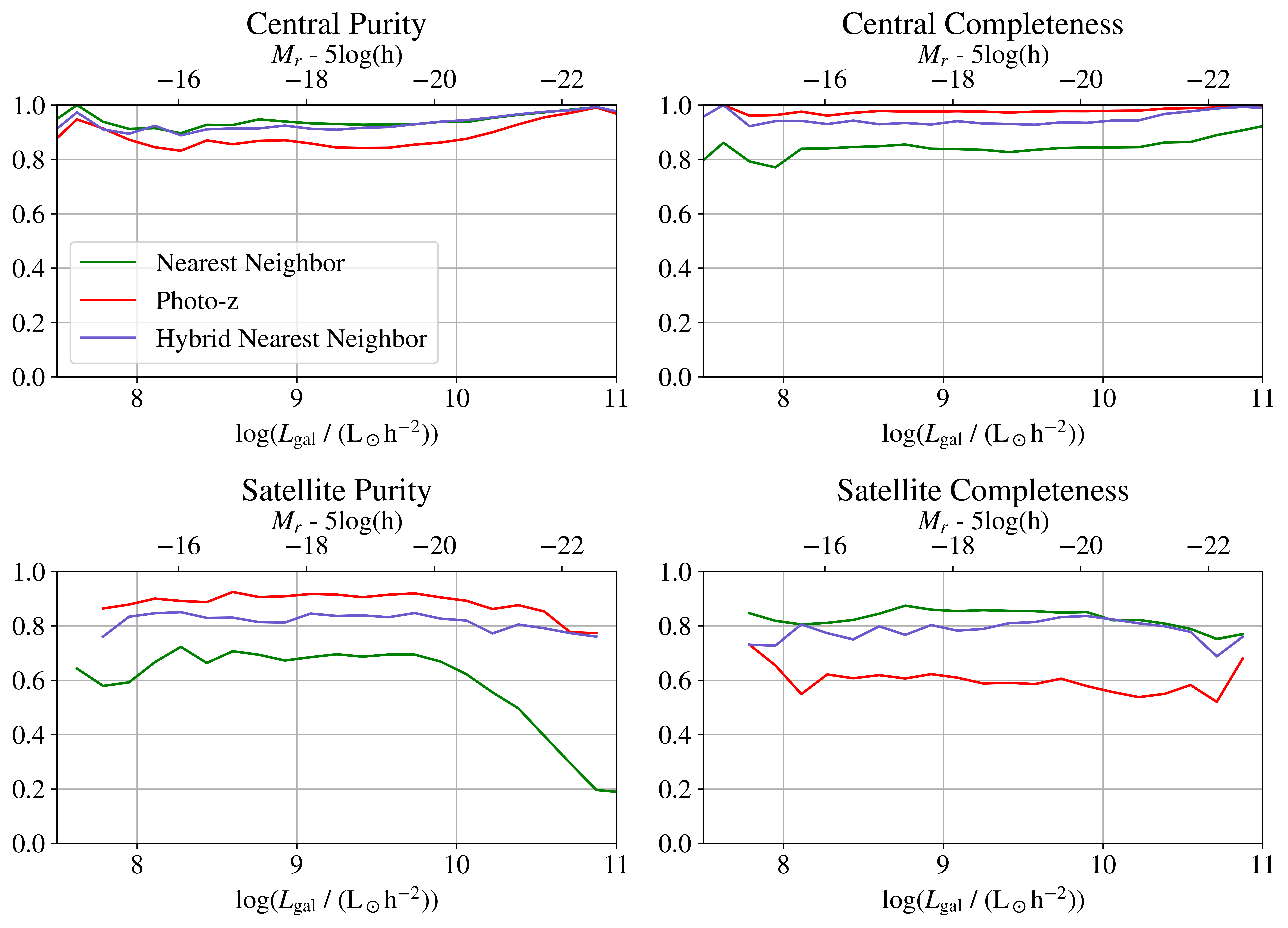}
    \caption{\textbf{Effect of fiber incompleteness on purity and completeness.} Using many neighbor redshifts, as in the green curves, is good for central purity and satellite completeness, but poor for satellite purity and central completeness. Using $\zphot$ only has the opposite effect. Our HNN approach balances these effects. The overall satellite purity and completeness for HNN is $82.3\%$ and $81.2\%$ respectively. The overall purity and completeness of centrals is 94.1\% and 94.5\%. Note that this comparison is to our SV3 ``Truth'' catalog, not a mock catalog, and so the true purity and completeness would include these fiber-incompleteness systematics and general group finding systematics as shown in Paper I, which lowers these numbers somewhat.}
    \label{fig:purity-completeness}
\end{figure*}

The purity and completeness with respect to the truth catalog is shown in Fig.~\ref{fig:purity-completeness}, alongside catalogs constructed using the naive lost galaxy handling methods. When a galaxy is given a neighbor's redshift incorrectly, the group finder will erroneously categorize it as a satellite, lowering satellite purity (if it is brighter and becomes the central, the true central will become a satellite and lower both the central and satellite purity). This is the inevitable trade-off in attempting to use neighbors to assign redshifts. Matching lost galaxies to neighbors instead of using $z_\mathrm{phot}$ improves satellite completeness but hurts the purity. On the other hand, using more photometric redshifts results in low satellite completeness as the galaxies are placed in semi-random locations (likely away from other structure), which eventually also shows up in lowered purity of the centrals. The chosen parameters balance these effects.

\begin{figure*}
    \centering
    \includegraphics[width=1\linewidth]{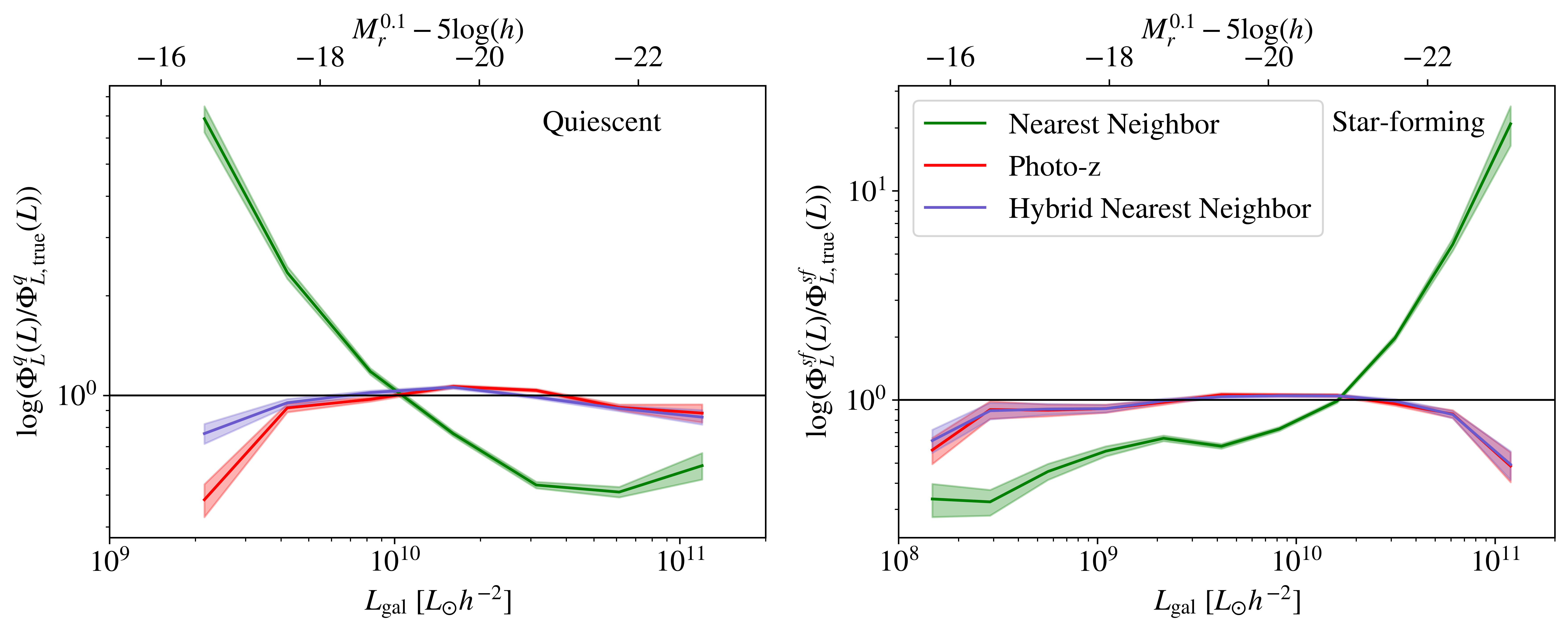}
    \caption{\textbf{Effect on Lost Galaxy Luminosity Function}. The left (right) plot shows the change in the luminosity function of lost quiescent (star-forming) galaxies. The shaded regions are bootstrapped error bars on the lost galaxy sample. The HNN method preserves the luminosity distribution of the lost galaxies to about 10\% within most of the heavily populated parts of $\Phi_L(L)$. It does at least as well as $\zphot$ does; and much better than basic nearest neighbor, which goes to over $10\times \Phi_L(L)$ in some regimes. Preserving $\Phi_L(L)$ helps the abundance matching step and minimizes effects on the LHMR.} 
    \label{fig:altered-lum-func}
\end{figure*}

Because halo-based group finding uses inverse abundance matching, the luminosity function of galaxies is important. The luminosity function of the lost galaxies, $\Phi_L(L)$, is different from the overall luminosity function $\Phi(L)$, as is the case for red and blue subsamples $\Phi_{L}^{\rm q}(L)$ and $\Phi_{L}^{\rm sf}(L)$. In Fig.~\ref{fig:altered-lum-func}, we show how $\Phi_{L}^{\rm q}(L)$ and $\Phi_{L}^{\rm sf}(L)$ are transformed from the measured values in SV3 due to our redshift assignments. While imperfect, our technique is no worse than using $z_\mathrm{phot}$, and is mostly biased on the tails where there are far fewer galaxies. A 20\% difference at some $L$ here translates to a 4\%  difference overall (in an 80\% complete sample), and so our processing has a modest impact on abundance matching even in the biased regimes.

\begin{figure}
    \centering
    \includegraphics[width=1.0\linewidth]{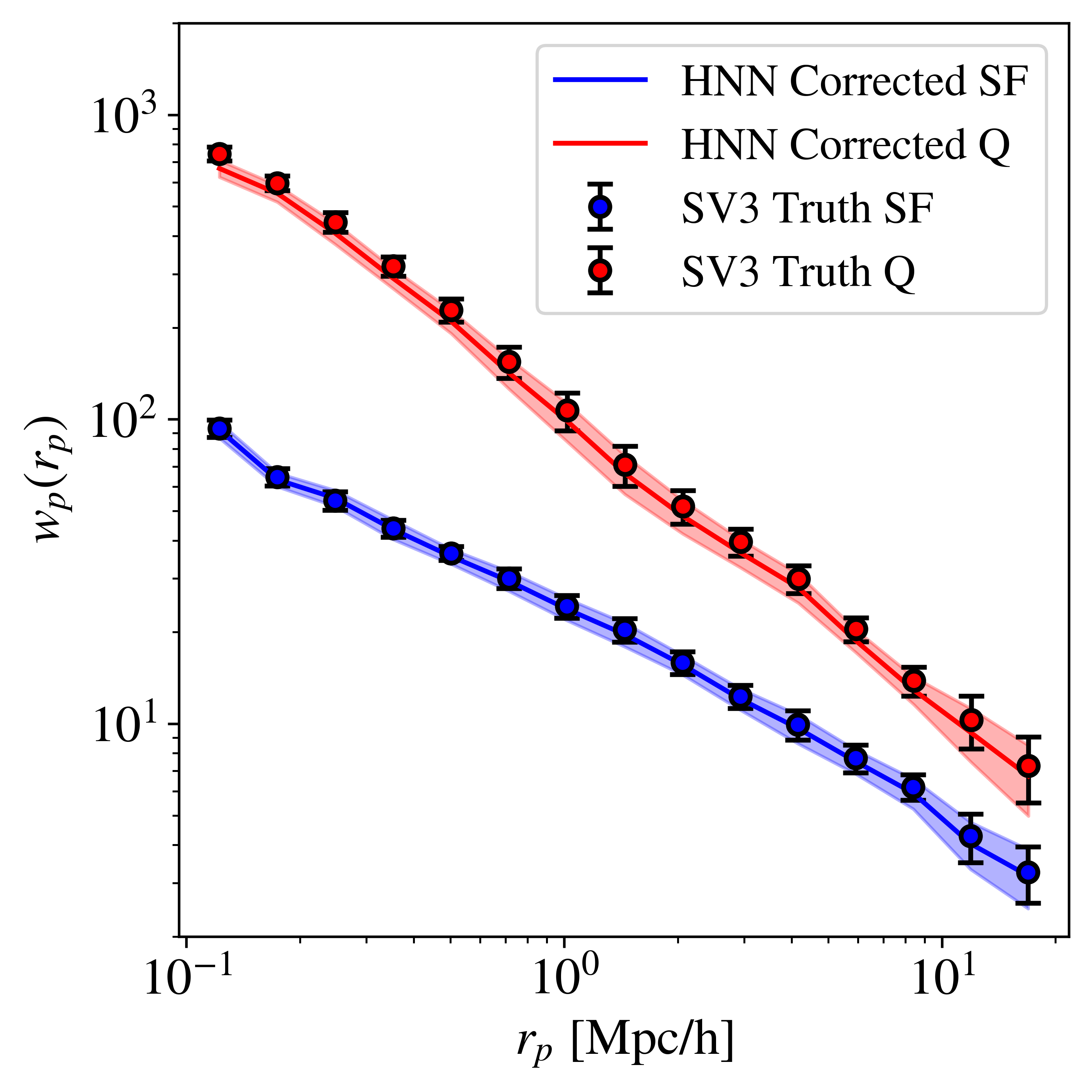}
    \caption{\textbf{Effect of lost galaxy handling on clustering.} The projected clustering is shown for star-forming and quiescent galaxy samples for the 98\% complete truth catalog (points with error bars) and the 80\% complete test catalog with redshifts assigned HNN (lines with shaded regions). The error estimates are from jackknife resampling.}
    \label{fig:fiber-on-clustering}
\end{figure}

A final test of the method is the effect on color-dependent projected clustering. Because HNN repositions lost galaxies in redshift space, clustering measurements will generally be sensitive to such changes. In Fig.~\ref{fig:fiber-on-clustering} we compare the clustering for quiescent and star-forming samples for the truth and test catalogs using HNN to assign missing redshifts. No additional weights are used in these calculations. The results are within the error bars of each other. Note that any small shifts in the clustering amplitude induced by HNN would not be directly problematic for this work because in our MCMC process we use the clustering of the spectroscopic sample only in our $\chi^2$ (using pair-inverse-probability weights to account for fiber incompleteness), and compare to the clustering of \textit{mocks} built from the group catalog's galaxy-halo connection.

Note that the truth catalog shown in all of these results still has 2.1\% lost galaxies. We use HNN for this small residual incompleteness in the truth catalog. Because it is such a small percent, we find that the various strategies for handling incompleteness within the truth catalog barely affect the summary statistics we study.

\section{BGS-SDSS Comparisons}
\label{apdx:sdss_compare}

Galaxy clustering measurements are part of the calibration data in this work. To aid in the comparison between results here and Paper II, we compare our results to the SDSS clustering measurements from Paper II and \cite{zehavi_etal:11}.

\begin{figure}
    \centering
    \includegraphics[width=1.0\linewidth]{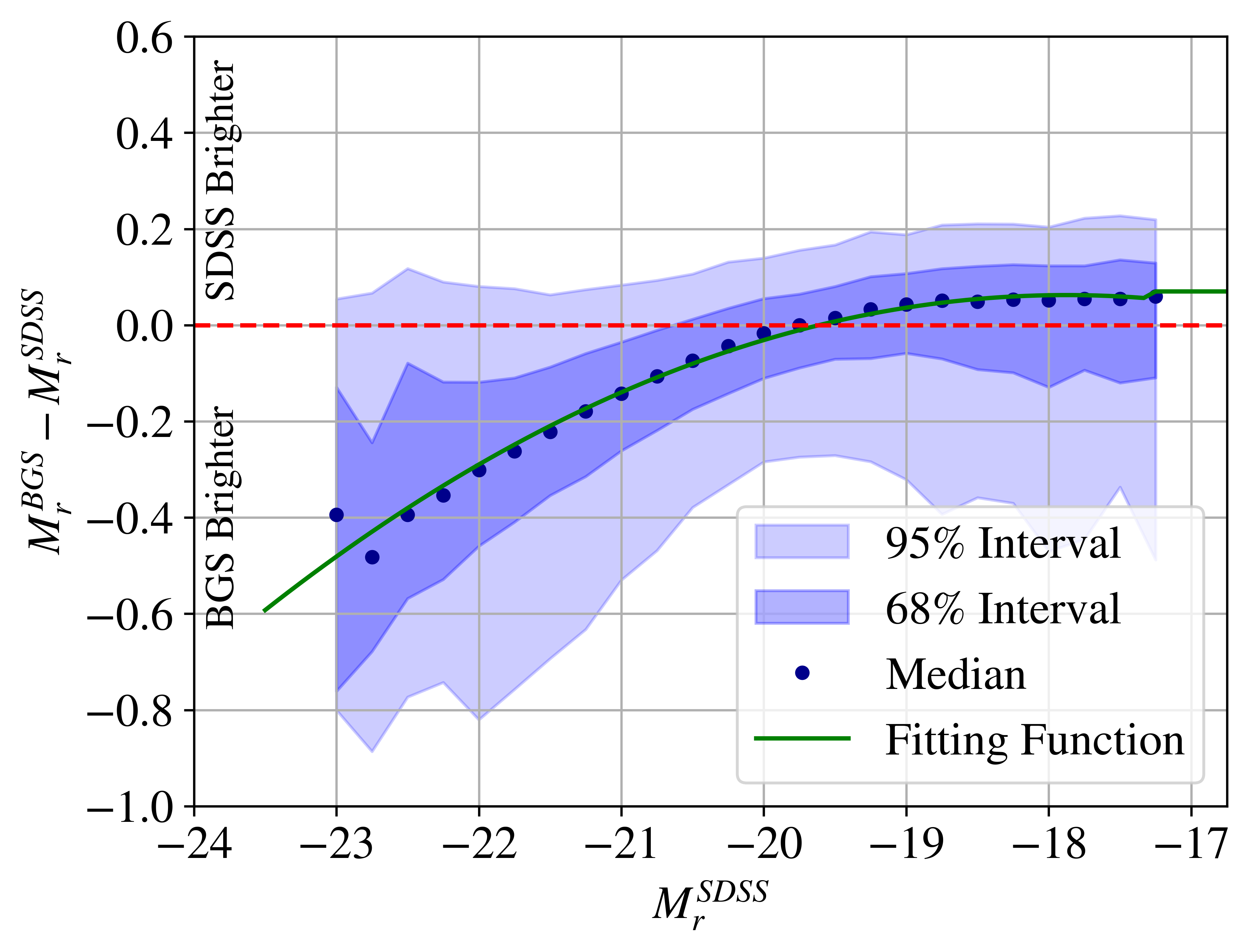}
    \caption{\textbf{Differences in $M_r^{0.1}$ from SDSS to BGS.} For galaxies in both the SDSS and BGS, we show the differences $K$-corrected absolute magnitudes from NYU-VAGC versus \texttt{FastSpecFit}. The shaded regions give the 68\% and 95\% scatter of values within the bin, with the dot showing the median. The differences are stronger for bright galaxies, with wide scatter everywhere. These differences complicate DESI-SDSS comparisons when binned by luminosity.}
    \label{fig:bgs_sdss_mag_compare}
\end{figure}

There are several complications in comparing these measurements to the SDSS ones. Firstly, increased noise in $\dn$ relative to SDSS reduces the purity of the samples in BGS, reducing the clustering amplitude, most pronounced in the faint quiescent galaxies where the quiescent fraction is lowest. Secondly, differences in the measured luminosities affect the clustering amplitudes when binned by luminosity, especially at high $\lgal$. Fig.~\ref{fig:bgs_sdss_mag_compare} shows how $M_r^{0.1}$ differs from BGS to SDSS for a sample of 71,499 overlapping targets (recall that by SDSS we specifically mean the magnitudes in NYU-VAGC, \cite{blanton_etal:05_vagc}). We find that bright galaxies in SDSS are even brighter in BGS, while fainter ones are mildly fainter in BGS, with $\sim0.1-0.2$ dex scatter. These differences have not been systematically studied in detail, but are reasonable given the software modeling differences between the Sloan and Legacy imaging surveys (Dustin Lang, private correspondence, and \citealt{LS.Overview.Dey.2019}). A third effect is evolutionary: as BGS goes deeper than SDSS, the population of galaxies within an otherwise equivalently defined luminosity bin is different, potentially leading to clustering changes.

\begin{figure*}
    \centering
    \includegraphics[width=1.0\linewidth]{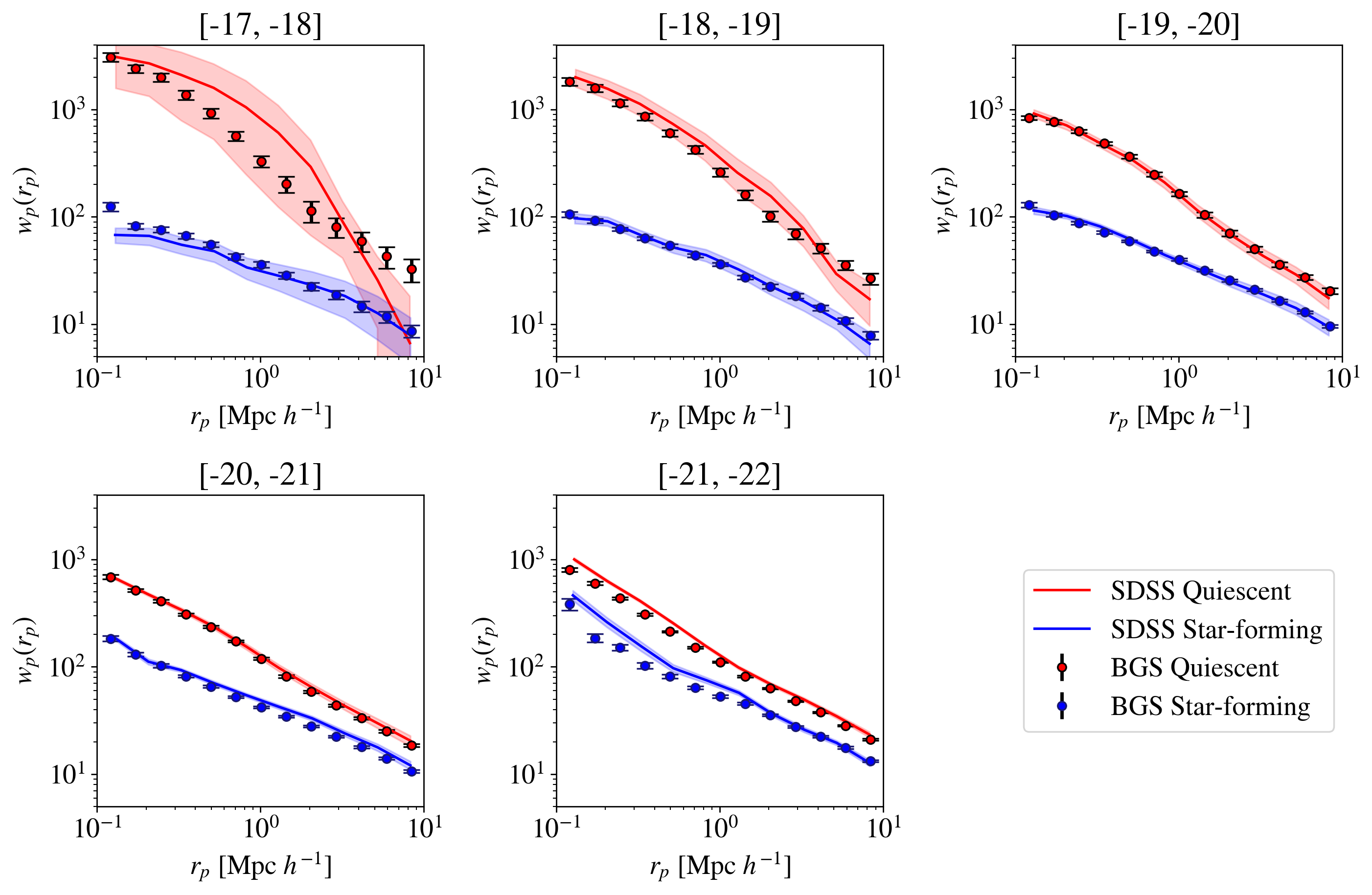}
    \caption{\textbf{Clustering comparison from BGS to SDSS.} The lines and shaded regions are the SDSS Paper II measurements, and the dots with error bars are the new BGS measurements, all with jackknife error bars. The quiescent clustering in the fainter bins is lower in BGS due to difficulty separating quiescent from star-forming galaxies. On the bright end, SDSS measurements have higher amplitude than the BGS ones because of the difference in the luminosities shown in the previous figure. Over half of the difference is removed by shifting the magnitude range of the BGS sample according to the fitting function shown in Fig.~\ref{fig:bgs_sdss_mag_compare}.}
    \label{fig:bgs_sdss_clustering_compare}
\end{figure*}

Some of these effects can be seen in Fig.~\ref{fig:bgs_sdss_clustering_compare}. The lines and shaded regions are the SDSS Paper II measurements, and the dots with error bars are the new BGS measurements. In the faintest two bins, the quiescent clustering is likely lower in BGS due to contamination from star-forming galaxies, although the measurements are within the SDSS error bars. In the brightest bin the SDSS measurements have higher amplitude than the BGS ones because of the difference in the luminosities; shifting the definition of the bins reduces this effect.

\begin{figure}
    \centering
    \includegraphics[width=1.0\linewidth]{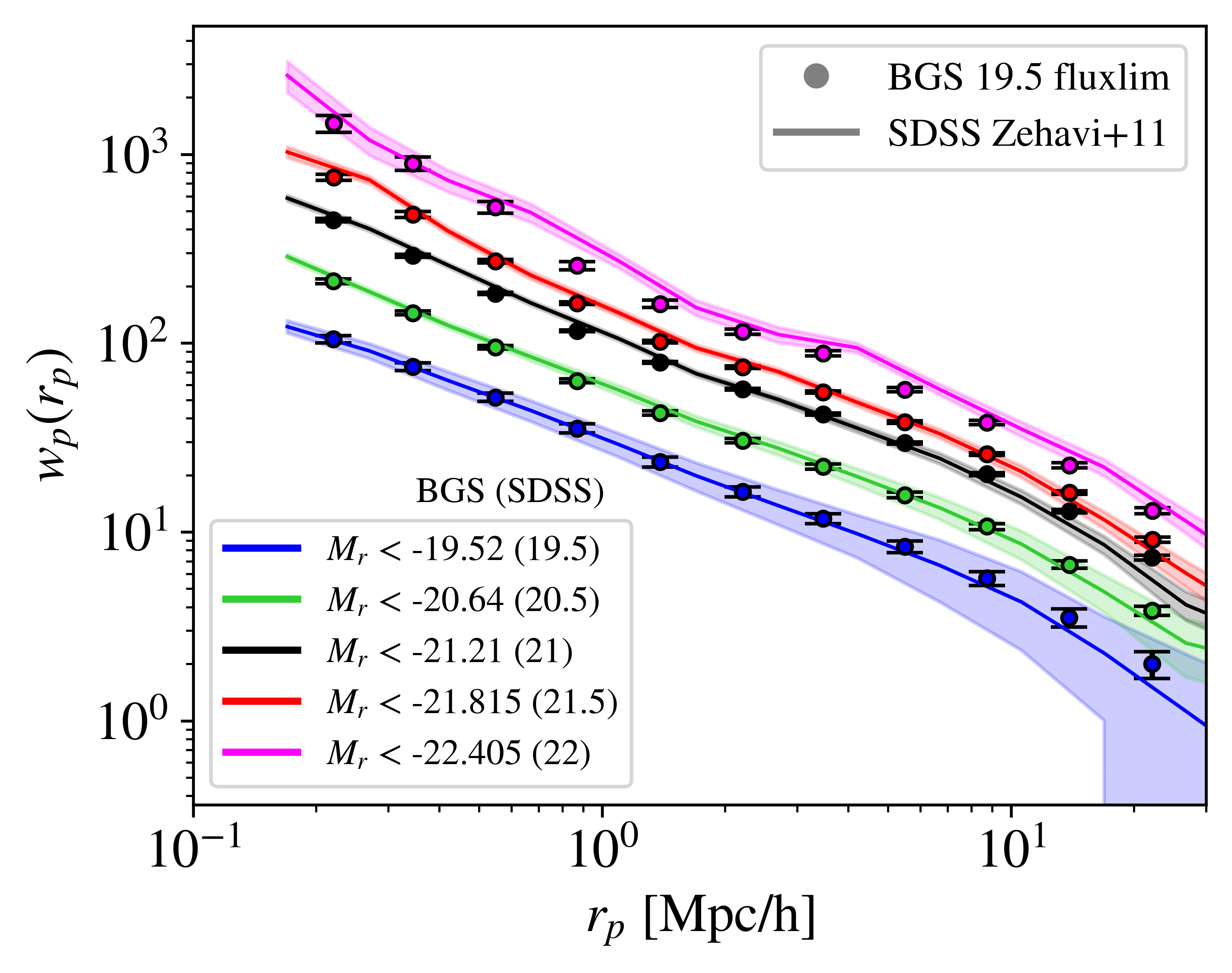}
    \caption{\textbf{Clustering comparison in $M_r^{0.1}$ thresholds.} To compare BGS projected clustering to the SDSS-based Luminosity threshold measurements of \cite{zehavi_etal:11}, we adjust the magnitude threshold until the galaxy density over the same depth matches. Each color shows a different threshold (in parenthesis is the SDSS threshold). The dots show the new BGS measurements with jackknife error bars and the lines with shaded regions are the \cite{zehavi_etal:11} measurements. }
    \label{fig:wp_bgs_sdss_thresh_compare}
\end{figure}

We also show luminosity threshold measurements on the color-combined sample. Here we have taken the SDSS luminosity thresholds definitions of \cite{zehavi_etal:11} and increased the threshold value until the number density in BGS (over the same volumes) matches the measurements in \cite{zehavi_etal:11}. This allows a comparison between BGS and SDSS where the first effect is fully removed and the second effect is partially ameliorated. The results are shown in Fig.~\ref{fig:wp_bgs_sdss_thresh_compare}, which match fairly well. The remaining small differences might be evolutionary, from sample variance, or from more complicated photometric differences uncontrolled by the simple threshold shift performed here. 

Fig.~\ref{fig:bgs_sdss_mag_compare}, showing the photometric differences between BGS and SDSS, also shows an empirical fit to the mean differences in magnitudes. Here $M_{\rm SDSS}$ is the SDSS $r$-band magnitude $K$-corrected to $z=0.1$ and $M_{\rm DESI}$ is the DESI \texttt{FastSpecFit} $r$-band magnitude $K$-corrected to $z=0.1$, the function used to convert an SDSS-like magnitude to a BGS-like one is $M_{\rm DESI} = M_{\rm SDSS} - C$, where the correction $C$ is:

\begin{equation}
C=
\begin{cases} 
  -0.07 &;~ -17.3 < M_{\rm SDSS}  \\
  6.549 +0.740 M_{\rm SDSS} + 0.021 M_{\rm SDSS}^2 &;~ -17.3 > M_{\rm SDSS} \\
\end{cases}
\end{equation}

\noindent Due to the scatter in the relation, these conversions are imperfect but can be useful when making comparisons across the surveys. In this work, the fitting function is only used to aid in comparing to the Paper II quiescent classification method (Fig.~\ref{fig:quiescent_classification}).

\section{Halo Occupation Distributions}
\label{apdx:hods}

\begin{figure*}
    \centering
    \includegraphics[width=1.0\linewidth]{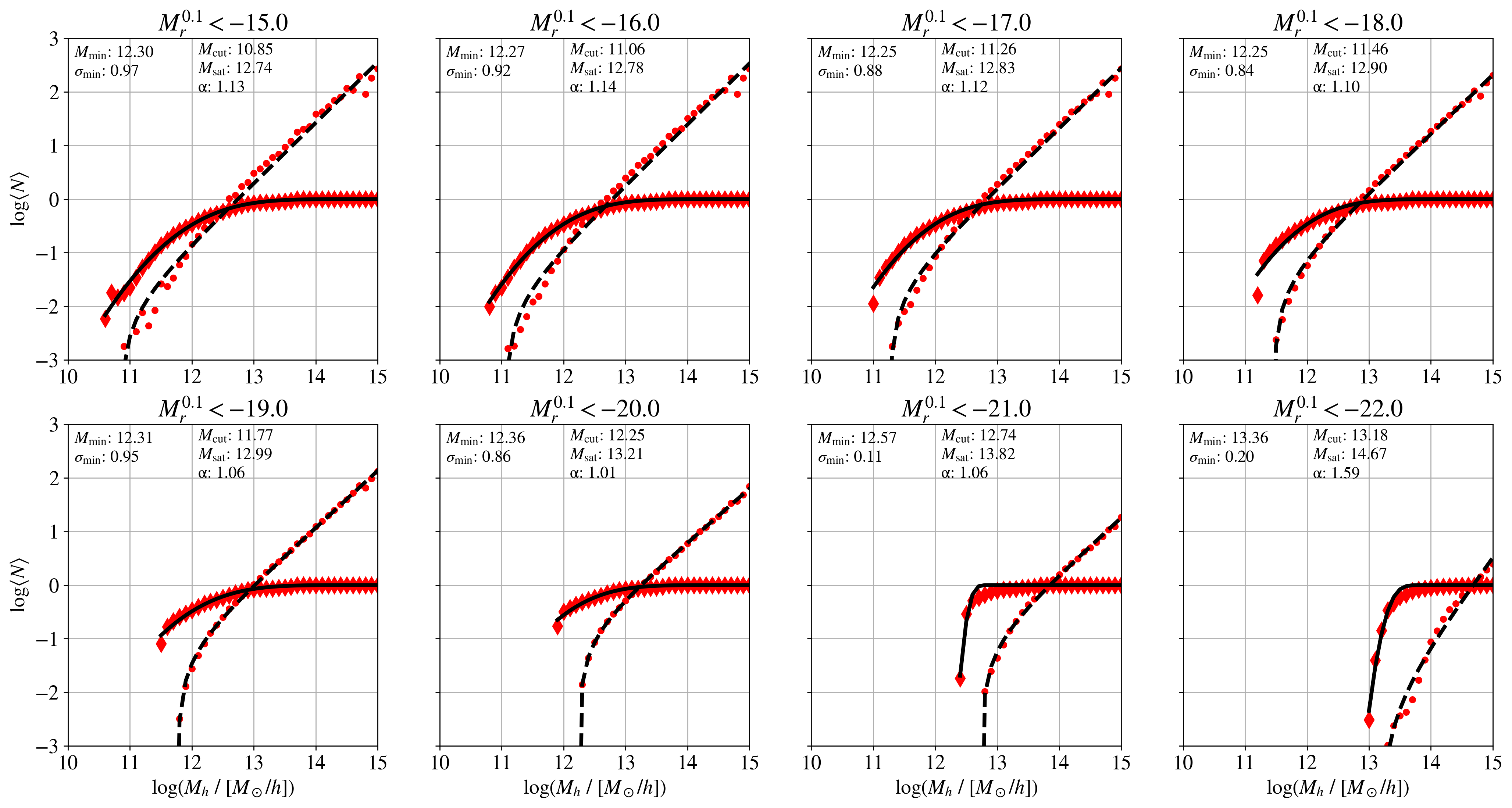}
    \caption{\textbf{Quiescent HOD.} Shown is the halo occupancy of quiescent galaxies in the final catalog. The red diamonds show the percent of halos at a given mass that host a central quiescent galaxy brighter than the indicated magnitude. The red dots show the mean number of quiescent satellites brighter than the threshold in halos at the given mass. The black lines are fits to these values using the model described in the main text with parameter values shown in each plot.}
    \label{fig:hod_th_q}
\end{figure*}

\begin{figure*}
    \centering
    \includegraphics[width=1.0\linewidth]{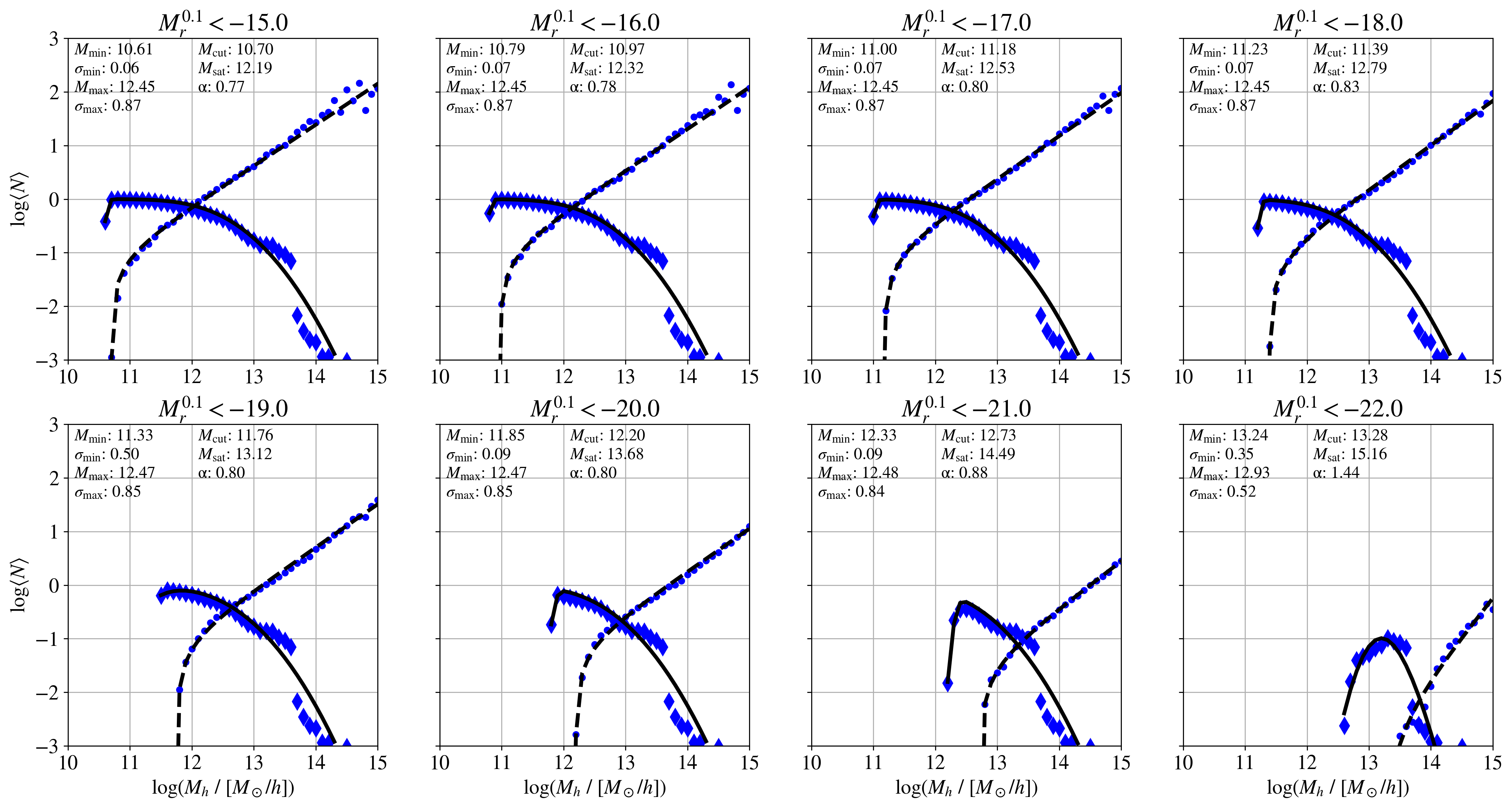}
    \caption{\textbf{Star-forming HOD.} Similar to Fig.~\ref{fig:hod_th_q}, except for star-forming galaxies. The star-forming samples uses a extra error function in the $\langle N_{\rm cen}^{\rm sf} \rangle$ model, which is necessary to capture the falling occupancy towards high halo masses. The satellite occupancy powerlaw is significantly shallower than for the quiescent sample. }
    \label{fig:hod_th_sf}
\end{figure*}

\begin{figure*}
    \centering
    \includegraphics[width=1.0\linewidth]{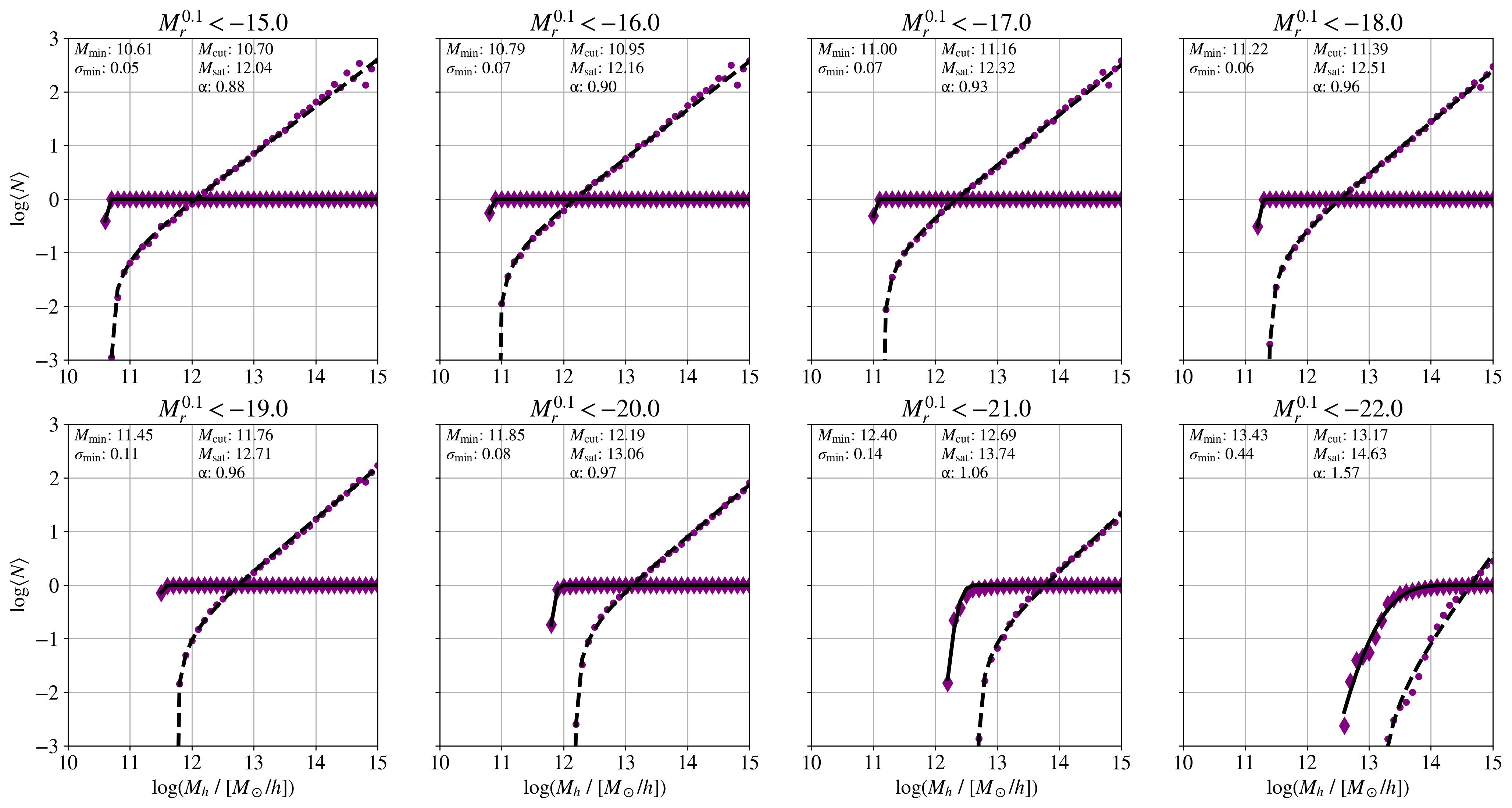}
    \caption{\textbf{Combined Sample HOD.} From the final catalog, the purple diamonds show the fraction of halos at a given mass that hose a galaxy in brighter than the threshold value shown. The dots show the satellite occupation. The group finder results are well-fit by the HOD. In the faintest two threshold samples there may be more satellites in very massive halos than the model can fit without an additional freedom, but this regime is also noisy as there are few $\log \Mh > 14$ halos in the volumes of those samples.}
    \label{fig:hod_th_all}
\end{figure*}

The group finder provides a non-parametric model of the galaxy-halo connection. However, by tabulating the halo occupancy, the model can then fit to common HOD parameterizations. In Fig.~\ref{fig:hod_th_q} and Fig.~\ref{fig:hod_th_sf} we show HODs for quiescent and star-forming galaxies in luminosity threshold samples, as well as Fig.~\ref{fig:hod_th_all} which shows the combined sample.

The colored diamonds show the tabulated occupation for central galaxies; the dots are satellites. The black lines show the best fit model. For the quiescent centrals, we adopt:

\begin{equation}
    \langle N_{\rm cen}^{\rm q} \rangle = \frac{1}{2}\Bigg[1+\mathrm{erf}\bigg( \frac{\log M - \log M_{\rm min}^{\rm q} }{\sigma_{\rm min}^{\rm q}} \bigg) \Bigg],
    \label{eq:ncen_q}
\end{equation} 

\noindent the standard 2-parameter model of \cite{zheng_etal:05}, with the modification that $\langle N_{\rm cen}^{\rm q} \rangle =0$ for $\Mh<M_0$, where $M_0$ is the smallest non-zero halo mass bin. We include this sharp turn-off because the number density of the fitting function and it's implied $\fsat$ should match the group finder results, which requires the function to have no occupancy where the group finder has no occupancy. The model fits the quiescent samples well, with noticeable differences only in the brightest two samples. The combined sample also uses this form.

Because star-forming galaxies are rarely centrals in large halos, we modify this to

\begin{equation}
\begin{split}
    \langle N_{\rm cen}^{\rm sf} \rangle = \frac{1}{4}\bigg(1+\mathrm{erf}\Big( \frac{\log \Mh - \log M_{\rm min}^{\rm sf} }{\sigma_{\rm min}^{\rm sf}} \Big) \bigg) \times \\
    \bigg(1-\mathrm{erf}\Big( \frac{\log \Mh - \log M_{\rm max}^{\rm sf} }{\sigma_{\rm max}^{\rm sf}} \Big) \bigg),
\end{split}
\label{eq:ncen_sf}
\end{equation} 

\noindent introducing 2 more parameters in the form of another error function to attenuate the central occupancy at high halo masses. Again, $\langle N_{\rm cen}^{\rm sf} \rangle =0$ for $\Mh<M_0$. The model shape does not quite fit the high halo mass occupancy, but these are noisy bins for blue centrals. Otherwise the model fits well.

For the satellites, we use the 3-parameter form of \cite{zheng_etal:05},

\begin{equation}
    \langle N_{\rm sat} \rangle = \bigg( \frac{\Mh-M_{\rm cut}}{M_{\rm sat}} \bigg)^{\alpha}
    \label{eq:nsat}
\end{equation} 

\noindent which fits well for all the satellite samples. The fainter quiescent samples show some small differences in the shape that might require an additional soft exponential turn-off to fit more closely. The satellite slope parameter $\alpha$ is $\sim 1.1$ for most of the quiescent samples, and $\sim 0.8$ for most of the star-forming ones.

\begin{figure*}
    \centering
    \includegraphics[width=1.0\linewidth]{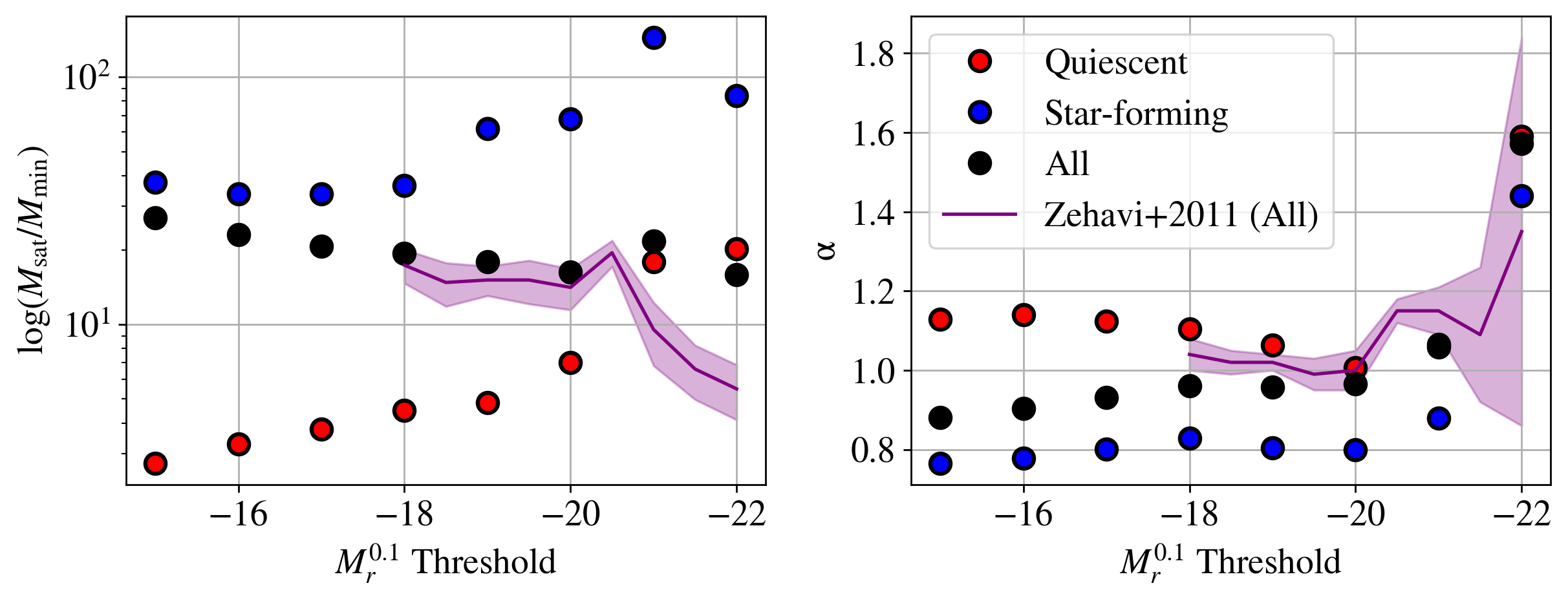}
    \caption{\textbf{Evolution in HOD Threshold Parameters.} \textit{Left panel:} We show the log ratio of $M_{\rm min}$ to $M_{\rm sat}$, which shows the relative halo mass scales where centrals vs. satellites become relevant. The results match that of \cite{zehavi_etal:11} except in the brightest samples. \textit{Right panel:} We show the satellite $\alpha$ parameter evolution. The HOD parameters are significantly different for the star-forming vs. quiescent samples.}
    \label{fig:hod_sat_params}
\end{figure*}

The parameters for all fits are shown in the plots and summarized in Table~\ref{tab:hod-thresh-params}. We also show the evolution in the parameters in Fig.~\ref{fig:hod_sat_params}, where we compare to \cite{zehavi_etal:11}. The left panel shows the log ratio of $M_{\rm min}$ to $M_{\rm sat}$, which shows the relative halo mass scales where centrals vs. satellites become relevant. The results match that of \cite{zehavi_etal:11} except in the brightest samples. The right panel shows the $\alpha$ parameter evolution. When broken down into the quiescent and star-forming subsamples, the parameters shown significant differences across luminosities. For comparison across surveys, recall from the previous Appendix that differences in luminosities exist, especially on the bright end, and matching number densities might be preferred.

\begin{table*}[]
\centering
\begin{tabular}{ccc|ccccccc}
\hline
$M_r$ & c & $\log (n_{\rm gal}/$[Mpc$^{-3}h^3$]) & $M_{\rm min}$ & $\sigma_{\rm min}$  & $M_{\rm max}$ & $\sigma_{\rm max}$ & $M_{\rm cut}$ & $M_{\rm sat}$ &$\alpha$ \\
\hline
$< -15$ & Q & -1.860 & 12.299 & 0.974 & -- & -- & 10.849 & 12.739 & 1.129 \\
$< -15$ & SF & -1.064 & 10.613 & 0.065 & 12.451 & 0.868 & 10.698 & 12.188 & 0.765 \\
$< -15$ & All & -1.000 & 10.610 & 0.053 & -- & -- & 10.696 & 12.041 & 0.880 \\
$< -16$ & Q & -1.906 & 12.265 & 0.920 & -- & -- & 11.059 & 12.781 & 1.141 \\
$< -16$ & SF & -1.235 & 10.794 & 0.072 & 12.449 & 0.871 & 10.973 & 12.322 & 0.779 \\
$< -16$ & All & -1.151 & 10.793 & 0.070 & -- & -- & 10.955 & 12.158 & 0.904 \\
$< -17$ & Q & -1.970 & 12.249 & 0.882 & -- & -- & 11.260 & 12.825 & 1.124 \\
$< -17$ & SF & -1.442 & 11.002 & 0.075 & 12.452 & 0.868 & 11.176 & 12.531 & 0.801 \\
$< -17$ & All & -1.329 & 11.001 & 0.066 & -- & -- & 11.163 & 12.318 & 0.932 \\
$< -18$ & Q & -2.054 & 12.246 & 0.843 & -- & -- & 11.462 & 12.900 & 1.104 \\
$< -18$ & SF & -1.683 & 11.226 & 0.071 & 12.453 & 0.867 & 11.395 & 12.788 & 0.828 \\
$< -18$ & All & -1.529 & 11.220 & 0.057 & -- & -- & 11.391 & 12.508 & 0.961 \\
$< -19$ & Q & -2.176 & 12.307 & 0.951 & -- & -- & 11.771 & 12.991 & 1.063 \\
$< -19$ & SF & -1.995 & 11.332 & 0.497 & 12.472 & 0.852 & 11.763 & 13.124 & 0.805 \\
$< -19$ & All & -1.775 & 11.455 & 0.107 & -- & -- & 11.756 & 12.708 & 0.957 \\
$< -20$ & Q & -2.423 & 12.364 & 0.865 & -- & -- & 12.246 & 13.211 & 1.006 \\
$< -20$ & SF & -2.456 & 11.853 & 0.094 & 12.467 & 0.855 & 12.196 & 13.683 & 0.799 \\
$< -20$ & All & -2.138 & 11.849 & 0.077 & -- & -- & 12.194 & 13.062 & 0.965 \\
$< -21$ & Q & -2.908 & 12.566 & 0.114 & -- & -- & 12.741 & 13.819 & 1.065 \\
$< -21$ & SF & -3.170 & 12.334 & 0.095 & 12.481 & 0.841 & 12.730 & 14.494 & 0.879 \\
$< -21$ & All & -2.719 & 12.402 & 0.136 & -- & -- & 12.691 & 13.741 & 1.058 \\
$< -22$ & Q & -3.757 & 13.365 & 0.197 & -- & -- & 13.184 & 14.671 & 1.590 \\
$< -22$ & SF & -4.292 & 13.237 & 0.349 & 12.932 & 0.517 & 13.282 & 15.161 & 1.440 \\
$< -22$ & All & -3.645 & 13.426 & 0.444 & -- & -- & 13.175 & 14.628 & 1.573 \\
\hline
\end{tabular}
\caption{\textbf{Threshold HOD Parameters.} Shown are the 5 or 7 parameters used fits to the HOD threshold samples defined by the $M_r$ (really $M_r^{0.1}-5\log h$) and $c$ columns (Q is quiescent, SF is star-forming). The comoving log number density of the sample, calculated using the $1/V_{\rm max}$ method is also shown.}
\label{tab:hod-thresh-params}
\end{table*}

We also examine the HODs in luminosity bins instead of thresholds. To model $\langle N_{\rm cen} \rangle$, we reuse the form of Eq.~\ref{eq:ncen_sf} which allows the occupation to turn on and off smoothly and asymmetrically. As with the central occupancy for thresholds, the model is 0 above and below the maximum and minimum halo bin that measured any occupancy, as is required for the model's number density and $\fsat$ to agree with the group catalog. However, for visual clarity we do not truncate the model (black line) in the plots here. For$\langle N_{\rm sat} \rangle$, we use the same form as in the thresholds, Eq.~\ref{eq:nsat}. 

\begin{figure*}
    \centering
    \includegraphics[width=1.0\linewidth]{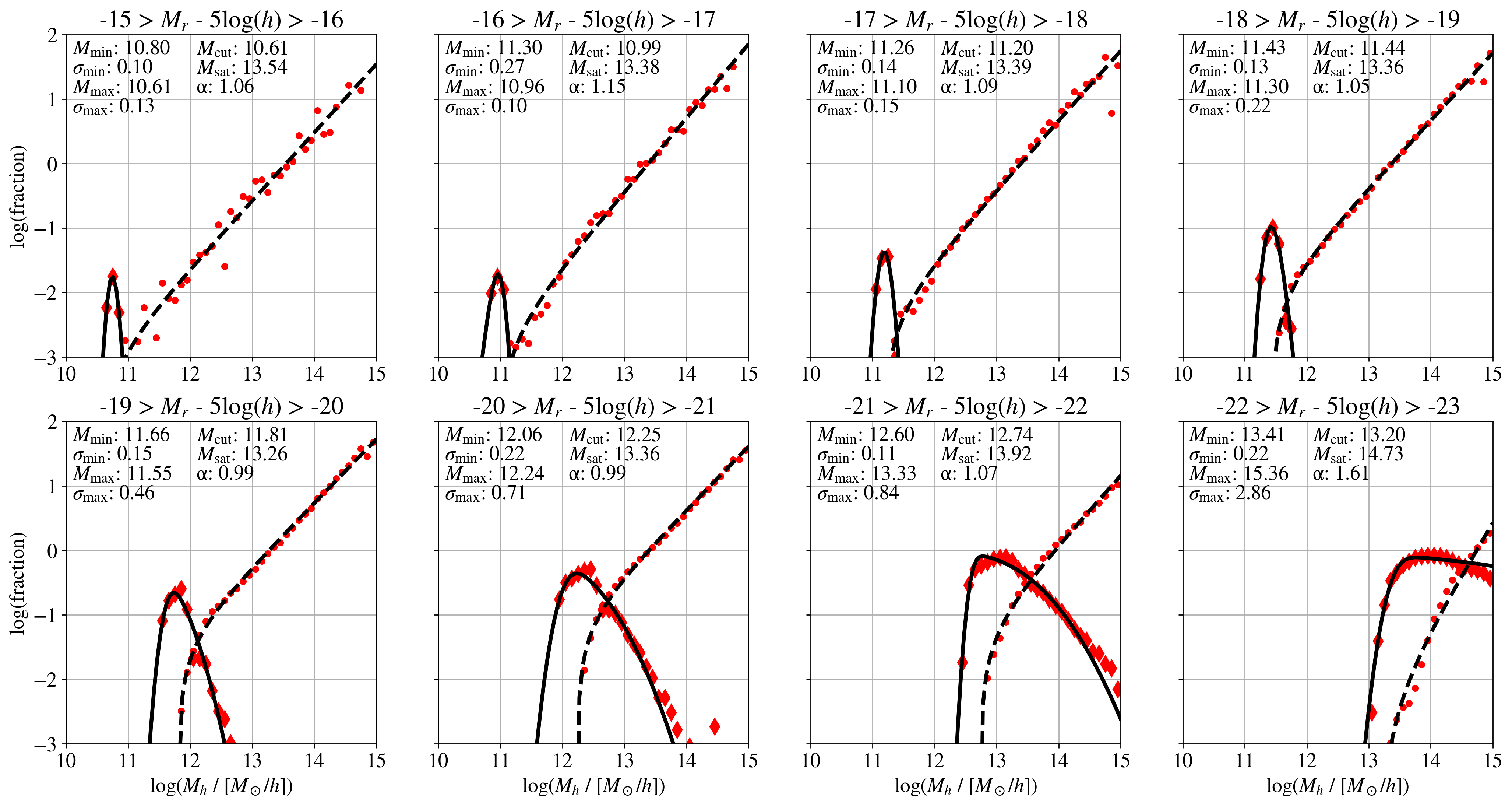}
    \caption{\textbf{Quiescent HOD.} Taken from the final catalog, the red diamonds show the percent of halos at a given mass that host a quiescent galaxy within the labeled luminosity range. The red dots show the satellite occupancy. The black lines are fits to these tabulated values using standard HOD parameterizations (see text). The raw data are noisy for faint galaxies in large, rare halos; when mock building we fit the slope of the high mass end for satellites and use that instead. }
    \label{fig:hod_q}
\end{figure*}

\begin{figure*}
    \centering
    \includegraphics[width=1.0\linewidth]{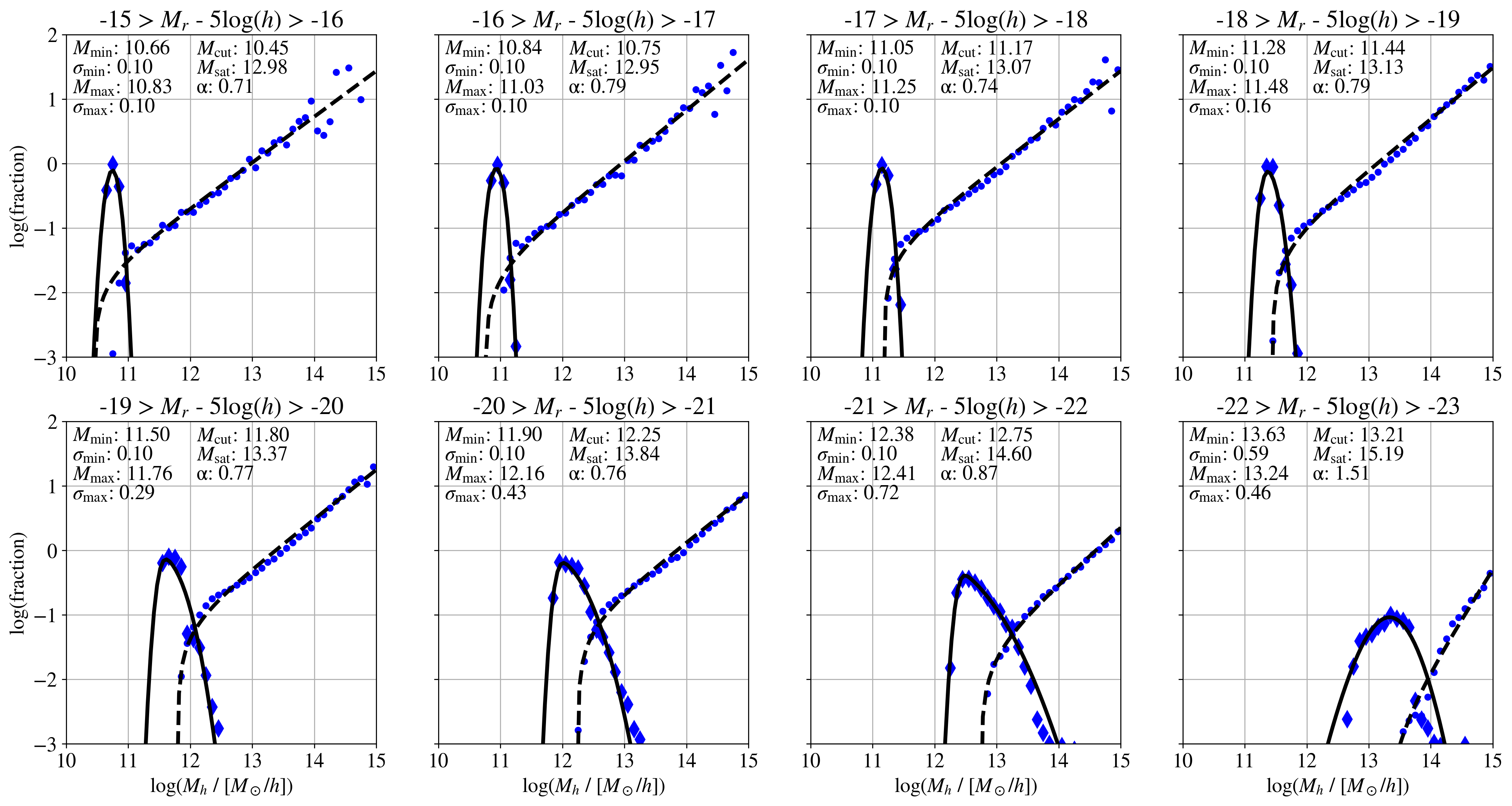}
    \caption{\textbf{Star-forming HOD.} Halo occupancy for star-forming galaxies,  from the final catalog. See Fig.~\ref{fig:hod_q} for a description.}
    \label{fig:hod_sf}
\end{figure*}

Fig.~\ref{fig:hod_q} and Fig.~\ref{fig:hod_sf} show these results. The central HOD fits appear to miss the peak in several bins. However, the fits are more realistic than the raw tabulated values because of unrealistically low scatter described in \S\ref{sec:lhmr}, which leads to artificially sharp peaks in central occupation. The satellite model fits well. The parameters for the binned HOD fits are shown in the plots and summarized in Table~\ref{tab:hod-bins-params}.

\begin{table*}[]
    \centering
\begin{tabular}{ccc|ccccccc}
\hline
$M_r$ & c & $\log (n_{\rm gal}/$[Mpc$^{-3}h^3$]) & $M_{\rm min}$ & $\sigma_{\rm min}$  & $M_{\rm max}$ & $\sigma_{\rm max}$ & $M_{\rm cut}$ & $M_{\rm sat}$ &$\alpha$ \\
\hline
-15 to -16 & Q & -2.858 & 10.804 & 0.102 & 10.615 & 0.135 & 10.611 & 13.543 & 1.057 \\
-15 to -16 & SF & -1.552 & 10.658 & 0.100 & 10.831 & 0.100 & 10.451 & 12.976 & 0.711 \\
-16 to -17 & Q & -2.767 & 11.295 & 0.269 & 10.960 & 0.100 & 10.988 & 13.384 & 1.147 \\
-16 to -17 & SF & -1.656 & 10.837 & 0.100 & 11.035 & 0.100 & 10.750 & 12.949 & 0.786 \\
-17 to -18 & Q & -2.726 & 11.258 & 0.135 & 11.095 & 0.148 & 11.202 & 13.388 & 1.085 \\
-17 to -18 & SF & -1.814 & 11.051 & 0.100 & 11.251 & 0.103 & 11.169 & 13.069 & 0.744 \\
-18 to -19 & Q & -2.665 & 11.427 & 0.127 & 11.303 & 0.219 & 11.440 & 13.363 & 1.047 \\
-18 to -19 & SF & -1.973 & 11.281 & 0.100 & 11.481 & 0.158 & 11.442 & 13.128 & 0.794 \\
-19 to -20 & Q & -2.539 & 11.655 & 0.146 & 11.549 & 0.460 & 11.815 & 13.255 & 0.989 \\
-19 to -20 & SF & -2.179 & 11.499 & 0.100 & 11.762 & 0.291 & 11.798 & 13.366 & 0.766 \\
-20 to -21 & Q & -2.595 & 12.059 & 0.218 & 12.239 & 0.709 & 12.246 & 13.365 & 0.987 \\
-20 to -21 & SF & -2.550 & 11.902 & 0.100 & 12.156 & 0.430 & 12.246 & 13.835 & 0.759 \\
-21 to -22 & Q & -2.975 & 12.603 & 0.112 & 13.331 & 0.838 & 12.742 & 13.917 & 1.072 \\
-21 to -22 & SF & -3.204 & 12.379 & 0.100 & 12.405 & 0.723 & 12.745 & 14.595 & 0.872 \\
-22 to -23 & Q & -3.772 & 13.406 & 0.216 & 15.360 & 2.858 & 13.200 & 14.726 & 1.606 \\
-22 to -23 & SF & -4.300 & 13.626 & 0.588 & 13.236 & 0.455 & 13.210 & 15.192 & 1.508 \\
\hline
\end{tabular}
    \caption{\textbf{Binned HOD Parameters.} Shown are the 7 parameters used for fitting the binned HOD, defined by the $M_r$ (really $M_r^{0.1}-5\log h$) and $c$ columns (Q is quiescent, SF is star-forming). The comoving log number density of the sample, calculated using the $1/V_{\rm max}$ method is also shown.}
    \label{tab:hod-bins-params}
\end{table*}

Compared to the binned results of \cite{zehavi_etal:11}, we find that the central occupancy of the brighter bins is more massive halos for quiescent than star-forming galaxies, a result that was not allowed in that work (comparing the -20 to -21 bin specifically). This is driven by the $\lsat$ calibration data. The asymmetry of central occupation is a noticeable difference as well, but this is largely driven by insufficient scatter on the left (low-mass) side of the central's halo occupancy distributions. On the satellite side, different $\alpha$ values are required from the standard value of 1.0 \cite{zehavi_etal:11}, especially for the star-forming sample where we find $\alpha =0.7-0.8$.

\end{document}